\documentclass[11pt,a4paper]{article}
\usepackage{amssymb}
\usepackage{graphicx}
\usepackage{amsmath}
\usepackage{makeidx}
\usepackage{indentfirst}
\usepackage[T1]{fontenc}
\usepackage[utf8]{inputenc}

\newcounter{assumptionnum}[section]
\newcounter{resultnum}[section]
\newcounter{consequencenum}[section]
\newcounter{conclusionnum}[section]
\newcounter{conditionnum}[section]
\newcounter{conjecturenum}[section]
\newcounter{examplenum}[section]
\newcounter{exercisenum}[section]
\newcounter{lemmanum}[section]
\newcounter{notationnum}[section]
\newcounter{theoremnum}[section]
\newcounter{definitionnum}[section]
\newcounter{corollarynum}[section]
\newcounter{remarknum}[section]
\newcounter{propositionnum}[section]
\newcounter{acknowledgementnum}[section]
\newcounter{algorithmnum}[section]
\newcounter{axiomnum}[section]
\newcounter{casenum}[section]
\newcounter{claimnum}[section]
\newcounter{summarynum}[section]
\newcounter{problemnum}[section]
\begin{document}

\title{Deforming Black Hole and Cosmological Solutions by Quasiperiodic
and/or Pattern Forming Structures in Modified and Einstein Gravity}
\date{May 7, 2018}
\author{ \vspace{.1 in}{\ \textbf{Lauren\c{t}iu Bubuianu}}\thanks{%
email: laurentiu.bubuianu@tvr.ro } \\
{\small \textit{TVR Ia\c{s}i, \ 33 Lasc\v{a}r Catargi street; and University
Apollonia, 2 Muzicii street, 700107 Ia\c{s}i, Romania }}\\
${}$ \\
\vspace{.1 in} \textbf{Sergiu I. Vacaru}\thanks{%
emails: sergiu.vacaru@gmail.com ; sergiuvacaru@mail.fresnostate.edu \newline
\textit{Address for post correspondence:\ } 67 Lloyd Street South,
Manchester, M14 7LF, the UK}\\
{\small \textit{Physics Department, California State University Fresno,
Fresno, CA 93740, USA}} \\
{\small and } \ {\small \textit{\ Project IDEI, University "Al. I. Cuza"
Ia\c si, Romania}} }
\maketitle

\begin{abstract}
We elaborate on the anholonomic frame deformation method, AFDM, for
constructing exact solutions with quasiperiodic structure in modified
gravity theories, MGTs, and general relativity, GR. Such solutions are
described by generic off-diagonal metrics, nonlinear and linear connections
and (effective) matter sources with coefficients depending on all spacetime
coordinates via corresponding classes of generation and integration
functions and (effective) matter sources. There are studied effective free
energy functionals and nonlinear evolution equations for generating
off-diagonal quasiperiodic deformations of black hole and/or homogeneous
cosmological metrics. The physical data for such functionals are stated by
different values of constants and prescribed symmetries for defining
quasiperiodic structures at cosmological scales, or astrophysical objects in
nontrivial gravitational backgrounds some similar forms as in condensed
matter physics. It is shown how quasiperiodic structures determined by
general nonlinear, or additive, functionals for generating functions and
(effective) sources may transform black hole like configurations into
cosmological metrics and inversely. We speculate on possible implications of
quasiperiodic solutions in dark energy and dark matter physics. Finally, it
is concluded that geometric methods for constructing exact solutions consist
an important alternative tool to numerical relativity for investigating
nonlinear effects in astrophysics and cosmology.

\vskip7pt

\textbf{Keywords:} Geometric methods in physics; mathematical relativity and
cosmology; quasiperiodic solutions in gravity; Lyapunov type free energy
and/or entropy type functionals; deformed black hole solutions;
inhomogeneous and locally anisotropic solutions in gravity; dark energy and
dark matter.

\vskip3pt

PACS:\ 47.54.-r; 05.45.-a; 47.35.Pq; 47.52.+j; 61.50.Ah, 61.44.Br; 05.20.-y;
64.70.D-
\end{abstract}

\tableofcontents

\section{Introduction}

A series of recent works has been devoted to constructing black hole, BH,
and cosmological solutions encoding quasiperiodic gravitational and matter
fields interactions and geometric flow structures \cite%
{vplb16,grvvap16,rajpoot17a,biv16p,viepjc17,abvfirs16p,aabovw16p}. We
elaborated a new geometric and analytic method for generating exact and
parametric solutions of (modified) Einstein equations characterized by
generic off-diagonal metrics and (generalized) connections with coefficients
depending on all spacetime coordinates. It is called the anholonomic frame
deformation method, AFDM; for reviews of results, see \cite%
{vexsol98,vjhep01,vpcqg01,vjmp05,vmon06,vijgmmp07,vijgmmp11}. This geometric
formalism provides an alternative to numerical methods and simulations in
modern gravity \cite{pretorius05,lehner15,campanelli06,baker06}, see a
review of such results in \cite{macpherson17}. In our works, there were
considered various geometric applications in particle physics and cosmology
and heterotic superstring theory, and studied various models of commutative
and noncommutative and/or supersymmetric modified gravity theories, MGT, and
general relativity, GR, see \cite%
{vgrg12,vcsf12,vepjp12,vepjc13,vepjc14a,gvvepjc14,gvvcqg15} and references
therein.

The AFDM was formulated by developing a geometric techniques of decoupling
and integrating systems of nonlinear partial differential equations, PDEs.
The general goal was to study nonlinear dynamical and/or evolution
properties of (modified) Einstein equations, when various classes of
solutions are generated by nonholonomic (i.e. non-integrable, equivalently,
anholonomic) frame transforms with deformed symmetries and distortions of
metric and connection structures. In this paper, we consider that readers
are familiar with main concepts and methods of differential geometry and
topology, functional analysis and theory of partial differential equations,
PDEs, and ordinary differential equations, ODEs, with applications in
mathematical relativity, astrophysics and cosmology (summarized in \cite%
{misner73,hawking73,wald82,kramer03,griffith09}). The most important feature
of the AFDM is that it allows us to construct in direct form exact solutions
of nonlinear systems when there are not involving simplified diagonal metric
ansatz resulting in ordinary differential equations, ODEs. We can study
singular and nonsingular metrics and generalized connections, quasiperiodic
(non) stationary gravitational and matter field structures, nonlinear waves
and solitons, deformed horizons, nontrivial topological structures etc.
Additionally, one can be analysed the conditions when and if some classes of
"general solutions" (with less clear physical properties/ implications) may
have certain (non) smooth limits and/or nonholonomic constraints to some
well-known and more "simple" physically important solutions.

We emphasize that new classes of exact and parametric quasiperiodic and
pattern forming solutions in MGTs and GR generated following the AFDM (in
this and partner works \cite{biv16p,abvfirs16p,aabovw16p}) can not be
constructed by applying former methods with diagonal metric ansatz, for
instance, depending only on a "radial" like coordinate, for BHs, or on a
time like coordinate, for homogeneous cosmological solutions. Prescribing
from the very beginning some "simplified" diagonal ansatz, we can transform
the gravitational and matter field equations into certain systems of
nonlinear ODEs which can be solved in general forms with metrics
parameterized by integration constants. This way, we lose various
possibilities to find more general classes of nonlinear solutions, for
instance, with solitonic hierarchies, pattern forming structure etc. which
are determined by generating and integration functions. More realistic
descriptions and understanding of physical properties of realistic nonlinear
gravitational and matter field systems are possible in terms of generalized
ansatz for coefficients of metrics, frames, and (non) linear connections,
and off-diagonal solutions of PDEs depending on a maximal possible number of
spacetime coordinates and various (non) commutative continuous/ discrete
parameters.

The solutions for gravitational and matter fields with quasiperiodic
structures are defined by generic off-diagonals metrics, nonlinear and
(generalized) connections, and (effective) matter sources with coefficients
determined by respective classes of generating and integration functions
depending on all spacetime coordinates.\footnote{%
A generic off-diagonal metric can not be diagonalized by coordinate
transforms on an open spacetime region. Generalized connections can be (non)
linear, metric (non) compatible and with nontrivial torsion. Additional
constraints can be imposed in order to extract Levi-Civita configurations
for metric compatible and zero torsion connections.} Prescribing for
physical and geometric objects corresponding smooth / singularity/ symmetry
conditions, solving corresponding Cauchy problems, and/ or satisfying
necessary boundary conditions, we can model various (non) commutative and/or
semi-classical and quantum stationary configurations and/or nonlinear
gravitational and matter field interactions. Such models of generalized
off-diagonal spacetimes are characterized by nonlinear / discrete /
continuous (semi) classical symmetries; there are computed locally
anisotropically polarized (physical) constants and stated values of
integration parameters; in general, such constructions can be with (non)
singular and/or topologically nontrivial structures. In mentioned partner
works \cite{biv16p,abvfirs16p,aabovw16p} (see also citations therein) and
related papers on applications of the AFDM, various examples of exact and
parametric solutions, for instance, with soliton distributions and nonlinear
waves, partial derivative and gravitational diffusions processes, geometric
flows, and quasicrystal like structures, locally anisotropic cosmological
configurations etc., have been considered.

The goal of this paper is to analyze the conditions when certain data for
generating and integration functions, (effective) sources, prescribed
symmetries for integration functions and various type parameters, define
off-diagonal deformations of physically important solutions. We study
nonholonomic deformations of BH and cosmological metrics into new classes of
exact solutions encoding quasiperiodic and pattern forming gravitational and
effective matter fields. Our work reconsiders a series of ideas, methods,
and results elaborated in condensed matter physics and develop them with
applications in modern gravity, cosmology, and astrophysics. On relevant
former results, we cite a series of papers for the physics of quasicrystals
and Penrose-like telings, see \cite%
{qsnp,qcalloys4,penrose1,penrose2,boyle,steinh1,steinh2,steinh3,achim,qcalloys1,qcalloys2,qcalloys3,rucklidge12,rucklidge16}%
). It is considered that various web like, filament, quasiperiodic,
aperiodic, singular and nonsigular cosmological evolution, solitonic like
distributions and nonlinear wave structures are distinguished /found in
modern cosmological observational data \cite%
{crystalinks,starob,rucklidge13,rucklidge15,ruchin17,rajpoot17,diemer17,gurzadyan13,cross09,kooistra17}%
. Such generic locally anisotropic and inhomogeneous cosmological
configurations can not be explained exhaustively using only higher symmetry
solutions (for instance, Friedmann-Lema\^{\i}tre-Roberstson-Walker, FLRW,
and/or Bianchi type metrics) depending only on time like coordinates and
constructed following methods of the theory of ODEs. Applying the AFDM, we
are able to construct more general classes of exact solutions in MGTs and GR
which allows us to model more rich spacetime structures and nonlinear field
interactions as solutions of systems of nonlinear PDEs. In this work, it is
shown how such nonlinear configurations are characterized by effective free
energy functionals and nonlinear evolutions equations, which can be applied
to mimic dark energy and dark matter effects, see also \cite%
{rucklidge12,rucklidge16,grvvap16,rajpoot17a,biv16p,viepjc17,abvfirs16p,aabovw16p}%
. We study quasiperiodic solutions not involving small deformation
parameters.

The content of the paper is organized as follows:\ \ In section \ref{safdm},
we provide a brief review of the AFDM (relevant details with necessary
N-adapted coefficient formulas and proofs are presented in Appendix \ref{as1}%
). Free energy functionals for generating functions resulting in
quasiperiodic and pattern forming structures for smooth, singular and BH
like structures, and cosmological solutions etc. are defined and analysed in
section \ref{sqpf}. We construct a series of examples of stationary and
nonstationary exact solutions describing quasiperiodic deformations of BH
solutions in MGTs and GR in section \ref{sbh}. There are analyzed different
classes of off-diagonal solutions determined by nonlinear or additive
functionals for generating functions and effective sources. Various classes
of locally anisotropic and/or inhomogeneous cosmological quasiperiodical
solutions are constructed in general form and studied in section \ref{scc}.
Such cosmological evolution models are determined by nonstationary
quasiperiodic generating functionals and effective sources. Finally, we
discuss and provide concluding remarks in section \ref{sconcl}.

\section{A Brief Review of the AFDM}

\label{safdm}We outline and reformulate the AFDM as a geometric formalism
for constructing exact generic off-diagonal stationary and cosmological
solutions in MGTs and GR determined by generation and integration functions
and effective matter sources, and related polarization functions. The method
and most important formulas are summarized in Tables 1-3. Such constructions
will be applied in order to extend the approach for solutions encoding
quasiperiodic and/or pattern forming, solitonic distributions and nonlinear
wave gravitational and matter field structures. Details, examples with small
parametric decompositions and rigorous mathematical proofs are contained in
\cite%
{vexsol98,vjhep01,vpcqg01,vjmp05,vmon06,vijgmmp07,vijgmmp11,vgrg12,vcsf12,vepjp12,vepjc13,vepjc14a,gvvepjc14,gvvcqg15}%
, see also a summary of necessary N-adapted formulas in Appendix \ref{as1}.

\subsection{Geometric preliminaries: Lorentz manifolds with nonholonomic 2+2
splitting}

Let us consider a 4-d Lorentzian manifold $V$ of signature $(+++-),$ when
the local coordinates are labeled in a form adapted to a conventional 2+2
splitting, when $u=(x,y)=\{u^{\alpha }=$ $(x^{i},y^{a})\}$, for $\alpha
=(i,a);\beta =(j,b),$ where $i,j,...=1,2$ and $a,b,...=3,4,$ with $y^{4}=t$
being a time like coordinate. For a 3+1 splitting, we can parameterize $%
u^{\alpha }=$ $(u^{\grave{\imath}},t)$ for spacelike coordinates $u^{\grave{%
\imath}}=(x^{i},y^{3})$ when $\grave{\imath},\grave{j},\grave{k},...=1,2,3$.
Such a spacetime manifold can be enabled with a pseudo-Riemannian metric
structure $\mathbf{g}=\widehat{\mathbf{g}}$,
\begin{eqnarray}
\mathbf{g} &=&g_{\alpha \beta }(x^{i},y^{a})du^{\alpha }\otimes du^{\beta },%
\mbox{  for dual frame coordinate basis }du^{\alpha };  \label{mcoord} \\
\widehat{\mathbf{g}} &=&\mathbf{g}_{\alpha \beta }(u)\mathbf{e}^{\alpha
}\otimes \mathbf{e}^{\beta }=\mathbf{g}_{i}(x^{k})dx^{i}\otimes dx^{i}+%
\mathbf{g}_{a}(x^{k},y^{b})\mathbf{e}^{a}\otimes \mathbf{e}^{b},  \label{dm}
\\
&&\mbox{for }\mathbf{e}^{\alpha }=(dx^{i},\mathbf{e}%
^{a}=dy^{a}+N_{i}^{a}(u^{\gamma })dx^{i})%
\mbox{ defining a N-adapted dual
frame basis}.  \label{ndfr}
\end{eqnarray}%
The 2+2 splitting (\ref{dm}) is nonholonomic (equivalently, non-integrable,
or anholonomic). This follows from the conditions that the co-basis $\mathbf{%
e}^{\alpha }=(dx^{i},\mathbf{e}^{a})$ (\ref{ndfr}) is dual to
\begin{equation}
\mathbf{e}_{\alpha }=(\mathbf{e}_{i},e_{a})=(\mathbf{e}_{i}=\partial
/\partial x^{i}-N_{i}^{a}(u)\partial /\partial y^{a},e_{a}=\partial
_{a}=\partial /\partial y^{a}),  \label{nfr}
\end{equation}%
which satisfies nonholonomy conditions
\begin{equation}
\mathbf{e}_{[\alpha }\mathbf{e}_{\beta ]}:=\mathbf{e}_{\alpha }\mathbf{e}%
_{\beta }-\mathbf{e}_{\beta }\mathbf{e}_{\alpha }=C_{\alpha \beta }^{\gamma
}(u)\mathbf{e}_{\gamma },  \label{anhr}
\end{equation}%
with anholonomy coefficients $\ C_{\alpha \beta }^{\gamma
}=\{C_{ia}^{b}=\partial _{a}N_{i}^{b},C_{ji}^{a}=\mathbf{e}_{j}N_{i}^{a}-
\mathbf{e}_{i}N_{j}^{a}\}.$\footnote{%
If the anholonomy coefficients are nontrivial (i.e. the N-adapted frame is
not integrable and can not be transformed into a coordinate one), the
respective equivalent metric parameterizations (\ref{mcoord}) and (\ref{dm})
are generic off-diagonal.} In above formulas, we can consider necessary type
frame transforms $\mathbf{e}^{\alpha }=\mathbf{e}_{\ \alpha
^{\prime}}^{\alpha }(u)du^{\alpha ^{\prime }},$ when $g_{\alpha ^{\prime
}\beta ^{\prime }}(u)=g_{\alpha \beta }\mathbf{e}_{\ \alpha ^{\prime
}}^{\alpha }\mathbf{e}_{\ \beta ^{\prime }}^{\beta }.$

A set of coefficients $\mathbf{N}=\{N_{i}^{a}\}$ from (\ref{dm}) and/or (\ref%
{nfr}) states a N--adapted decomposition of the tangent Lorentz bundle $T%
\mathbf{V}$ into conventional horizontal, h, and vertical, v, subspaces,
when the Whitney sum
\begin{equation}
\mathbf{N}:T\mathbf{V}=hT\mathbf{V}\oplus vT\mathbf{V}  \label{ncon}
\end{equation}%
define a nonlinear connection (N-connection) structure. A Lorentz manifold $%
V $ is called nonholonomic if it is enabled with a nonholonomic
distribution. In a particular case, we shall write $\mathbf{V}=(V,\mathbf{g,N%
})$ for a nonholonomic spacetime manifold enabled with N-connection
structure $\mathbf{N}$ (\ref{ncon}). This concept was introduced and studied
in details in generalized Finsler geometry and geometric and physical models
on (co) vector/ tangent bundles and their supersymmetric/noncommutative
generalizations, see \cite{vmon06} and references therein. For (pseudo)
Riemannian manifolds, a conventional $h$-/ $v$-splitting can be considered,
for instance, in order to state a fibered structure with $y^{a}$ coordinates
and N-adapted frames (\ref{nfr}). Such nonholonomic structures and
deformations of geometric/ physical objects (like metrics, linear
connections, (effective) sources etc.) can be introduced in such forms which
allows decoupling of gravitational and matter field equations in MGTs and
GR, \cite{vplb16,grvvap16,rajpoot17a,biv16p,viepjc17,abvfirs16p,aabovw16p}.%
\footnote{%
We shall follow the Einstein rule on summation on "up-low" cross indices if
there is not a contrary statement for an explicit formula. One shall not be
considered summation on repeating "low-low", or "up-up" indices. In our
works , we elaborated a system of "N--adapted notations" with boldface
symbols for manifolds and fiber bundles enables with a nontrivial
N--connection structure $\mathbf{N.}$ Details on nonholonomic differential
geometry and N-connections are explained in \cite%
{vjmp05,vmon06,vijgmmp07,vijgmmp11,vgrg12,vcsf12,vepjp12,vepjc13,vepjc14a,gvvepjc14,gvvcqg15}
and references therein. If the anholonomy coefficients $C_{\alpha
\beta}^{\gamma }$ in \ (\ref{anhr}) are nontrivial, a metric $g_{\alpha
\beta }$ (\ref{mcoord}) can not be diagonalized in a local finite, or
infinite, spacetime region with respect to coordinate frames. Such metrics
are called generally off-diagonal and characterized by six independent
nontrivial coefficients from a set $\mathbf{g}=\{g_{\alpha \beta }(u)\}.$ A
nonholonomic frame is holonomic if all corresponding anholonomy coefficients
are zero (for instance, the coordinate frames).}

It is possible to model a nonholonomic deformation with $\eta $-polarization
functions, \ $\mathbf{\mathring{g}\rightarrow }\widehat{\mathbf{g}},$ of a
'prime' metric, $\mathbf{\mathring{g}}$, into a 'target' metric $\mathbf{g}=%
\widehat{\mathbf{g}}$ (\ref{dm}), if
\begin{equation}
\widehat{\mathbf{g}}=\eta _{i}(x^{k})\mathring{g}_{i}dx^{i}\otimes
dx^{i}+\eta _{a}(x^{k},y^{b})\mathring{h}_{a}\mathbf{e}^{a}[\eta ]\otimes
\mathbf{e}^{a}[\eta ],  \label{dme}
\end{equation}%
where the N-elongated basis (\ref{nfr}) is represented for $%
N_{i}^{a}(u)=\eta _{i}^{a}(x^{k},y^{b})\mathring{N}_{i}^{a}(x^{k},y^{b})$,
i.e. in the form\footnote{%
in this paragraph, we do not consider summation on repeating indices if they
are not written as contraction of "up-low" ones} $\mathbf{e}^{\alpha
}[\eta]=(dx^{i},\mathbf{e}^{a}=dy^{a}+\eta _{i}^{a}\mathring{N}%
_{i}^{a}dx^{i}).$ We shall subject a $\widehat{\mathbf{g}}$ \ to the
condition that it defines a solution of (modified) Einstein equations. A
general prime metric in coordinate parametrization of type (\ref{mcoord}),
\begin{equation*}
\mathbf{\mathring{g}}=\mathring{g}_{\alpha \beta }(x^{i},y^{a})du^{\alpha
}\otimes du^{\beta }
\end{equation*}
can be also represented in N-adapted form
\begin{eqnarray}
\mathbf{\mathring{g}} &=&\mathring{g}_{\alpha }(u)\mathbf{\mathring{e}}%
^{\alpha }\otimes \mathbf{\mathring{e}}^{\beta }=\mathring{g}%
_{i}(x)dx^{i}\otimes dx^{i}+\mathring{g}_{a}(x,y)\mathbf{\mathring{e}}%
^{a}\otimes \mathbf{\mathring{e}}^{a},  \label{primedm} \\
&&\mbox{ for }\mathbf{\mathring{e}}^{\alpha }=(dx^{i},\mathbf{e}^{a}=dy^{a}+%
\mathring{N}_{i}^{a}(u)dx^{i}), \mbox{ and }\mathbf{\mathring{e}}_{\alpha }=(%
\mathbf{\mathring{e}}_{i}=\partial /\partial y^{a}-\mathring{N}%
_{i}^{b}(u)\partial /\partial y^{b},\ {e}_{a}=\partial /\partial y^{a}).
\notag
\end{eqnarray}%
It can be, or not, a solution of some gravitational field equations in a MGT
or GR.

In this work, we shall be interested in two physically important cases when $%
\mathbf{\mathring{g}}$ (\ref{primedm}) defines a BH solution (for instance,
a vacuum Kerr, or Schwarzschild, metric), or a Friedman--Lema\^{\i}%
tre--Robertson--Walker (FLRW) type metric. For such diagonalizable metrics
(the off-diagonal structure of the Kerr metric is determined by rotation
frames and coordinates), we can always find a coordinate system when $%
\mathring{N}_{i}^{b}=0.$ In order to avoid singular noholonomic
deformations, it is convenient to construct exact solutions of necessary
type gravitational equations with nontrivial conventional "polarization"
functions $\eta _{\alpha }=(\eta _{i},\eta _{a}),\eta _{i}^{a},$ and nonzero
coefficients $\mathring{N}_{i}^{b}(u).$ This can be achieved by considering
necessary type frame/ coordinate transforms. Having constructed an explicit
form a d-metric (\ref{dme}), we can study the existence and geometric/
physical properties of solutions, for instance, when $\eta
_{\alpha}\rightarrow 1$ and $N_{i}^{a}\rightarrow \mathring{N}_{i}^{a},$ or
if $\eta _{\alpha }=1$ and/or $\mathring{N}_{i}^{a}=0$ are imposed as some
noholonomic constraints.\footnote{\label{fnsmallp}It is possible to keep a
physical interpretation of a target metric $\widehat{\mathbf{g}}$ (\ref{dme}%
) with generic off-diagonal terms, as an "almost" BH, or FLRW cosmological,
like metric by constructing parametric solutions with small nonholonomic
deformations on some constant parameters $\eta _{\alpha }=(\eta _{i},\eta
_{a}),\eta _{i}^{a},$ for $0\leq \varepsilon _{\alpha },\varepsilon
_{i}^{b}\ll 1,$ when $\eta _{i}\simeq \check{\eta}_{i}(x^{k})[1+\varepsilon
_{i}\chi _{i}(x^{k})]\simeq 1+\varepsilon _{i}\chi _{i}(x^{k}),\eta
_{a}\simeq \check{\eta}_{a}(x^{k},y^{b})[1+\varepsilon _{a}\chi
_{a}(x^{k},y^{b})]\simeq 1+\varepsilon _{a}\chi _{a}(x^{k},y^{b})$, and $%
\eta _{i}^{a}\simeq \check{\eta}_{i}^{a}(x^{k},y^{b})[1+\varepsilon
_{i}^{a}\chi _{i}^{a}(x^{k},y^{b})]\simeq 1+\varepsilon _{i}^{a}\chi
_{i}^{a}(x^{k},y^{b})$. Such parametric $\varepsilon $-decompositions can be
performed in a self-consistent form by omitting quadratic and higher terms
after a class of solutions have been found for some general data $(\eta
_{\alpha },\eta _{i}^{a}).$ For certain subclasses of solutions, we can
consider that $\varepsilon _{i},\varepsilon _{a},\varepsilon _{i}^{a}\sim
\varepsilon $, when one small parameter is considered for all coefficients
of nonholonomic deformations. It is possible to works with mixed types of
solutions and model only small diagonal deformations $\varepsilon
_{i},\varepsilon _{a},\sim \varepsilon $ of metrics, for some general $\eta
_{i}^{a},$ or to work with some nontrivial $\eta _{\alpha }$ but $%
\varepsilon _{i}^{a}$ $\sim \varepsilon .$ In particular, one can be
generated various classes of nonholonomic small deformations of solutions
like in Refs. \cite%
{vplb16,grvvap16,rajpoot17a,biv16p,viepjc17,abvfirs16p,aabovw16p}. In this
work, our goal is to construct and study quasiperiodic solutions with
generation functions and sources without small parameters.} We denote
certain nonholonomic deformations of a prime d-metrics into a target one as $%
\mathbf{\mathring{g}}\rightarrow \widehat{\mathbf{g}}=[g_{\alpha }=\eta
_{\alpha }\mathring{g}_{\alpha },\ \eta _{i}^{a}\mathring{N}_{i}^{a}]. $
Here we emphasize that one constructs, in general, different classes of
solutions for nonlinear systems of PDEs, if such approximations are
considered in (modified) Einstein equations before finding solutions, or at
the end (after a class of solutions has been constructed in explicit form).
This is an important property of nonlinear dynamical / evolution systems
which can be subjected to additional nonholonomic constraints.

The standard formulation of (pseudo) Riemannian geometry is in terms of the
Levi-Civita, LC, connection $\nabla ,$ which (by definition) is metric
compatible and with zero torsion. Nevertheless, any Lorentzian manifold $V$
can be enabled additionally (following certain geometric/ physical
principles) with other types of linear connection structures considering any
$D=\nabla +Z,$ characterized by a respective distortion tensor $Z$. A
general linear connection $D$ can be metric noncompatible and/or with
nontrivial torsion. On a nonholonomic $\mathbf{V,}$ we can consider a
N-adapted variant of linear connection structure, called a \textsf{%
distinguished connection, d--connection}, $\mathbf{D}=(h\mathbf{D,}v\mathbf{%
D),}$ which is defined as a metric--affine (linear) connection preserving
under parallel transports the $N$--connection splitting into $h$- and $v$
-subspaces. We can define and compute for any d-connection $\mathbf{D}$ (in
standard form) the torsion tensor, $\mathcal{T}=\{\mathbf{T}_{\beta \gamma
}^{\alpha }\},$ and the curvature tensor, $\mathcal{R}=\{\mathbf{R}_{\ \beta
\gamma \delta }^{\alpha }\}$, where the coefficients can be written in
N-adapted form with respect to necessary tensor products of bases (\ref{nfr}%
) and their duals.

On a nonholonomic $\mathbf{V,}$ we can work in equivalent form with two
different linear connections:
\begin{equation}
(\mathbf{g,N})\rightarrow \left\{
\begin{array}{cc}
\mathbf{\nabla :} & \mathbf{\nabla g}=0;\ ^{\nabla }\mathcal{T}=0,%
\mbox{\
for  the LC--connection } \\
\widehat{\mathbf{D}}: & \widehat{\mathbf{D}}\mathbf{g}=0;\ h\widehat{%
\mathcal{T}}=0,v\widehat{\mathcal{T}}=0,hv\widehat{\mathcal{T}}\neq 0,%
\mbox{
for the canonical d--connection  },%
\end{array}%
\right.  \label{twocon}
\end{equation}%
where $\widehat{\mathbf{D}}=(h\widehat{\mathbf{D}},v\widehat{\mathbf{D}})$
is completely defined by $\mathbf{g}$ for any prescribed N--connection
structure $\mathbf{N.}$ In these formulas, we denote (respectively, for $%
\widehat{\mathbf{D}}$ and $\nabla )$ the torsions, $\widehat{\mathcal{T}}$
and $\ ^{\nabla }\mathcal{T}=0,$ and curvatures, $\widehat{\mathcal{R}}=\{%
\widehat{\mathbf{R}}_{\ \beta \gamma \delta }^{\alpha }\}$ and $\ ^{\nabla }%
\mathcal{R}=\{R_{\ \beta \gamma \delta }^{\alpha }\}$, which can be defined
and computed in coordinate free and/or coefficient forms.\footnote{%
It should be noted that the well known LC-connection $\nabla $ is a linear
one but not a d--connection because it does not preserve, under general
frame/coordinate transforms, a h-v--splitting. $\widehat{\mathcal{T}}$ is a
nonholonomically induced torsion determined by nontrivial values $(C_{\alpha
\beta }^{\gamma },\partial _{\beta }N_{i}^{a},g_{\alpha \beta })$ which is
different from the Einstein--Cartan, or string theory, when additional field
equations and sources for torsion fields are considered.} There is a
canonical distortion relation
\begin{equation}
\widehat{\mathbf{D}}=\nabla +\widehat{\mathbf{Z}}.  \label{candistr}
\end{equation}%
The distortion distinguished tensor, d-tensor, $\widehat{\mathbf{Z}}=\{%
\widehat{\mathbf{Z}}_{\ \beta \gamma }^{\alpha }[\widehat{\mathbf{T}}_{\
\beta \gamma }^{\alpha }]\},$ is an algebraic combination of the
coefficients of the corresponding torsion d-tensor $\widehat{\mathcal{T}}=\{%
\widehat{\mathbf{T}}_{\ \beta \gamma }^{\alpha }\}$ of $\widehat{\mathbf{D}}%
. $ Readers may consult Appendix \ref{as1} for a summary of relevant
formulas and some details on the geometry Lorentz manifolds with N-adapted
2+2 variables.

We can define two different Ricci tensors, $\ \widehat{\mathcal{R}}ic=\{%
\widehat{\mathbf{R}}_{\ \beta \gamma }:=\widehat{\mathbf{R}}_{\ \alpha \beta
\gamma }^{\gamma }\}$ and $Ric=\{R_{\ \beta \gamma }:=R_{\ \alpha \beta
\gamma }^{\gamma }\},$ when $\widehat{\mathcal{R}}ic$ is characterized by $h$
-$v$ N-adapted coefficients,
\begin{equation}
\widehat{\mathbf{R}}_{\alpha \beta }=\{\widehat{R}_{ij}:=\widehat{R}_{\
ijk}^{k},\ \widehat{R}_{ia}:=-\widehat{R}_{\ ika}^{k},\ \widehat{R}_{ai}:=%
\widehat{R}_{\ aib}^{b},\ \widehat{R}_{ab}:=\widehat{R}_{\ abc}^{c}\}.
\label{candricci}
\end{equation}%
Respectively, there are two different scalar curvatures, $\ R:=\mathbf{g}%
^{\alpha \beta }R_{\alpha \beta }$ and $\widehat{\mathbf{R}}:=\mathbf{g}%
^{\alpha \beta }\widehat{\mathbf{R}}_{\alpha \beta }=g^{ij}\widehat{R}%
_{ij}+g^{ab}\widehat{R}_{ab}.$ Following the two connection approach (\ref%
{twocon}), we conclude that the (pseudo) Riemannian geometry can be
equivalently described by two different geometric data $\left(\mathbf{%
g,\nabla }\right) $ and $(\mathbf{g,N,}\widehat{\mathbf{D}}).$ Using the
canonical distortion relation (\ref{candistr}), we can compute respective
distortions of curvature and Ricci tensors,
\begin{equation}
\widehat{\mathcal{R}}=\ ^{\nabla }\mathcal{R+}\ ^{\nabla }\mathcal{Z}%
\mbox{
and }\widehat{\mathcal{R}}ic=Ric+\widehat{\mathcal{Z}}ic,
\label{candriccidist}
\end{equation}%
with corresponding distortion tensors $\ ^{\nabla }\mathcal{Z}$ and $%
\widehat{\mathcal{Z}}ic.$ Such formulas motivate application in GR and MGTs
of nonholonomic geometric methods with multiple metric and connection
structures, and adapted frames. For certain well defined nonholonomic
configurations, various types of gravitational and matter field equations
rewritten in nonholonomic variables $(\mathbf{g,N},\widehat{\mathbf{D}})$
can be decoupled and integrated in some general forms.\footnote{%
It is not possible to decouple the gravitational and filed equations for
generic off-diagonal metrics if we work from the very beginning with the
data $\left( \mathbf{g,\nabla }\right) $. Having constructed certain classes
of generalized solutions (following the AFDM), necessary type
LC-configurations can be extracted if the condition $\widehat{\mathcal{T}}=0$
is imposed at the end.}

\subsection{MGTs in N-adapted variables and decoupling of (modified)
Einstein equations}

We show how the gravitational field equations can be written in nonholonomic
variables and consider the key steps for generating exact solutions in
Tables 1, 2, and 3.

\subsubsection{Gravitational field equations for the canonical d-connection}

Various models of modified gravity and acceleration cosmology (see \cite%
{vplb16,grvvap16,rajpoot17a,biv16p,viepjc17,abvfirs16p,aabovw16p} and
references therein) have bee elaborated for the so-called Starobinsky type $%
R^{2}$ gravity \cite{starob}. For well defined conditions on a class of
conformal transforms, the gravitational field equations for the so-called
quadratic gravity are is equivalent to the Einstein gravity with scalar
field sources. In nonholonomic variables $(\mathbf{g,N,}\widehat{\mathbf{D}}%
) $ and for interactions with scalar fields defined by Lagrange density $%
~^{m}\mathcal{L}(\mathbf{g,N,\phi })$, the action for such theories can be
written in the form
\begin{equation}
\mathcal{S}=M_{P}^{2}\int d^{4}u\sqrt{|\mathbf{g}|}[\widehat{\mathbf{R}}%
^{2}+~^{m}\mathcal{L[\mathbf{\phi }]}].  \label{actqmg}
\end{equation}%
In this formula the Planck mass $M_{P}$ is determined by the gravitational
constant; the Lagrange density $\ ^{m}\mathcal{L}[\phi ]$ and the action $\
^{m}\mathcal{S}=\int d^{4}u\sqrt{|\mathbf{g}|}~^{m}\mathcal{L}$ are
postulated in such forms which, for simplicity, allow to find explicit
solutions of (modified) Einstein equations. This is possible if there are
considered $\ ^{m}\mathcal{L}[\phi ]$ depending only on the coefficients of
a metric field and not on their derivatives. Applying a N--adapted
variational calculus, the energy--momentum d--tensor is computed $\ ^{m}%
\mathbf{T}_{\alpha \beta }:=-\frac{2}{\sqrt{|\mathbf{g}_{\mu \nu }|}}\frac{%
\delta (\sqrt{|\mathbf{g}_{\mu \nu }|}\ \ ^{m}\mathcal{L})}{\delta \mathbf{g}%
^{\alpha \beta }}=\ ^{m}\mathcal{L}\mathbf{g}_{\alpha \beta }+2\frac{\delta
(\ ^{m}\mathcal{L})}{\delta \mathbf{g}^{\alpha \beta }}.$ Following
N-adapted variations of $\mathcal{S}$ (\ref{actqmg}), one derives such
gravitational field equations
\begin{eqnarray}
\widehat{\mathbf{R}}_{\mu \nu } &=&\mathbf{\Upsilon }_{\mu \nu }=~^{m}%
\mathbf{\Upsilon }_{\mu \nu }+~\widehat{\mathbf{\Upsilon }}_{\mu \nu },%
\mbox{ for }  \label{qgreq} \\
&&\ ^{m}\mathbf{\Upsilon }_{\alpha \beta }=\frac{1}{2M_{P}^{2}}(\ ^{m}%
\mathbf{T}_{\alpha \beta }-\frac{1}{2}\mathbf{g}_{\alpha \beta }\mathbf{g}%
^{\gamma \tau }\ ^{m}\mathbf{T}_{\gamma \tau })\mbox{ and }\widehat{\mathbf{%
\Upsilon }}_{\mu \nu }=(\frac{1}{4}\widehat{\mathbf{R}}-\frac{\widehat{%
\square }\ \widehat{\mathbf{R}}}{~\widehat{\mathbf{R}}})\mathbf{g}_{\mu \nu
}+\frac{\widehat{\mathbf{D}}_{\mu }\widehat{\mathbf{D}}_{\nu }\ \widehat{%
\mathbf{R}}}{\widehat{\mathbf{R}}},  \label{sourc}
\end{eqnarray}%
where $\widehat{\square }:=\widehat{\mathbf{D}}^{2}=\mathbf{g}^{\mu \nu }%
\widehat{\mathbf{D}}_{\mu }\widehat{\mathbf{D}}_{\nu }$ and certain
constraints/ conditions with $\widehat{\mathbf{D}}_{\mid \widehat{\mathcal{T}%
}\rightarrow 0}=\mathbf{\nabla }$ can be imposed as in GR. The equations (%
\ref{qgreq}) \ for MGTs can be considered as modified Einstein equations
when $\mathbf{\nabla }$ is changed into $\widehat{\mathbf{D}}$ and the
standard energy momentum tensors are nonholonomically deformed into
N-adapted sources (\ref{sourc}).

The key steps for constructing exact solutions using two different methods
[1) with reduction to ODEs or 2) with integration of PDEs] of constructing
exact solutions of gravitational and matter field equations in MGTs and GR
are briefly reviewed in Appendix and summarized in the Tables 1, 2, and 3.
In references \cite{misner73,hawking73,wald82,kramer03,griffith09}, there
are provided details and main examples for constructing physically important
solutions by using diagonal ansatz, for instance, with spherical/cylinder
symmetries reducing the (generalized) Einstein equations to certain systems
of nonlinear ODEs. Proofs of results and a number of examples how the AFDM
should be applied in order to generate exact solutions of gravitational,
matter field and evolution nonlinear systems of PDEs are considered in \cite%
{vexsol98,vjhep01,vpcqg01,vjmp05,vmon06,vijgmmp07,vijgmmp11,vgrg12,vcsf12,vepjp12,vepjc13,vepjc14a,gvvepjc14,gvvcqg15}%
.

\subsubsection{(Off-) diagonal metric ansatz and ODEs and PDEs in (non)
holonomic variables}

We explain how using holonomic 3+1 and 2+2 nonholonomic variables and
corresponding ansatz it is possible to transform gravitational field
equations in MGTs and GR into respective systems of nonlinear ODEs and PDEs
is summarized in Table 1. N-adapted formulas are provided in Appendix \ref%
{as1}.


{\scriptsize
\begin{eqnarray*}
&&%
\begin{tabular}{l}
\hline\hline
\begin{tabular}{lll}
& {\ \textsf{Table 1:\ (Modified) Einstein eqs as systems of nonlinear PDEs}
and} &  \\
& the Anholonomic Frame Deformation Method, \textbf{AFDM}, &  \\
& \textit{for constructing generic off-diagonal exact, parametric, and
physically important solutions} &
\end{tabular}%
\end{tabular}
\\
&&{%
\begin{tabular}{lll}
\hline
diagonal ansatz: PDEs $\rightarrow $ \textbf{ODE}s &  & AFDM: \textbf{PDE}s
\textbf{with decoupling; \ generating functions} \\
radial coordinates $u^{\alpha }=(r,\theta ,\varphi ,t)$ & $u=(x,y):$ &
\mbox{ nonholonomic 2+2
splitting, } $u^{\alpha }=(x^{1},x^{2},y^{3},y^{4}=t)$ \\
LC-connection $\mathring{\nabla}$ & [connections] & $%
\begin{array}{c}
\mathbf{N}:T\mathbf{V}=hT\mathbf{V}\oplus vT\mathbf{V,}\mbox{ locally }%
\mathbf{N}=\{N_{i}^{a}(x,y)\} \\
\mbox{ canonical connection distortion }\widehat{\mathbf{D}}=\nabla +%
\widehat{\mathbf{Z}}%
\end{array}%
$ \\
$%
\begin{array}{c}
\mbox{ diagonal ansatz  }g_{\alpha \beta }(u) \\
=\left(
\begin{array}{cccc}
\mathring{g}_{1} &  &  &  \\
& \mathring{g}_{2} &  &  \\
&  & \mathring{g}_{3} &  \\
&  &  & \mathring{g}_{4}%
\end{array}%
\right)%
\end{array}%
$ & $\mathbf{g}\Leftrightarrow $ & $%
\begin{array}{c}
g_{\alpha \beta }=%
\begin{array}{c}
g_{\alpha \beta }(x^{i},y^{a})\mbox{ general frames / coordinates} \\
\left[
\begin{array}{cc}
g_{ij}+N_{i}^{a}N_{j}^{b}h_{ab} & N_{i}^{b}h_{cb} \\
N_{j}^{a}h_{ab} & h_{ac}%
\end{array}%
\right] ,\mbox{ 2 x 2 blocks }%
\end{array}
\\
\mathbf{g}_{\alpha \beta }=[g_{ij},h_{ab}],\mathbf{g}=\mathbf{g}%
_{i}(x^{k})dx^{i}\otimes dx^{i}+\mathbf{g}_{a}(x^{k},y^{b})\mathbf{e}%
^{a}\otimes \mathbf{e}^{b}%
\end{array}%
$ \\
$\mathring{g}_{\alpha \beta }=\left\{
\begin{array}{cc}
\mathring{g}_{\alpha }(r) & \mbox{ for BHs} \\
\mathring{g}_{\alpha }(t) & \mbox{ for FLRW }%
\end{array}%
\right. $ & [coord.frames] & $g_{\alpha \beta }=\left\{
\begin{array}{cc}
g_{\alpha \beta }(r,\theta ,y^{3}=\varphi ) & \mbox{ stationary conf. } \\
g_{\alpha \beta }(r,\theta ,y^{4}=t) & \mbox{ cosm. conf. }%
\end{array}%
\right. $ \\
&  &  \\
$%
\begin{array}{c}
\mbox{coord.tranfsorms }e_{\alpha }=e_{\ \alpha }^{\alpha ^{\prime
}}\partial _{\alpha ^{\prime }}, \\
e^{\beta }=e_{\beta ^{\prime }}^{\ \beta }du^{\beta ^{\prime }},\mathring{g}%
_{\alpha \beta }=\mathring{g}_{\alpha ^{\prime }\beta ^{\prime }}e_{\ \alpha
}^{\alpha ^{\prime }}e_{\ \beta }^{\beta ^{\prime }} \\
\begin{array}{c}
\mathbf{\mathring{g}}_{\alpha }(x^{k},y^{a})\rightarrow \mathring{g}_{\alpha
}(r),\mbox{ or }\mathring{g}_{\alpha }(t), \\
\mathring{N}_{i}^{a}(x^{k},y^{a})\rightarrow 0.%
\end{array}%
\end{array}%
$ & [N-adapt. fr.] & $\left\{
\begin{array}{cc}
\begin{array}{c}
\mathbf{g}_{i}(r,\theta ),\mathbf{g}_{a}(r,\theta ,\varphi ), \\
\mbox{ or }\mathbf{g}_{i}(r,\theta ),\mathbf{g}_{a}(r,\theta ,t),%
\end{array}
& \mbox{ d-metrics } \\
\begin{array}{c}
N_{i}^{3}=w_{i}(r,\theta ,\varphi ),N_{i}^{4}=n_{i}(r,\theta ,\varphi ), \\
\mbox{ or }N_{i}^{3}=n_{i}(r,\theta ,t),N_{i}^{4}=w_{i}(r,\theta ,t),%
\end{array}
&
\end{array}%
\right. $ \\
$\mathring{\nabla},$ $Ric=\{\mathring{R}_{\ \beta \gamma }\}$ & Ricci tensors
& $\widehat{\mathbf{D}},\ \widehat{\mathcal{R}}ic=\{\widehat{\mathbf{R}}_{\
\beta \gamma }\}$ \\
$~^{m}\mathcal{L[\mathbf{\phi }]\rightarrow }\ ^{m}\mathbf{T}_{\alpha \beta }%
\mathcal{[\mathbf{\phi }]}$ & sources & $%
\begin{array}{cc}
\mathbf{\Upsilon }_{\ \nu }^{\mu }=\mathbf{e}_{\ \mu ^{\prime }}^{\mu }%
\mathbf{e}_{\nu }^{\ \nu ^{\prime }}\mathbf{\Upsilon }_{\ \nu ^{\prime
}}^{\mu ^{\prime }}[\ ^{m}\mathcal{L}(\mathbf{\varphi }+\mathbf{\psi ),}\
\widehat{\mathbf{\Upsilon }}_{\mu \nu }] &  \\
=diag[~\ _{h}\Upsilon (x^{i})\delta _{j}^{i},\Upsilon (x^{i},\varphi )\delta
_{b}^{a}], & \mbox{stationary conf.} \\
=diag[~\ _{h}\overline{\Upsilon }(x^{i})\delta _{j}^{i},\overline{\Upsilon }%
(x^{i},t)\delta _{b}^{a}], & \mbox{ cosmol. conf.}%
\end{array}%
$ \\
trivial equations for $\mathring{\nabla}$-torsion & LC-conditions & $%
\widehat{\mathbf{D}}_{\mid \widehat{\mathcal{T}}\rightarrow 0}=\mathbf{%
\nabla }\mbox{extracting new classes of solutions in GR}$ \\ \hline\hline
\end{tabular}%
}
\end{eqnarray*}%
}


\subsubsection{Decoupling and integration of gravitational filed equations
and stationary solutions}

The key steps of AFDM for generating stationary off-diagonal exact solutions
of (modified) Einstein equations described in appendices \ref{asst}, \ref%
{ass3} are summarized in Table 2. Following a nonholonomic deformation
procedure, for instance, for a generating function $h_{4}({r,\theta }%
,\varphi ),$ we construct a class of off--diagonal stationary solutions with
Killing symmetry on $\partial _{t}$ determined by sources$~(\ _{h}\Upsilon
,\Upsilon )$ and effective cosmological constant $\Lambda ,$
\begin{eqnarray*}
ds^{2} &=&e^{\ \psi (x^{k})}[(dx^{1})^{2}+(dx^{2})^{2}]-\frac{%
(h_{4}^{\diamond }{})^{2}}{|\int d\varphi \ \Upsilon (h_{4})^{\diamond }|\
h_{4}}[dy^{3}+\frac{\partial _{i}(\int d\varphi \ \Upsilon \ h_{4}^{\diamond
}{}])}{\Upsilon \ h_{4}^{\diamond }{}}dx^{i}] \\
&&+h_{4}[dt+(\ _{1}n_{k}+4\ _{2}n_{k}\int d\varphi \frac{(h_{4}^{\diamond
}{})^{2}}{|\int dy^{3}\ \Upsilon (h_{4}^{\diamond })|\ (h_{4})^{5/2}}%
)dx^{k}].
\end{eqnarray*}%
Such solutions are, in general, with nontrivial nonholonomically induced
torsion (\ref{dtors}). They can be re-defined equivalently in terms of
generating functions $\Psi ({r,\theta },\varphi )$ or $\Phi ({r,\theta }%
,\varphi ),$ see (\ref{gensolstat}).

LC-configurations in GR can be extracted for additional zero torsion
constraints with a more special class of "integrable" generating functions $(%
\check{h}_{4},$ and $\check{\Psi}({r,\theta },\varphi )$ and/or $\check{\Phi}%
({r,\theta },\varphi ))$ for respective sources $\check{\Upsilon}$ and $%
\Lambda $ (\ref{lcsolstat}), \
\begin{equation*}
ds^{2}=e^{\ \psi (x^{k})}[(dx^{1})^{2}+(dx^{2})^{2}]-\left( (\check{h}%
_{4}^{\diamond }{})^{2}/[|\int d\varphi \ \Upsilon \check{h}_{4}^{\diamond
}|\ \check{h}_{4}]\right) [d\varphi +(\partial _{i}\check{A})dx^{i}]+\check{h%
}_{4}\left[ dt+(\partial _{k}n)dx^{k}\right] .
\end{equation*}%
In terms of $\eta $--polarization functions, such d-metrics and
N-connections, can be parameterized to describe nonholonomic deformations of
a primary (for instance, BH) d-metric $\mathbf{\mathring{g}}$ into target
generic off diagonal stationary solutions $\widehat{\mathbf{g}},$ see (\ref%
{dme}), as $\mathbf{\mathring{g}}\rightarrow \widehat{\mathbf{g}}=[g_{\alpha
}=\eta _{\alpha }\mathring{g}_{\alpha },\ \eta _{i}^{a}\mathring{N}_{i}^{a}]$%
.

{\scriptsize
\begin{eqnarray*}
&&%
\begin{tabular}{l}
\hline\hline
\begin{tabular}{lll}
& {\large \textsf{Table 2:\ Off-diagonal stationary configurations}} &  \\
& Exact solutions of $\widehat{\mathbf{R}}_{\mu \nu }=\mathbf{\Upsilon }%
_{\mu \nu }$ (\ref{qgreq}) transformed into a system of nonlinear PDEs (\ref%
{e1a})-(\ref{e4a}) &
\end{tabular}
\\
\end{tabular}
\\
&&%
\begin{tabular}{lll}
\hline\hline
&  &  \\
$%
\begin{array}{c}
\mbox{d-metric ansatz with} \\
\mbox{Killing symmetry }\partial _{4}=\partial _{t}%
\end{array}%
$ &  & $%
\begin{array}{c}
ds^{2}=g_{i}(x^{k})(dx^{i})^{2}+g_{a}(x^{k},y^{3})(dy^{a}+N_{i}^{a}(x^{k},y^{3})dx^{i})^{2},%
\mbox{ for } \\
g_{i}=e^{\psi {(r,\theta )}},\,\,\,\,g_{a}=h_{a}({r,\theta },\varphi ),\
N_{i}^{3}=w_{i}({r,\theta },\varphi ),\,\,\,\,N_{i}^{4}=n_{i}({r,\theta }%
,\varphi ),%
\end{array}%
$ \\
&  &  \\
Effective matter sources &  & $\mathbf{\Upsilon }_{\ \nu }^{\mu }=[~\
_{h}\Upsilon ({r,\theta })\delta _{j}^{i},\Upsilon ({r,\theta },\varphi
)\delta _{b}^{a}];x^{1}=r,x^{2}=\theta ,y^{3}=\varphi ,y^{4}=t$ \\ \hline
Nonlinear PDEs (\ref{e1a})-(\ref{e4a}) &  & $%
\begin{array}{c}
\psi ^{\bullet \bullet }+\psi ^{\prime \prime }=2~\ \ _{h}\Upsilon ; \\
\varpi ^{\diamond }\ h_{4}^{\diamond }=2h_{3}h_{4}\Upsilon ; \\
\beta w_{i}-\alpha _{i}=0; \\
n_{k}^{\diamond \diamond }+\gamma n_{k}^{\diamond }=0;%
\end{array}%
$ for $%
\begin{array}{c}
\varpi {=\ln |\partial _{3}h_{4}/\sqrt{|h_{3}h_{4}|}|,} \\
\alpha _{i}=(\partial _{\varphi }h_{4})\ (\partial _{i}\varpi ), \\
\ \beta =(\partial _{\varphi }h_{4})\ (\partial _{\varphi }\varpi ), \\
\ \gamma =\partial _{\varphi }\left( \ln |h_{4}|^{3/2}/|h_{3}|\right) , \\
\partial _{1}q=q^{\bullet },\partial _{2}q=q^{\prime },\partial
_{3}q=\partial q/\partial \varphi =q^{\diamond }%
\end{array}%
$ \\ \hline
$%
\begin{array}{c}
\mbox{ Generating functions:}\ h_{3}({r,\theta },\varphi ), \\
\Psi ({r,\theta },\varphi )=e^{\varpi },\Phi ({r,\theta },\varphi ); \\
\mbox{integration functions:}\ h_{3}^{[0]}(x^{k}),\  \\
_{1}n_{k}(x^{i}),\ _{2}n_{k}(x^{i}); \\
\mbox{\& nonlinear symmetries}%
\end{array}%
$ &  & $%
\begin{array}{c}
\ (\Psi ^{2})^{\diamond }=-\int dy^{3}\ \Upsilon h_{4}^{\ \diamond }, \\
\Phi ^{2}=-4\Lambda h_{4},\mbox{ see }(\ref{nsym1a}); \\
h_{4}=h_{4}^{[0]}-\Phi ^{2}/4\Lambda ,h_{4}^{\diamond }\neq 0,\Lambda \neq
0=const%
\end{array}%
$ \\ \hline
Off-diag. solutions, $%
\begin{array}{c}
\mbox{d--metric} \\
\mbox{N-connec.}%
\end{array}%
$ &  & $%
\begin{array}{c}
\ g_{i}=e^{\ \psi (x^{k})}\mbox{ as a solution of 2-d Poisson eqs. }\psi
^{\bullet \bullet }+\psi ^{\prime \prime }=2~\ _{h}\overline{\Upsilon }; \\
h_{3}=-(\Psi ^{\diamond })^{2}/4\Upsilon ^{2}h_{4},\mbox{ see }(\ref%
{offdstat}); \\
h_{4}=h_{4}^{[0]}-\int dy^{3}(\Psi ^{2})^{\diamond }/4\Upsilon
=h_{4}^{[0]}-\Phi ^{2}/4\Lambda ; \\
\\
w_{i}=\partial _{i}\ \Psi /\ \partial _{\varphi }\Psi =\partial _{i}\ \Psi
^{2}/\ \partial _{\varphi }\Psi ^{2}|; \\
n_{k}=\ _{1}n_{k}+\ _{2}n_{k}\int dy^{3}(\Psi ^{\diamond })^{2}/\Upsilon
^{2}|h_{4}^{[0]}-\int dy^{3}(\Psi ^{2})^{\diamond }/4\Upsilon |^{5/2}. \\
\\
\end{array}%
$ \\ \hline
LC-configurations (\ref{lcconda}) &  & $%
\begin{array}{c}
\partial _{\varphi }w_{i}=(\partial _{i}-w_{i}\partial _{\varphi })\ln \sqrt{%
|h_{3}|},(\partial _{i}-w_{i}\partial _{\varphi })\ln \sqrt{|h_{4}|}=0, \\
\partial _{k}w_{i}=\partial _{i}w_{k},\partial _{\varphi }n_{i}=0,\partial
_{i}n_{k}=\partial _{k}n_{i}; \\
\\
\Psi =\check{\Psi}(x^{i},\varphi ),(\partial _{i}\check{\Psi})^{\diamond
}=\partial _{i}(\check{\Psi}^{\diamond })\mbox{ and } \\
\Upsilon (x^{i},\varphi )=\Upsilon \lbrack \check{\Psi}]=\check{\Upsilon},%
\mbox{ or }\Upsilon =const. \\
\end{array}%
$ \\ \hline
N-connections, zero torsion &  & $%
\begin{array}{c}
w_{i}=\partial _{i}\check{A}=\left\{
\begin{array}{c}
\partial _{i}(\int d\varphi \ \check{\Upsilon}\ \check{h}_{4}{}^{\diamond
}])/\check{\Upsilon}\ \check{h}_{4}{}^{\diamond }; \\
\partial _{i}\check{\Psi}/\check{\Psi}^{\diamond }; \\
\partial _{i}(\int dy^{3}\ \check{\Upsilon}(\check{\Phi}^{2})^{\diamond })/(%
\check{\Phi})^{\diamond }\check{\Upsilon};%
\end{array}%
\right. \\
\mbox{ and }n_{k}=\check{n}_{k}=\partial _{k}n(x^{i}).%
\end{array}%
$ \\ \hline
$%
\begin{array}{c}
\mbox{polarization functions} \\
\mathbf{\mathring{g}}\rightarrow \widehat{\mathbf{g}}\mathbf{=}[g_{\alpha
}=\eta _{\alpha }\mathring{g}_{\alpha },\ \eta _{i}^{a}\mathring{N}_{i}^{a}]%
\end{array}%
$ &  & $%
\begin{array}{c}
\\
ds^{2}=\eta _{1}(r,\theta )\mathring{g}_{1}(r,\theta )[dx^{1}(r,\theta
)]^{2}+\eta _{2}(r,\theta )\mathring{g}_{2}(r,\theta )[dx^{2}(r,\theta
)]^{2}+ \\
\eta _{3}(r,\theta ,\varphi )\mathring{g}_{3}(r,\theta )[d\varphi +\eta
_{i}^{3}(r,\theta ,\varphi )\mathring{N}_{i}^{3}(r,\theta )dx^{i}(r,\theta
)]^{2}+ \\
\eta _{4}(r,\theta ,\varphi )\mathring{g}_{4}(r,\theta )[dt+\eta
_{i}^{4}(r,\theta ,\varphi )\mathring{N}_{i}^{4}(r,\theta )dx^{i}(r,\theta
)]^{2}, \\
\end{array}%
$ \\ \hline
Prime metric defines a BH &  & $%
\begin{array}{c}
\\
\lbrack \mathring{g}_{i}(r,\theta ),\mathring{g}_{a}=\mathring{h}%
_{a}(r,\theta );\mathring{N}_{k}^{3}=\mathring{w}_{k}(r,\theta ),\mathring{N}%
_{k}^{4}=\mathring{n}_{k}(r,\theta )] \\
\mbox{diagonalizable by frame/ coordinate transforms.} \\
\end{array}%
$ \\
Example of a prime metric &  & $%
\begin{array}{c}
\\
\mathring{g}_{1}=(1-r_{g}/r)^{-1},\mathring{g}_{2}=r^{2},\mathring{h}%
_{3}=r^{2}\sin ^{2}\theta ,\mathring{h}_{4}=(1-r_{g}/r),r_{g}=const \\
\mbox{the Schwarzschild solution, or any BH solution.} \\
\\
\end{array}%
$ \\ \hline
Solutions for polarization funct. &  & $%
\begin{array}{c}
\eta _{i}=e^{\ \psi (x^{k})}/\mathring{g}_{i};\eta _{3}\mathring{h}_{3}=-%
\frac{4[(|\eta _{4}\mathring{h}_{4}|^{1/2})^{\diamond }]^{2}}{|\int dy^{3}%
\widehat{\Upsilon }[(\eta _{4}\mathring{h}_{4})]^{\diamond }|\ }; \\
\eta _{4}=\eta _{4}(r,\theta ,\varphi )\mbox{ as a generating
function}; \\
\ \eta _{i}^{3}\ \mathring{N}_{i}^{3}=\frac{\partial _{i}\ \int dy^{3}%
\widehat{\Upsilon }(\eta _{4}\ \mathring{h}_{4})^{\diamond }}{\widehat{%
\Upsilon }\ (\eta _{4}\ \mathring{h}_{4})^{\diamond }}; \\
\eta _{k}^{4}\ \mathring{N}_{k}^{4}=\ _{1}n_{k}+16\ \ _{2}n_{k}\int dy^{3}%
\frac{\left( [(\eta _{4}\mathring{h}_{4})^{-1/4}]^{\diamond }\right) ^{2}}{%
|\int dy^{3}\widehat{\Upsilon }[(\eta _{4}\ \mathring{h}_{4})]^{\diamond }|\
}%
\end{array}%
\mbox{ see ap. }$\ref{ass3}. \\ \hline
Polariz. funct. with zero torsion &  & $%
\begin{array}{c}
\eta _{i}=e^{\ \psi (x^{k})}/\mathring{g}_{i};\eta _{4}=\check{\eta}%
_{4}(r,\theta ,\varphi )\mbox{ as a generating function}; \\
\eta _{3}=-\frac{4[(|\eta _{4}\mathring{h}_{4}|^{1/2})^{\diamond }]^{2}}{%
\mathring{g}_{3}|\int dy^{3}\widehat{\Upsilon }[(\check{\eta}_{4}\mathring{h}%
_{4})]^{\diamond }|\ };\eta _{i}^{3}=\partial _{i}\check{A}/\mathring{w}%
_{k},\eta _{k}^{4}=\frac{\ \partial _{k}n}{\mathring{n}_{k}}, \\
\\
\end{array}%
\mbox{ see ap. }$\ref{ass3}. \\ \hline\hline
\end{tabular}%
\end{eqnarray*}%
%
%
%
%
%
%
%
%
%
%
%
%
%
} 

\subsubsection{Decoupling and integration of gravitational PDEs generating
cosmological solutions}

In Table 3, we state the key steps of the AFDM for generating off-diagonal
locally anisotropic solutions of (modified) Einstein equations described in
appendices \ref{ascs} and \ref{ass3}.

{\scriptsize
\begin{eqnarray*}
&&%
\begin{tabular}{l}
\hline\hline
\begin{tabular}{lll}
& {\large \textsf{Table 3:\ Off-diagonal locally anisotropic cosmological
solutions}} &  \\
& Exact solutions of $\widehat{\mathbf{R}}_{\mu \nu }=\mathbf{\Upsilon }%
_{\mu \nu }$ (\ref{qgreq}) transformed into a system of nonlinear PDEs (\ref%
{e1})-(\ref{e4}) &
\end{tabular}
\\
\end{tabular}
\\
&&%
\begin{tabular}{lll}
\hline\hline
$%
\begin{array}{c}
\mbox{d-metric ansatz with} \\
\mbox{Killing symmetry }\partial _{3}=\partial _{\varphi }%
\end{array}%
$ &  & $%
\begin{array}{c}
ds^{2}=g_{i}(x^{k})(dx^{i})^{2}+g_{a}(x^{k},y^{4})(dy^{a}+N_{i}^{a}(x^{k},y^{4})dx^{i})^{2},%
\mbox{ for } \\
g_{i}=e^{\psi {(x}^{k}{)}},\,\,\,\,g_{a}=\overline{h}_{a}({x}^{k},t),\
N_{i}^{3}=\overline{n}_{i}({x}^{k},t),\,\,\,\,N_{i}^{4}=\overline{w}_{i}({x}%
^{k},t),%
\end{array}%
$ \\
&  &  \\
Effective matter sources &  & $\overline{\mathbf{\Upsilon }}_{\ \nu }^{\mu
}=[~\ _{h}\overline{\Upsilon }({x}^{k})\delta _{j}^{i},\overline{\Upsilon }({%
x}^{k},t)\delta _{b}^{a}];x^{1},x^{2},y^{3},y^{4}=t$ \\ \hline
Nonlinear PDEs (\ref{e1})-(\ref{e4}) &  & $%
\begin{array}{c}
\psi ^{\bullet \bullet }+\psi ^{\prime \prime }=2~\ \ _{h}\overline{\Upsilon
}; \\
\overline{\varpi }^{\diamond }\ \overline{h}_{3}^{\diamond }=2\overline{h}%
_{3}\overline{h}_{4}\overline{\Upsilon }; \\
\overline{n}_{k}^{\ast \ast }+\overline{\gamma }\overline{n}_{k}^{\ast }=0;
\\
\overline{\beta }\overline{w}_{i}-\overline{\alpha }_{i}=0;%
\end{array}%
$ for $%
\begin{array}{c}
\overline{\varpi }{=\ln |\partial _{t}}\overline{{h}}{_{3}/\sqrt{|\overline{h%
}_{3}\overline{h}_{4}|}|,} \\
\overline{\alpha }_{i}=(\partial _{t}\overline{h}_{3})\ (\partial _{i}%
\overline{\varpi }), \\
\ \overline{\beta }=(\partial _{t}\overline{h}_{3})\ (\partial _{t}\overline{%
\varpi }), \\
\ \overline{\gamma }=\partial _{t}\left( \ln |\overline{h}_{3}|^{3/2}/|%
\overline{h}_{4}|\right) , \\
\partial _{1}q=q^{\bullet },\partial _{2}q=q^{\prime },\partial
_{4}q=\partial q/\partial t=q^{\ast }%
\end{array}%
$ \\ \hline
$%
\begin{array}{c}
\mbox{ Generating functions:}\ h_{4}({x}^{k},t), \\
\overline{\Psi }(x^{k},t)=e^{\overline{\varpi }},\overline{\Phi }({x}^{k},t);
\\
\mbox{integration functions:}\ h_{4}^{[0]}(x^{k}),\  \\
_{1}n_{k}(x^{i}),\ _{2}n_{k}(x^{i}); \\
\mbox{\& nonlinear symmetries}%
\end{array}%
$ &  & $%
\begin{array}{c}
\ (\overline{\Psi }^{2})^{\ast }=-\int dt\ \overline{\Upsilon }\overline{h}%
_{3}^{\ast }, \\
\overline{\Phi }^{2}=-4\overline{\Lambda }\overline{h}_{3},\mbox{
see }(\ref{nsym1b}); \\
\overline{h}_{3}=\overline{h}_{3}^{[0]}-\overline{\Phi }^{2}/4\overline{%
\Lambda },\overline{h}_{3}^{\ast }\neq 0,\overline{\Lambda }\neq 0=const%
\end{array}%
$ \\ \hline
Off-diag. solutions, $%
\begin{array}{c}
\mbox{d--metric} \\
\mbox{N-connec.}%
\end{array}%
$ &  & $%
\begin{array}{c}
\ g_{i}=e^{\ \psi (x^{k})}\mbox{ as a solution of 2-d Poisson eqs. }\psi
^{\bullet \bullet }+\psi ^{\prime \prime }=2~\ _{h}\overline{\Upsilon }; \\
\overline{h}_{4}=-(\overline{\Psi }^{2})^{\ast }/4\overline{\Upsilon }^{2}%
\overline{h}_{3},\mbox{ see }(\ref{offdcosm}); \\
\overline{h}_{3}=h_{3}^{[0]}-\int dt(\overline{\Psi }^{2})^{\ast }/4%
\overline{\Upsilon }=h_{3}^{[0]}-\overline{\Phi }^{2}/4\overline{\Lambda };
\\
\\
\overline{n}_{k}=\ _{1}n_{k}+\ _{2}n_{k}\int dt(\overline{\Psi }^{\ast
})^{2}/\overline{\Upsilon }^{2}|h_{3}^{[0]}-\int dt(\overline{\Psi }%
^{2})^{\ast }/4\overline{\Upsilon }|^{5/2}; \\
\overline{w}_{i}=\partial _{i}\ \overline{\Psi }/\ \partial _{t}\overline{%
\Psi }=\partial _{i}\ \overline{\Psi }^{2}/\ \partial _{t}\overline{\Psi }%
^{2}|. \\
\\
\end{array}%
$ \\ \hline
LC-configurations (\ref{lccondb}) &  & $%
\begin{array}{c}
\partial _{t}\overline{w}_{i}=(\partial _{i}-\overline{w}_{i}\partial
_{t})\ln \sqrt{|\overline{h}_{4}|},(\partial _{i}-\overline{w}_{i}\partial
_{4})\ln \sqrt{|\overline{h}_{3}|}=0, \\
\partial _{k}\overline{w}_{i}=\partial _{i}\overline{w}_{k},\partial _{t}%
\overline{n}_{i}=0,\partial _{i}\overline{n}_{k}=\partial _{k}\overline{n}%
_{i}; \\
\\
\Psi =\overline{\check{\Psi}}(x^{i},t),(\partial _{i}\overline{\check{\Psi}}%
)^{\ast }=\partial _{i}(\overline{\check{\Psi}}^{\ast })\mbox{ and } \\
\overline{\Upsilon }(x^{i},t)=\overline{\Upsilon }[\overline{\check{\Psi}}]=%
\overline{\check{\Upsilon}},\mbox{ or }\overline{\Upsilon }=const. \\
\end{array}%
$ \\ \hline
N-connections, zero torsion &  & $%
\begin{array}{c}
\overline{n}_{k}=\overline{\check{n}}_{k}=\partial _{k}\overline{n}(x^{i})
\\
\mbox{ and }\overline{w}_{i}=\partial _{i}\overline{\check{A}}=\left\{
\begin{array}{c}
\partial _{i}(\int dt\ \overline{\check{\Upsilon}}\ \overline{\check{h}}%
_{3}^{\ast }])/\overline{\check{\Upsilon}}\ \overline{\check{h}}_{3}^{\ast
}{}; \\
\partial _{i}\overline{\check{\Psi}}/\overline{\check{\Psi}}^{\ast }; \\
\partial _{i}(\int dt\ \overline{\check{\Upsilon}}(\overline{\check{\Phi}}%
^{2})^{\ast })/(\overline{\check{\Phi}})^{\ast }\overline{\check{\Upsilon}};%
\end{array}%
\right. .%
\end{array}%
$ \\ \hline
$%
\begin{array}{c}
\mbox{polarization functions} \\
\mathbf{\mathring{g}}\rightarrow \overline{\widehat{\mathbf{g}}}\mathbf{=}[%
\overline{g}_{\alpha }=\overline{\eta }_{\alpha }\mathring{g}_{\alpha },\
\overline{\eta }_{i}^{a}\mathring{N}_{i}^{a}]%
\end{array}%
$ &  & $%
\begin{array}{c}
\\
ds^{2}=\overline{\eta }_{i}(x^{k},t)\mathring{g}_{i}(x^{k},t)[dx^{i}]^{2}+
\\
\overline{\eta }_{3}(x^{k},t)\mathring{h}_{3}(x^{k},t)[dy^{3}+\overline{\eta
}_{i}^{3}(x^{k},t)\mathring{N}_{i}^{3}(x^{k},t)dx^{i}]^{2}+ \\
\overline{\eta }_{4}(x^{k},t)\mathring{h}_{4}(x^{k},t)[dt+\overline{\eta }%
_{i}^{4}(x^{k},t)\mathring{N}_{i}^{4}(x^{k},t)dx^{i}]^{2}, \\
\end{array}%
$ \\ \hline
$%
\begin{array}{c}
\mbox{ Prime metric defines } \\
\mbox{ a cosmological solution}%
\end{array}%
$ &  & $%
\begin{array}{c}
\\
\lbrack \mathring{g}_{i}(x^{k},t),\mathring{g}_{a}=\mathring{h}_{a}(x^{k},t);%
\mathring{N}_{k}^{3}=\mathring{w}_{k}(x^{k},t),\mathring{N}_{k}^{4}=%
\mathring{n}_{k}(x^{k},t)] \\
\mbox{diagonalizable by frame/ coordinate transforms.} \\
\end{array}%
$ \\
$%
\begin{array}{c}
\mbox{Example of a prime } \\
\mbox{ cosmological metric }%
\end{array}%
$ &  & $%
\begin{array}{c}
\\
\mathring{g}_{1}=a^{2}(t)/(1-kr^{2}),\mathring{g}_{2}=a^{2}(t)r^{2}, \\
\mathring{h}_{3}=a^{2}(t)r^{2}\sin ^{2}\theta ,\mathring{h}%
_{4}=c^{2}=const,k=\pm 1,0; \\
\mbox{ any frame transform of a FLRW or a Bianchi metrics} \\
\end{array}%
$ \\ \hline
Solutions for polarization funct. &  & $%
\begin{array}{c}
\eta _{i}=e^{\ \psi (x^{k})}/\mathring{g}_{i};\overline{\eta }_{4}\mathring{h%
}_{4}=-\frac{4[(|\overline{\eta }_{3}\mathring{h}_{3}|^{1/2})^{\ast }]^{2}}{%
|\int dt\overline{\Upsilon }[(\overline{\eta }_{3}\mathring{h}_{3})]^{\ast
}|\ }; \\
\overline{\eta }_{3}=\overline{\eta }_{3}(x^{i},t)%
\mbox{ as a generating
function}; \\
\overline{\eta }_{k}^{3}\ \mathring{N}_{k}^{3}=\ _{1}n_{k}+16\ \
_{2}n_{k}\int dt\frac{\left( [(\overline{\eta }_{3}\mathring{h}%
_{3})^{-1/4}]^{\ast }\right) ^{2}}{|\int dt\overline{\Upsilon }[(\overline{%
\eta }_{3}\mathring{h}_{3})]^{\ast }|\ }; \\
\ \overline{\eta }_{i}^{4}\ \mathring{N}_{i}^{4}=\frac{\partial _{i}\ \int dt%
\overline{\Upsilon }(\overline{\eta }_{3}\mathring{h}_{3})^{\ast }}{%
\overline{\Upsilon }(\overline{\eta }_{3}\mathring{h}_{3})^{\ast }},%
\end{array}%
\mbox{ see ap. }$\ref{ass3}. \\ \hline
Polariz. funct. with zero torsion &  & $%
\begin{array}{c}
\overline{\eta }_{i}=e^{\ \psi }/\mathring{g}_{i};\overline{\eta }_{4}=-%
\frac{4[(|\overline{\eta }_{3}\mathring{h}_{3}|^{1/2})^{\ast }]^{2}}{%
\mathring{g}_{4}|\int dt\overline{\Upsilon }[(\overline{\eta }_{3}\mathring{h%
}_{3})]^{\ast }|\ }; \\
\overline{\eta }_{3}=\overline{\check{\eta}}_{3}({x}^{i},t)%
\mbox{  as
a generating function}; \\
\overline{\eta }_{k}^{4}=\partial _{k}\overline{\check{A}}/\mathring{w}_{k};%
\overline{\eta }_{k}^{3}=(\partial _{k}\overline{n})/\mathring{n}_{k}, \\
\end{array}%
\mbox{ see ap. }$\ref{ass3}. \\ \hline\hline
\end{tabular}%
\end{eqnarray*}%
%
%
%
%
%
%
%
%
%
%
%
%
%
} 

Applying the nonholonomic deformation procedure (for simplicity, we consider
metrics determined by a generating function $\overline{h}_{4}({x}^{k},t),$
we construct a class of generic off--diagonal cosmological solutions with
Killing symmetry on $\partial _{\varphi }$ determined by sources, $~\ _{h}%
\overline{\Upsilon }$ and $\overline{\Upsilon },$ and an effective
nontrivial cosmological constant $\overline{\Lambda },$
\begin{eqnarray*}
ds^{2} &=&e^{\ \psi (x^{k})}[(dx^{1})^{2}+(dx^{2})^{2}]+\overline{h}%
_{3}[dy^{3}+(\ _{1}n_{k}+4\ _{2}n_{k}\int dt\frac{(\overline{h}_{3}{}^{\ast
})^{2}}{|\int dt\ \overline{\Upsilon }\overline{h}_{3}{}^{\ast }|(\overline{h%
}_{3})^{5/2}})dx^{k}] \\
&&-\frac{(\overline{h}_{3}{}^{\ast })^{2}}{|\int dt\ \overline{\Upsilon }%
\overline{h}_{3}{}^{\ast }|\ \overline{h}_{3}}[dt+\frac{\partial _{i}(\int
dt\ \overline{\Upsilon }\ \overline{h}_{3}{}^{\ast }])}{\ \overline{\Upsilon
}\ \overline{h}_{3}{}^{\ast }}dx^{i}],
\end{eqnarray*}

Such locally anisotropic and inhomogeneous cosmological solutions are, in
general, with nontrivial nonholonomically induced torsion (\ref{dtors}).
This class of solutions can be re-defined equivalently in terms of
generating functions $\overline{\Psi }({x}^{k},t)$ and/or $\overline{\Phi }({%
x}^{k},t),$ see (\ref{gensolcosm}).

Cosmological configurations in GR can be extracted by imposing additional
constraints for zero torsion by restricting the class of "integrable"
generating functions $(\overline{h}_{4}=\overline{\check{h}}_{4},$ and $%
\overline{\check{\Psi}}({x}^{k},t)$ and/or $\overline{\check{\Phi}}({x}%
^{k},t)),$ for respective types of sources $\overline{\check{\Upsilon}}$ and
$\Lambda ,$ as in (\ref{lcsolcosm}), \
\begin{equation*}
ds^{2}=e^{\ \psi (x^{k})}[(dx^{1})^{2}+(dx^{2})^{2}]+\overline{\check{h}}_{3}%
\left[ dy^{3}+(\partial _{k}\overline{n})dx^{k}\right] -\left( (\overline{%
\check{h}}_{3}{}^{\ast })^{2}/[|\int dt\ \overline{\Upsilon }\overline{%
\check{h}}_{3}{}^{\ast }|\ \overline{\check{h}}_{3}]\right) [dt+(\partial
_{i}\overline{\check{A}})dx^{i}.
\end{equation*}%
\

In terms of $\eta $--polarization functions, the coefficients of
cosmological d-metrics and N-connections can be parameterized to describe
nonholonomic deformations of a primary (for instance, a FLRW) d-metric $%
\mathbf{\mathring{g}}$ into target generic off diagonal cosmological
solutions $\overline{\widehat{\mathbf{g}}}(x^{i},t)\rightarrow \overline{%
\widehat{\mathbf{g}}}(t),$ see (\ref{dme}), as $\mathbf{\mathring{g}}%
\rightarrow \overline{\widehat{\mathbf{g}}}\mathbf{=}[g_{\alpha }=\overline{%
\eta }_{\alpha }\mathring{g}_{\alpha },\ \overline{\eta }_{i}^{a}\mathring{N}%
_{i}^{a}]$.

\section{Quasiperiodic/ Pattern Forming Generating Functions and Sources}

\label{sqpf} The goal of this section is to analyse a series of examples of
quasiperiodic and/or pattern forming distributions and related free energy
functionals and systems of PDEs evolution equations on a nonholonomic
Lorentzian manifold $\mathbf{V}$, which have certain analogy in condensed
matter physics (see, for instance, \cite%
{rucklidge12,rucklidge16,rucklidge13,rucklidge15}). For small parametric
dependencies, such gravitational and matter field (non) stationary
distributions and nonlinear field interactions were studied in our partner
works \cite{biv16p,abvfirs16p,aabovw16p}. In this paper, the approach is
elaborated for generating and integration polarization functions and
(effective) sources stated in certain general forms without additional
assumptions on decompositions on small parameters.

\subsection{Nonholonomic 3+1 distributions with quasiperiodic/ pattern
forming structures}

Let us consider necessary smooth classes of functions $q=q(x^{i},y^{3}),$
for space like distributions, and $\overline{q}=\overline{q}(x^{i},y^{4}=t),$
for locally anisotropic cosmological configurations, defined respectively in
N-adapted coordinates on open regions of $U\subset \mathbf{V}$ and $%
\overline{U}\subset \mathbf{V}$. Such values will be used as generating
functions and/or (effective) sources for different models of quasiperiodic
and/or pattern forming spacetime structures. Additionally to a nonholonomic
2+2 splitting which allows us to decouple systems of nonlinear PDEs and
construct exact solutions of (modified) gravitational equations, we shall
consider a 3+1 splitting with local coordinates parameterized in the form $%
u^{\alpha }=(u^{\grave{\imath}},t)$, where the space like coordinates are $%
u^{\grave{\imath}}=(x^{i},y^{3})$, with $\grave{\imath},\grave{j},\grave{k}%
,...=1,2,3$. In GR, models with 3+1 spacetime decomposition were elaborated
with the aim to introduce values similar to the energy and momentum and
thermodynamical like characteristics for gravitational and scalar fields,
see details in Refs. \cite{misner73,wald82}). In our recent works \cite%
{biv16p,abvfirs16p,aabovw16p,ruchin17,rajpoot17}, double 2+2 and 3+1
spacetime decompositions (and various respective extra dimension
generalizations) were considered for constructing generic off-diagonal
solutions in MGTs and GR encoding quasi-periodic and aperiodic structures.

Any metric and/or equivalent d-metric structures parameterized in a form (%
\ref{dm}) or (\ref{dme}) can be re-written in certain forms with
nonholonomic 3+1 splitting,
\begin{eqnarray}
\mathbf{g} &=&b_{i}(x^{k})dx^{i}\otimes dx^{i}+b_{3}(x^{k},y^{3},t)\mathbf{e}%
^{3}\otimes \mathbf{e}^{3}-\breve{N}^{2}(x^{k},y^{3},t)\mathbf{e}^{4}\otimes
\mathbf{e}^{4},  \label{lapsnonh} \\
\mathbf{e}^{3} &=&dy^{3}+N_{i}^{3}(x^{k},y^{3},t)dx^{i},\,\,\,\,\mathbf{e}%
^{4}=dt+N_{i}^{4}(x^{k},y^{3},t)dx^{i}.  \notag
\end{eqnarray}%
For such configurations, a 4--d metric $\mathbf{g}$ can be considered as an
extension of a 3--d metric $b_{\grave{\imath}\grave{j}}=diag(b_{\grave{\imath%
}})=(b_{i},b_{3})$ on a family of 3-d hypersurfaces $\widehat{\Xi }_{t} $
parameterized by coordinate $t$ considered as a parameter, and when $%
b_{3}=h_{3}$ and $\breve{N}^{2}(u)=-h_{4}$ is defined by a lapse function $%
\breve{N}(u).$ We can impose additional conditions in order to transform
stationary d-metrics (\ref{gensolstat}), or locally anisotropic cosmological
d-metrics (\ref{gensolcosm}), into respective \ 3+1 versions (\ref{lapsnonh}
). Correspondingly, we shall work with a Killing symmetry on $\partial _{4}$
and lapse functions of type $\breve{N}(x^{k},y^{3}),$ or with a Killing
symmetry on $\partial _{3}$ and lapse functions of type $\breve{N}(x^{k},t).$
Such a decomposition results in a representation $\widehat{\mathbf{D}}=(\
^{3}\widehat{\mathbf{D}},\ ^{t}\widehat{\mathbf{D}}),$ where $\ ^{3}\widehat{%
\mathbf{D}}$ defines the action of the canonical d-connection covariant
derivative on space like coefficients and $\ ^{t}\widehat{\mathbf{D}}$ of
time like coefficients. For LC-configurations, the covariant derivative
operator splits as $\nabla =(\ ^{3}\nabla ,\ ^{t}\nabla ),$ when the action
on a scalar field $q(u)$ can be parameterized via frame/ coordinate
transforms as $\nabla q=(\ ^{3}\nabla q,\ ^{t}\nabla q=\partial
_{t}q=q^{\ast }).$

\subsubsection{Many pattern-forming nonlinear gravitational and matter
fields systems}

In condensed matter physics, models with tree-waves interactions, 3WIs, for
many pattern-forming systems were elaborated for explaining experimental
observations of certain microscopic and quasi-classical quantum systems, see
\cite{rucklidge12} and references therein. Modern cosmological data show a
very complex web like quasiperiodic and/or aperiodic like structure
formation, geometric anisotropic evolution and nonlinear gravitational and
matter field interactions, including dark energy and and dark matter
configurations \cite{crystalinks,diemer17}. We can apply very similar
mathematical methods for geometric modeling of quasi-crystal matter or
(super) galactic clusters and 3-d distributions of dark energy and dark
matter and generating solutions of systems on nonliner PDEs describing such
physical systems.

A prime pattern-forming field can be taken in the form
\begin{equation}
\overline{\mathring{q}}(x^{i},t)=\sum_{l=1,|\mathbf{k}_{l}|=1}^{\infty
}z_{l}(t)e^{i\mathbf{k}_{l}\cdot \mathbf{u}}+\sum_{l=1,|\mathbf{c}%
_{l}|=c}v_{l}(t)e^{i\mathbf{c}_{l}\cdot \mathbf{u}}+%
\mbox{ higher order
terms}  \label{primepatern}
\end{equation}%
defining in flat spaces 3WIs involving two comparable wavelengths. We
consider systems with two wave numbers $k=1$ and $k=c,$\footnote{%
we emphasize that in this work, the constant $c$ should be not confused with
the speed of light} for $0<c<1,$ when 3WIs are modelled in two forms:

\begin{enumerate}
\item two waves (with wave number 1, on the outer circle) interact
nonlinearly with a wave on the inner circle, with wave number $c$ (for
instance, wave vectors configurations $\mathbf{k}_{1}$ and $\mathbf{k}_{2}$
interact with $\mathbf{c}_{1}=\mathbf{k}_{1}+$ $\mathbf{k}_{2});$

\item for $1/2<c<\,1,$ two waves on the inner circle interact with another
wave on the outer circle (for instance, wave vectors configurations $\mathbf{%
c}_{2}$ and $\mathbf{c}_{3}$ interact with $\mathbf{k}_{1}=\mathbf{c}_{2}+%
\mathbf{c}_{3}).$
\end{enumerate}

For (\ref{primepatern}), we can parameterize the coefficients and respective
values in such forms that there are modelled two types of 3WIs. There are
involved (in the case 1 above) a triad of wave vectors $\mathbf{k}_{1},%
\mathbf{k}_{2}$ and $\mathbf{c}_{1}=\mathbf{k}_{1}+\mathbf{k}_{2},$ when
amplitudes are subjected to equations
\begin{equation}
z_{1}^{\ast } =\mu z_{1}+Q_{zv}\overline{z}_{2}v_{1}+\mbox{cubic terms},
z_{2}^{\ast }=\mu z_{2}+Q_{zv}\overline{z}_{1}v_{1}+\mbox{cubic terms}, %
\mbox{ and }v_{1}^{\ast } = \xi z_{1}+Q_{zz}z_{1}z_{2}+\mbox{cubic terms}.
\label{pat1a}
\end{equation}%
Other configuration can be defined (in the case 2) for wave vectors $\mathbf{%
c}_{2},\mathbf{c}_{3}$ and $\mathbf{k}_{1}=\mathbf{c}_{2}+\mathbf{c}_{3}$
with respective equations for amplitudes%
\begin{equation}
v_{2}^{\ast } = \xi v_{2}+Q_{vz}\overline{v}_{2}z_{1}+\mbox{cubic terms},
v_{3}^{\ast }=\xi v_{3}+Q_{vz}\overline{v}_{1}z_{1}+\mbox{cubic terms}, %
\mbox{ and }z_{1}^{\ast } =\mu z_{1}+Q_{zz}v_{2}v_{3}+\mbox{cubic terms}.
\label{pat1b}
\end{equation}%
In these formulas, the coefficients $\mu $ and $\xi $ determine,
respectively, the growth rates of amplitudes corresponding to wave numbers $%
1 $ and $c;$ and, for instance, $Q_{zv}$ and $Q_{zz}$ are quadratic
elements. For simplicity, we omit the cubic terms even they play also an
important role in the nonlinear dynamics of such waves and result in
observable effects both in condensed matter physics and at cosmological
distances. Similar equations can be replicated for all possible combinations
of modes describing 3WIs. Different stationary and dynamical models with
nonlinear wave interactions are generated for certain ranges of values of
coefficients $\mu ,$ $\xi,Q_{zv},Q_{zz},...$ and the signs of quadratic
coefficients for products of type $Q_{zv}Q_{zz}$ etc. were analyzed in \cite%
{rucklidge12} (on experimental data in condensed matter physics, see
references therein).

In a more general context, we can consider nonlinear deformations of prime
waves (\ref{primepatern}), $\overline{\mathring{q}}(x^{i},t)\rightarrow \
^{P}\overline{q}(x^{i},t),$ when the target field $\ ^{P}\overline{q}=%
\overline{q}$ describes pattern forming configurations as solutions of
nonlinear PDE,
\begin{equation}
\overline{q}^{\ast }=\mathcal{L}\overline{q}+\ ^{1}Q\overline{q}^{2}+\
^{2}Q(\ ^{3}\widehat{\mathbf{D}}_{\grave{\imath}}\ ^{3}\widehat{\mathbf{D}}^{%
\grave{\imath}}\overline{q})+\ ^{3}Q|(\ ^{3}\widehat{\mathbf{D}}_{\grave{%
\imath}}\overline{q})|^{2}-\overline{q}^{3},  \label{patternfev}
\end{equation}%
with summation on up-low index $\grave{\imath}.$ This equation can be
parameterized with a linear part $\mathcal{L}$ acting on a mode $e^{ikx}$
with an eigenvalue $\sigma (k)$ specified by $\sigma (1)=\mu $ and $%
\sigma(c)=\xi .$ Such a specification controls growth rates of the modes of
interest; with $d\sigma /dx=0,$ for $k=1$ and $k=c,$ and $\sigma (0)=\sigma
_{0}<0,$ one controls the depths of minimum between $k=1$ and $k=c.$ For
even functions $\sigma $ of $k,$ we can consider a 4th order polynomial on $%
k^{2},$%
\begin{eqnarray*}
\sigma (k) &=&\frac{[\mu A(k)+\xi B(k)]k^{2}}{(1-c^{2})^{3}c^{4}}+\frac{%
\sigma _{0}}{c^{4}}(1-k^{2})(c^{2}-k^{2})^{2},\mbox{ where } \\
A(k) &=&[(c^{2}-3)k^{2}-2c^{2}+4](k^{2}-c^{2})^{2}c^{4}\mbox{ and } B(k) =
[(3c^{2}-1)k^{2}+2c^{2}-4c^{4}](k^{2}-1)^{2}.
\end{eqnarray*}%
The linear operator $\mathcal{L}$ in the linear part of (\ref{patternfev})
is defined by replacing $k^{2}$ by $-\ ^{3}\widehat{\mathbf{D}}_{\grave{%
\imath}}\ ^{3}\widehat{\mathbf{D}}^{\grave{\imath}}.$ The nonlinear terms in
that PDE can be boosted as quadratic and cubic combinations of $\overline{q}$
and its derivatives, which can re-parameterized as certain nonlinear
deformations of $\overline{\mathring{q}}$ (\ref{primepatern}). A standard
weakly nonlinear theory expresses the values $\ ^{1}Q,\ ^{2}Q,$ and $\ ^{3}Q$
as certain products $Q_{zv}$ $Q_{zz}$ etc. when the solutions of (\ref{pat1a}%
) and (\ref{pat1b}) are nonlinearly mixed which results in different signs
and constant coefficients.

Using $\ ^{P}\overline{q}=\overline{q}$ $\ $\ as a solution of (\ref%
{patternfev}), we can generate various type pattern forming configurations
in a nonholonomic spacetime. For instance, there are possible 3 important
results (for simplicity, we take values of constants reproducing the results
explained in details in \cite{rucklidge12} and references therein; there are
cited also some papers analyzing similar structures in modern gravity,
cosmology and astrophysics):

\begin{enumerate}
\item A bifurcation pattern can be seen for $c=0.66,\sigma _{0}=-2,\
^{1}Q=0.3,\ ^{2}Q=1.3,$ and $\ ^{3}Q=1.7.$ There are: $z$ hexagons, for $%
k=1;v$ hexagons for $k=c;$ certain spatiotemporal chaos, STC, with mixed
patterns defined by $v$ stripes with patches of $z$ rectangles; two super
lattice patterns (the first one is with 6 modes on the outer circle and with
12 modes on the inner; the second one is with 6 modes on the inner circle
and with 12 ones on the outer). Here it should be noted that the scales
stated by $\mu $ and $\xi $ are not uniform and that we can model
additionally various diffusion and additional nonlinear wave interactions
\cite{vjhep01,vpcqg01}.

\item One can be found a STC pattern with $\mu =\xi =0.000707,$ when the
correlation length is about 1-2 wave-lengths. We can ad new terms with
fractional chaos and diffusion \cite{vcsf12,vepjp12}.

\item Another interesting configuration with two critical circles can be
reproduced in the power section \cite{gurzadyan13}.
\end{enumerate}

Solutions of (\ref{patternfev}) constructed as nonholonomic deformations of
prime generating functions (\ref{primepatern}) with nonlinear superpositions
of (\ref{pat1a}) and (\ref{pat1b}) reproduce, for explicit parameterizations
of constants, various type of nonlinear interactions and pattern forming
configurations. In condensed matter physics, such interactions between two
waves of one wavelength with a third wave of the other wavelength are known
both experimentally and theoretically. Both in condensed matter physics and
at cosmological scales, that they play a key role in producing a rich
variety of interesting phenomena such as web structures \cite{diemer17}. In
this and partner works \cite{biv16p,abvfirs16p,aabovw16p}, we show that one
can be reproduced also quasipatterns, superlattice patterns, and STC. The
geometric methods and mechanism we elaborate in our works can be applied to
any systems in which such 3WIs can occur at microscopic or cosmological
scales. In condensed matter physics, such effects are confirmed via Faraday
wave experiments, for coupled Turing systems, and some optical systems, see
details in \cite{rucklidge12} and references therein. At cosmological
scales, similar configurations are observed and modeled theoretically \cite%
{crystalinks,diemer17,gurzadyan13}.

\subsubsection{Quasicrystal like configurations in MGTs and GR}

We can consider other types of quasi-periodic, or aperiodic, generating
functions (not) related to 3WIs.

\paragraph{Formation of quasicrystalline structures and analogous dynamic
phase field crystal models: \newline
}

Quasicrystal, QC, structures can be modeled by generating functions $\ ^{QC}%
\overline{q}=\overline{q}(x^{i},t)$ defined as solutions of an evolution
equation with conserved dynamics,
\begin{equation}
\frac{\partial \overline{q}}{\partial t}=\ ^{b}\widehat{\Delta }\left[ \frac{%
\delta F}{\delta \overline{q}}\right] =-\ ^{b}\widehat{\Delta }(\Theta
\overline{q}+Q\overline{q}^{2}-\overline{q}^{3}).  \label{evoleq}
\end{equation}%
In this formula, the canonically nonholonomically deformed hypersurface
Laplace operator $\ ^{b}\widehat{\Delta }:=(\ ^{b}\widehat{D})^{2}=b^{\grave{%
\imath}\grave{j}}\widehat{D}_{\grave{\imath}}\widehat{D}_{\grave{j}}$ is
defined by (\ref{lapsnonh}) as a distortion of $\ ^{b}\Delta :=(\
^{b}\nabla)^{2}.$ Such operators can be computed for any family of
hypersurfaces $\widehat{\Xi }_{t}$ using formulas (\ref{candistr}).

The functional $F$ in (\ref{evoleq}) defines an effective free energy
\begin{equation}
F[\overline{q}]=\int \left[ -\frac{1}{2}\overline{q}\Theta \overline{q}-%
\frac{Q}{3}\overline{q}^{3}+\frac{1}{4}\overline{q}^{4}\right] \sqrt{b}%
dx^{1}dx^{2}\delta y^{3},  \label{dener}
\end{equation}%
where $b=\det |b_{\grave{\imath}\grave{j}}|,\delta y^{3}=\mathbf{e}^{3}$ and
the operator $\Theta $ will be defined below. Such nonlinear interactions
can be stabilized by the cubic term when the second order resonant
interactions are varied by setting the value of $Q$. The average value $<%
\overline{q}>$ of the generating function $\overline{q}$ is conserved for
any fixed $t.$ In result, we can consider $\overline{q}$ as an effective
parameter of the system and that we can choose $<\overline{q}>_{|t=t_{0}}=0$
since other values can be redefined and accommodated by altering $\Theta $
and $Q.$ Using the functional (\ref{dener}), we can elaborate on models of
dark energy, DE, for certain locally anisotropic and inhomogeneous
cosmological configurations \cite{biv16p,abvfirs16p,aabovw16p}. Similarly to
QCs in condensed matter physics \cite{rucklidge16}, we concluded that the
effective free energy $F[\overline{q}]$ characterizes a 3-d phase
gravitational field crystal model, when modulations are generated with two
length scales for off--diagonal cosmological structures. This model consists
of a nonlinear PDE with conserved nonholonomic dynamics resulting in
evolution equation. It describes a time evolution of $\overline{q}$ over
diffusive time scales.

\paragraph{3-d phase field like quasicrystal structures and evolution:
\newline
}

In cosmological theories, there are studied scalar fields potentials $V(%
\mathbf{\varphi })$ modified by effective quasicrystal structures, $\mathbf{%
\varphi \rightarrow \varphi =\varphi }_{0}\mathbf{+\psi ,}$ where $\mathbf{%
\psi (}x^{i},y^{3},t)$ with (quasi) crystal like phases described by
periodic or quasi-periodic modulations. Such modifications can be modelled
in dynamical phase field crystal, PFC, like form \cite{cross09,rucklidge16}.
Applying such mathematical methods in modern cosmology \cite%
{biv16p,abvfirs16p,aabovw16p}, we can elaborate models of 3-d
nonrelativistic dynamics which determined by Laplace like operators $\
^{3}\triangle =(\ ^{3}\nabla )^{2},$ or $\ ^{b}\widehat{\Delta }$ (the left
label 3 emphasizes that such an operator is for a 3-d hypersurface). We
write $\mathbf{\psi }$ instead of $\overline{q}$ in order to distinguish
such QC structures (which can be generated both by gravitational and matter
field with two length scales) from the class of models considered above.

In N--adapted frames with 3+1 splitting the equations for a local minimum
conserving dynamics,
\begin{equation}
\partial _{t}\mathbf{\psi =}\ ^{3}\triangle \left[ \frac{\delta F[\mathbf{%
\psi }]}{\delta \mathbf{\psi }}\right] ,\mbox{ or in a nonholonomic variant }%
\partial _{t}\mathbf{\psi =}\ \ ^{b}\widehat{\Delta }\left[ \frac{\delta F[%
\mathbf{\psi }]}{\delta \mathbf{\psi }}\right] ,  \label{qcevol}
\end{equation}%
with two length scales $l_{\underline{i}}=2\pi /k_{\underline{i}},$ for $%
\underline{i}=1,2.$ We can elaborate on local diffusion processes determined
by a free energy functional%
\begin{eqnarray}
F[\mathbf{\psi }] &=&\int \sqrt{\mid \ ^{3}g\mid }dx^{1}dx^{2}dy^{3}[\frac{1%
}{2}\mathbf{\psi \{-\epsilon +}\prod\limits_{\underline{i}=1,2}(k_{%
\underline{i}}^{2}+\ ^{3}\triangle )^{2}\mathbf{\}\psi +}\frac{1}{4}\mathbf{%
\psi }^{4}],  \label{efunctqc} \\
\mbox{ or }\ ^{b}F[\mathbf{\psi }] &=&\int \sqrt{\mid \ ^{3}g\mid }%
dx^{1}dx^{2}dy^{3}[\frac{1}{2}\mathbf{\psi \{-\epsilon +}\prod\limits_{%
\underline{i}=1,2}(k_{\underline{i}}^{2}+\ ^{b}\widehat{\Delta })^{2}\mathbf{%
\}\psi +}\frac{1}{4}\mathbf{\psi }^{4}],  \notag
\end{eqnarray}%
where $\mid \ ^{3}g\mid $ is the determinant of the 3-d space metric and $%
\mathbf{\epsilon }$ is a constant. For simplicity, we can restrict our
constructions to only non-relativistic diffusion processes, see \cite%
{vepjp12} for relativistic and N--adapted generalizations. The functional $\
^{b}F[\mathbf{\psi }]$ is defined by a nonholonomic deformation of the
Laplace operator, $\ ^{3}\triangle \rightarrow \ ^{b}\widehat{\Delta },$
resulting in a nonholonomic distortion of $F[\mathbf{\psi }].$

\subsubsection{Solitonic space like distributions and nonlinear waves}

Off-diagonal interactions determined by generating and integration functions
and nontrivial effective sources in MGTs and GR heterotic string gravity may
result in various effects with solitonic like distributions, cosmological
and geometric evolution models \cite%
{vcsf12,grvvap16,rajpoot17a,biv16p,abvfirs16p,aabovw16p}.

\paragraph{Stationary solitonic distributions: \newline
}

We shall use distributions $\ ^{sd}q=q(r,\vartheta ,\varphi )$ as solutions
of a respective class of solitonic 3-d equations
\begin{eqnarray}
\partial _{rr}^{2}q+\epsilon \partial _{\varphi }(\partial _{\vartheta
}q+6q\partial _{\varphi }q+\partial _{\varphi \varphi \varphi }^{3}q) =0, \
\partial _{rr}^{2}q+\epsilon \partial _{\vartheta }(\partial _{\varphi
}q+6q\partial _{\vartheta }q+\partial _{\vartheta \vartheta \vartheta
}^{3}q) &=&0,  \label{solitdistr} \\
\partial _{\vartheta \vartheta }^{2}q+\epsilon \partial _{\varphi }(\partial
_{r}q+6q\partial _{\varphi }q+\partial _{\varphi \varphi \varphi }^{3}q) =
0, \partial _{\vartheta \vartheta }^{2}q+\epsilon \partial _{r}(\partial
_{\varphi }q+6q\partial _{r}q+\partial _{rrr}^{3}q) &=&0,  \notag \\
\partial _{\varphi \varphi }^{2}q+\epsilon \partial _{r}(\partial
_{\vartheta }q+6q\partial _{r}q+\partial _{rrr}^{3}q) =0,\ \partial
_{\varphi \varphi }^{2}q+\epsilon \partial _{\vartheta }(\partial
_{r}q+6q\partial _{\vartheta }q+\partial _{\vartheta \vartheta \vartheta
}^{3}q) &=&0,  \notag
\end{eqnarray}%
for $\epsilon =\pm 1$. The left label $sd$ states that such a function is
defined as a "solitonic distribution" when in N-adapted frames a function $\
^{sd}q$ does not depend on time coordinate. The equations (\ref{solitdistr})
and their solutions can be redefined via frame/coordinate transforms for
stationary generating functions parameterized in non-spherical coordinates, $%
\ ^{sd}q=q(x^i,y^3)$.

\paragraph{Generating nonlinear solitonic waves: \newline
}

3-d solitonic waves with explicit dependence on time coordinate $t$ are
solutions of such nonlinear PDEs:{\small
\begin{equation}
\ \ ^{sw}\overline{q}=\left\{
\begin{array}{ccc}
\ \overline{q}(t,\vartheta ,\varphi ) & \mbox{ as a solution of } & \partial
_{tt}^{2}\ \ \overline{q}+\epsilon \frac{\partial }{\partial \varphi }%
[\partial _{\vartheta }\ \ \overline{q}+6\ \ \overline{q}\frac{\partial }{%
\partial \varphi }\ \ \overline{q}+\frac{\partial ^{3}}{(\partial \varphi
)^{3}}\ \ \overline{q}]=0; \\
\ \overline{q}(\vartheta ,t,\varphi ) & \mbox{ as a solution of } & \partial
_{\vartheta \vartheta }^{2}\ \overline{q}+\epsilon \frac{\partial }{\partial
\varphi }[\partial _{t}\ \overline{q}+6\ \overline{q}\frac{\partial }{%
\partial \varphi }\ \overline{q}+\frac{\partial ^{3}}{(\partial \varphi )^{3}%
}\ \overline{q}]=0; \\
\ \overline{q}(t,r,\varphi ) & \mbox{ as a solution of } & \partial
_{tt}^{2}\ \overline{q}+\epsilon \frac{\partial }{\partial \varphi }%
[\partial _{r}\ \overline{q}+6\ \overline{q}\frac{\partial }{\partial
\varphi }\ \overline{q}+\frac{\partial ^{3}}{(\partial \varphi )^{3}}\
\overline{q}]=0; \\
\overline{q}(r,t,\varphi ) & \mbox{ as a solution of } & \partial _{rr}^{2}%
\overline{q}+\epsilon \frac{\partial }{\partial \varphi }[\partial _{t}%
\overline{q}+6\overline{q}\frac{\partial }{\partial \varphi }\overline{q}+%
\frac{\partial ^{3}}{(\partial \varphi )^{3}}\overline{q}]=0; \\
\overline{q}(t,\varphi ,\vartheta ) & \mbox{ as a solution of } & \partial
_{tt}^{2}\overline{q}+\epsilon \frac{\partial }{\partial \vartheta }%
[\partial _{\varphi }\overline{q}+6\overline{q}\frac{\partial }{\partial
\vartheta }\overline{q}+\frac{\partial ^{3}}{(\partial \vartheta )^{3}}%
\overline{q}]=0; \\
\ \overline{q}(\varphi ,t,\vartheta ) & \mbox{ as a solution of } & \partial
_{\varphi \varphi }^{2}\ \overline{q}+\epsilon \frac{\partial }{\partial
\vartheta }[\partial _{t}\ \overline{q}+6\ \overline{q}\frac{\partial }{%
\partial \vartheta }\ \overline{q}+\frac{\partial ^{3}}{(\partial \vartheta
)^{3}}\ \overline{q}]=0.%
\end{array}%
\right.  \label{swaves}
\end{equation}%
} Applying general frame/coordinate transforms, solitonic waves of type $\
^{sw}\overline{q}=\overline{q}(x^{i},t),$ $=\overline{q}(x^{1},y^{3},t),$ or
$=\overline{q}(x^{2},y^{3},t),$ can be used as generating functions for
certain classes of nonholonomic deformations of stationary, or cosmological
metrics, and as generating sources.

\subsection{Effective sources with effective quasiperiodic free energy}

We can prescribe respective generating functions $\Phi ({r,\theta },\varphi
),$ or $\Psi ({r,\theta },\varphi ),$ (for stationary configurations with
nonlinear symmetry (\ref{nsym1a})), and $\overline{\Phi }(x^{i},t),$ or $%
\overline{\Psi }(x^{i},t),$ (for cosmological solutions with nonlinear
symmetry (\ref{nsym1b})) for quasiperiodic and/or aperiodic effective
sources (\ref{sourc}) in (\ref{qgreq}). Such configurations can be
determined by additive source functionals and effective cosmological
constants, or by nonlinear functionals.

\subsubsection{Additive effective sources and cosmological constants}

We shall be able to integrate in explicit form gravitational and matter
fields systems of nonlinear PDEs for parameterizations of N-adapted sources (%
\ref{dsourcparam}) as functionals of type
\begin{eqnarray}
\Upsilon ({r,\theta },\varphi ) &=&\ ^{P}\Upsilon \lbrack \ ^{P}\overline{q}%
_{0}]+\ ^{QC}\Upsilon \lbrack \ ^{QC}\overline{q}_{0}]+\ ^{\psi }\Upsilon
\lbrack \mathbf{\psi }_{0}]+\ ^{sd}\Upsilon \lbrack \ ^{sd}q]+\
^{sw}\Upsilon \lbrack \ ^{sw}\overline{q}_{0}],  \label{qpfuncts} \\
\overline{\Upsilon }(x^{i},t) &=&\ ^{P}\overline{\Upsilon }[\ ^{P}\overline{q%
}]+\ ^{QC}\overline{\Upsilon }[\ ^{QC}\overline{q}_{0}]+\ ^{\psi }\overline{%
\Upsilon }[\mathbf{\psi }]+\ ^{sd}\overline{\Upsilon }[\ ^{sd}q_{0}]+\ ^{sw}%
\overline{\Upsilon }[\ ^{sw}\overline{q}].  \label{qpfunctc}
\end{eqnarray}%
In these formulas, the left labels emphasize what types of effective
v-sources are considered. For simplicity, the h-sources can be taken any
general ones $~\ _{h}\Upsilon ({r,\theta }),$ or $~\ _{h}\overline{\Upsilon }%
(x^{i}).$ The functional dependence $[...]$ is parameterized for such
classes of functions: $\ ^{P}\overline{q}(x^{i},t)$ (\ref{patternfev}) with
stationary configurations $^{P}\overline{q}_{0}:=\ ^{P}\overline{q}%
(x^{i},t=t_{0})$ for a fixed value $t_{0};$ $\ ^{QC}\overline{q}=\overline{q}%
(x^{i},t)$ (\ref{evoleq}) with stationary configurations $\ ^{QC}\overline{q}
_{0}:=\ ^{QC}\overline{q}(x^{i},t_{0}); \mathbf{\psi (}x^{i},y^{3},t)$ (\ref%
{qcevol}) with stationary $\mathbf{\psi }_{0}:=\psi (x^{i},y^{3},t_{0});$ $\
^{sd}q(r,\vartheta ,\varphi )$ (\ref{solitdistr}) when cosmological
configurations are generated by any source $\ ^{sd}q_{0}:= \
^{sd}q(r,\vartheta ,\varphi _{0})$ for any fixed value $\varphi =\varphi
_{0};$ and $\ ^{sw}\overline{q}=\overline{q}(x^{i},t)$ (\ref{solitdistr})
with stationary $\ ^{sw}\overline{q}_{0}:=\ ^{sw}\overline{q}(x^{i},t_{0}).$

For additive sources (\ref{qpfuncts}) and (\ref{qpfunctc}), respective
stationary and cosmological configurations posses nonlinear symmetries:
{\small
\begin{eqnarray}
\Lambda \ \Psi ^{2} &=&\Phi ^{2}(|\ ^{P}\Upsilon |+|\ ^{QC}\Upsilon |+|\
^{\psi }\Upsilon |+|\ ^{sd}\Upsilon |+|\ ^{sw}\Upsilon |)-\int d\varphi \
\Phi ^{2}(|\ ^{P}\Upsilon |^{\diamond }+|\ ^{QC}\Upsilon |^{\diamond }+|\
^{\psi }\Upsilon |^{\diamond }+|\ ^{sd}\Upsilon |^{\diamond }+|\
^{sw}\Upsilon |^{\diamond }),  \notag \\
\Lambda &=&\ ^{P}\Lambda +\ ^{QC}\Lambda +\ ^{\psi }\Lambda +\ ^{sd}\Lambda
+\ ^{sw}\Lambda ,\mbox{ with effective cosmological constants };  \notag \\
&&  \label{nnsymads} \\
\overline{\Lambda }\overline{\ \Psi }^{2} &=&\overline{\Phi }^{2}(|\ ^{P}%
\overline{\Upsilon }|+|\ ^{QC}\overline{\Upsilon }|+|\ ^{\psi }\overline{%
\Upsilon }|+|\overline{\Upsilon }|+|\overline{\Upsilon }|)-\int dt\
\overline{\Phi }^{2}(|\ ^{P}\overline{\Upsilon }|^{\ast }+|\ ^{QC}\overline{%
\Upsilon }|^{\ast }+|\ ^{\psi }\overline{\Upsilon }|^{\ast }+|\ ^{sd}%
\overline{\Upsilon }|^{\ast }+|\ ^{sw}\overline{\Upsilon }|^{\ast }),  \notag
\\
\overline{\Lambda } &=&\ ^{P}\overline{\Lambda }+\ ^{QC}\overline{\Lambda }%
+\ ^{\psi }\overline{\Lambda }+\ ^{sd}\overline{\Lambda }+\ ^{sw}\overline{%
\Lambda },\mbox{ with effective cosmological constants }.  \label{nnsymadc}
\end{eqnarray}%
} Such formulas are of type (\ref{nsym1a}) and (\ref{nsym1b}) and can be
stated separately for all sources. The generating functions are chosen in a
general form $\Psi ,$ or $\Phi ,$ and, correspondingly, $\overline{\Psi },$
or $\overline{\Phi }.$ The QC like components of such quasiperiodic/
aperiodic structures are characterized by free energy functionals (\ref%
{dener}), for $\ ^{QC}\overline{q},$ and (\ref{efunctqc}), for $\mathbf{\psi
.}$

\subsubsection{Nonlinear functionals for effective sources and cosmological
constants}

The modified Einstein equations can be integrated in explicit form for
general nonlinear functionals%
\begin{eqnarray}
\ ^{qp}\Upsilon ({r,\theta },\varphi ) &=&\Upsilon \lbrack \ ^{P}\overline{q}%
_{0},\ ^{QC}\overline{q}_{0},\mathbf{\psi }_{0},\ ^{sd}q,\ ^{sw}\overline{q}%
_{0}]\mbox{ and }  \label{nonfqps} \\
\ ^{qp}\overline{\Upsilon }(x^{i},t) &=&\overline{\Upsilon }[\ ^{P}\overline{%
q},\ ^{QC}\overline{q}_{0},\mathbf{\psi ,}\ ^{sd}q_{0},\ ^{sw}\overline{q}].
\label{nonfqpc}
\end{eqnarray}%
The respective nonlinear symmetries (\ref{nsym1a}) and/or (\ref{nsym1b}) are
parameterized%
\begin{eqnarray}
\Lambda \ \Psi ^{2} &=&\Phi ^{2}|\ ^{qp}\Upsilon |-\int d\varphi \ \Phi
^{2}|\ ^{qp}\Upsilon |^{\diamond },\mbox{ for }\Lambda =\Lambda (\
^{P}\Lambda ,\ ^{QC}\Lambda ,\ ^{\psi }\Lambda ,\ ^{sd}\Lambda ,\
^{sw}\Lambda );  \label{nnsyms} \\
\overline{\Lambda }\ \overline{\Psi }^{2} &=&\overline{\Phi }^{2}|\ ^{qp}%
\overline{\Upsilon }|-\int dt\ \overline{\Phi }^{2}|\ ^{qp}\overline{%
\Upsilon }|^{\ast },\mbox{ for }\overline{\Lambda }=\overline{\Lambda }(\
^{P}\overline{\Lambda },\ ^{QC}\overline{\Lambda },\ ^{\psi }\overline{%
\Lambda },\ ^{sd}\overline{\Lambda },\ ^{sw}\overline{\Lambda }),
\label{nnsymc}
\end{eqnarray}%
resulting in functional dependencies of effective cosmological constants.

The formulas (\ref{nonfqps}) and (\ref{nonfqpc}) transform respectively in (%
\ref{qpfuncts}) and (\ref{qpfunctc}) for additional effective sources and
cosmological constants. We emphasize that nonlinear effects are very
important in structure formation and for multi-wave nonlinear (solitonic or
other types) matter fields and gravitational interactions. Nonlinear
dependencies and running of physical constants can be considered also in
quantum models. Such stationary and/or cosmological solutions (without
contributions of quasiperiodic fields) were studied in a series of our works
\cite{vplb16,vijgmmp07,vepjc14a,gvvcqg15}, see also recent results for
quasiperiodic structures \cite{biv16p,viepjc17,abvfirs16p,aabovw16p}.

For nonlinear effective sources (\ref{nonfqps}) and respective nonlinear
symmetries (\ref{nnsyms}), we can define additionally free energy
functionals (\ref{dener}), for $\ ^{QC}\overline{q},$ and (\ref{efunctqc}),
for $\psi$. Such values and constants have to be determined in explicit form
for astrophysical and/or cosmological configurations in order to describe
observational data for dark matter and dark energy distributions with
respective scales and/or quasiperiodic/ aperiodic structures.

\subsection{Quasiperiodic generating functions}

We can generate various types of gravitational field stationary and/or
cosmological configurations using respective classes of generating
functions. Such quasiperiodic/aperiodic configurations can be defined by
additive quadratic functionals, or in some general nonlinear forms.

\subsubsection{Additive quasiperiodic quadratic generating functions}

We can prescribe a nontrivial cosmological constant $\Lambda ,\ $\ or $\
\overline{\Lambda },$ and consider generating functions of type
\begin{eqnarray}
\Phi ^{2}({r,\theta },\varphi ) &=&\ ^{a}\Phi ^{2}=\ ^{P}\Phi ^{2}[\ ^{P}%
\overline{q}_{0}]+\ ^{QC}\Phi ^{2}[\ ^{QC}\overline{q}_{0}]+\ ^{\psi }\Phi
^{2}[\mathbf{\psi }_{0}]+\ ^{sd}\Phi ^{2}[\ ^{sd}q]+\ ^{sw}\Phi ^{2}[\ ^{sw}%
\overline{q}_{0}],\mbox{ or }  \label{adgfs} \\
\overline{\Phi }^{2}(x^{i},t) &=&\ ^{a}\overline{\Phi }^{2}=\ ^{P}\overline{%
\Phi }^{2}[\ ^{P}\overline{q}]+\ ^{QC}\overline{\Phi }^{2}[\ ^{QC}\overline{q%
}_{0}]+\ ^{\psi }\overline{\Phi }^{2}[\mathbf{\psi }]+\ ^{sd}\overline{\Phi }%
^{2}[\ ^{sd}q_{0}]+\ ^{sw}\overline{\Phi }^{2}[\ ^{sw}\overline{q}],
\label{adgfc}
\end{eqnarray}%
where the left label "a" emphasizes that we certain additions of
functionals. Nonlinear symmetries of type type (\ref{nsym1a}) and/or (\ref%
{nsym1b}) allow to compute respectively corresponding data $(\
^{a}\Psi,\Upsilon ),$ or $(\ ^{a}\overline{\Psi },\overline{\Upsilon }),$
for certain fixed effective sources $\Upsilon =\Upsilon ({r,\theta },\varphi
),$ or $\overline{\Upsilon }=\overline{\Upsilon }(x^{i},t).$ The respective
formulas are
\begin{eqnarray}
&&\Lambda \ (\ ^{P}\Psi ^{2}+\ ^{QC}\Psi ^{2}+\ ^{\psi }\Psi ^{2}+\
^{sd}\Psi ^{2}+\ ^{sw}\Psi ^{2})=  \label{adgfns1} \\
&&(\ ^{P}\Phi ^{2}+\ ^{QC}\Phi ^{2}+\ ^{\psi }\Phi ^{2}+\ ^{sd}\Phi ^{2}+\
^{sw}\Phi ^{2})|\Upsilon |-\int d\varphi \ (\ ^{P}\Phi ^{2}+\ ^{QC}\Phi
^{2}+\ ^{\psi }\Phi ^{2}+\ ^{sd}\Phi ^{2}+\ ^{sw}\Phi ^{2})|\Upsilon
|^{\diamond },  \notag \\
&&  \notag \\
&&\overline{\Lambda }\ (\ ^{P}\overline{\ \Psi }^{2}+\ ^{QC}\overline{\ \Psi
}^{2}+\ ^{\psi }\overline{\ \Psi }^{2}+\ ^{sd}\overline{\ \Psi }^{2}+\ ^{sw}%
\overline{\ \Psi }^{2})=  \label{adgfnc1} \\
&&(\ ^{P}\overline{\Phi }^{2}+\ ^{QC}\overline{\Phi }^{2}+\ ^{\psi }%
\overline{\Phi }^{2}+\ ^{sd}\overline{\Phi }^{2}+\ ^{sw}\overline{\Phi }%
^{2})|\overline{\Upsilon }|-\int dt\ (\ ^{P}\overline{\Phi }^{2}+\ ^{QC}%
\overline{\Phi }^{2}+\ ^{\psi }\overline{\Phi }^{2}+\ ^{sd}\overline{\Phi }%
^{2}+\ ^{sw}\overline{\Phi }^{2})|\overline{\Upsilon }|^{\ast }.  \notag
\end{eqnarray}%
The QC like components of such quasiperiodic/ aperiodic gravitational
structures are also characterized by respective effective free energy
functionals (\ref{dener}), for $\ ^{QC}\overline{q},$ and (\ref{efunctqc}),
for $\mathbf{\psi }$ encoding nontrivial vacuum structures with effective
cosmological constant.

\subsubsection{Nonlinear functionals for quasiperiodic quadratic generating
functions}

We shall be able to generate in explicit form solutions of modified Einstein
equations for nonlinear functionals
\begin{eqnarray}
\ \Phi ^{2}({r,\theta },\varphi ) &=&\ ^{qp}\Phi ^{2}=\Phi ^{2}[\ ^{P}%
\overline{q}_{0},\ ^{QC}\overline{q}_{0},\mathbf{\psi }_{0},\ ^{sd}q,\ ^{sw}%
\overline{q}_{0}]\mbox{ and }  \label{nadgfs} \\
\ ^{qp}\overline{\Phi }^{2}(x^{i},t) &=&\ ^{qp}\overline{\Phi }^{2}=%
\overline{\Phi }^{2}[\ ^{P}\overline{q},\ ^{QC}\overline{q},\mathbf{\psi ,}\
^{sd}q_{0},\ ^{sw}\overline{q}].  \label{nadgfc}
\end{eqnarray}%
Respective nonlinear symmetries (\ref{nsym1a}) and/or (\ref{nsym1b}) involve
correspondingly an effective cosmological constant, $\Lambda \ ,$ or $%
\overline{\Lambda },$ and nonholonomic constraints for v-sources, $\Upsilon (%
{r,\theta },\varphi ),$ or $\overline{\Upsilon }(x^{i},t),$
\begin{eqnarray}
\Lambda \ \ ^{qp}\Psi ^{2} &=&\ ^{qp}\Phi ^{2}|\Upsilon |-\int d\varphi \ \
^{qp}\Phi ^{2}|\Upsilon |^{\diamond }\mbox{ and/or  }  \label{nadgfns} \\
\overline{\Lambda }\ \ \ ^{qp}\overline{\Psi }^{2} &=&\ \ ^{qp}\overline{%
\Phi }^{2}|\overline{\Upsilon }|-\int dt\ \ \ ^{qp}\overline{\Phi }^{2}|%
\overline{\Upsilon }|^{\ast },  \label{nadgfnc}
\end{eqnarray}%
resulting in functional dependencies of effective cosmological constants.

The formulas (\ref{nadgfs})--(\ref{nadgfnc}) can be re-parameterized
respectively as (\ref{adgfs})--(\ref{adgfnc1}) when the values of effective
cosmological constant and matter sources are prescribed to be compatible
with experimental data. For such configurations, we can consider structures
described additionally by a free energy functional (\ref{dener}), for $\
^{QC}\overline{q},$ modeling QC like gravitational nonholonomic deformations.

\section{(Non)stationary Black Hole Deformations and Quasiperiodic Structures%
}

\label{sbh}Generic off-diagonal nonholonomic deformations of BH like
solutions were constructed in \cite%
{vexsol98,vjhep01,vjmp05,vmon06,vijgmmp07,vepjc13,gvvepjc14,grvvap16,rajpoot17a}
by applying the AFDM in various theories of (non) commutative, generalized
Finsler, supergravity and supersting MGTs and GR. For small parametric
decompositions, such solutions define black ellipsoid stationary
configurations, deformations of BH horizons and locally anisotropic
polarizations of physical constants, deformations of vacuum solutions into
nonvacuum ones. Different classes of solutions were generated with
nontrivial (non)commutative backgrounds, containing solitonic distributions
and/or describing propagation of black holes in extra dimensions, geometric
flows of black holes etc. The approach was generalized for various
quasiperiodic, quasicrystal and other type aperiodic solutions in (super)
string gravity and for nonholonomic Ricci soliton configurations \cite%
{biv16p,viepjc17}.

The goal of this section is to construct and study physical implications of
(non) stationary generic off-diagonal solutions describing deformations of
some prime BH solutions by quasiperiodic / aperiodic structures. The
necessary geometric formalism is summarized in Table 2 and appendices \ref%
{asst} and \ref{ass31}. In this work, the solutions are generated in general
forms (not depending on small parameters, see footnote \ref{fnsmallp}).

\subsection{Nonlinear PDEs for quasiperiodic / aperiodic stationary
configurations}

There are two possibilities to transform the (modified) Einstein equations (%
\ref{qgreq}) into systems of nonlinear PDEs (\ref{eq1a})--(\ref{eq4a}) with
quasiperiodic solutions. In the first case, one considers quasiperiodic
sources determined by some additive or general nonlinear functionals. In the
second case, additive/ general nonlinear functionals on quasiperiodic
solutions are prescribed for generating functions. It is possible also to
construct certain classes of solutions involving nonlinear functionals both
for generating functions and (effective) sources.

\subsubsection{Gravitational eqs and nonsingular solutions for stationary
quasiperiodic sources}

\paragraph{Stationary solutions with additive sources: \newline
}

Considering a source of type (\ref{qpfuncts}) (the left label $as$ is used
for "additive stationary"), when
\begin{eqnarray}
&&\ _{0}^{P}\Upsilon :=\ ^{P}\Upsilon \lbrack \ ^{P}\overline{q}_{0}],\
_{0}^{QC}\Upsilon :=\ ^{QC}\Upsilon \lbrack \ ^{QC}\overline{q}_{0}],\
_{0}^{\psi }\Upsilon :=\ ^{\psi }\Upsilon \lbrack \mathbf{\psi }_{0}]\ ,\
^{sd}\Upsilon :=\ ^{sd}\Upsilon \lbrack \ ^{sd}q],\ _{0}^{sw}\Upsilon :=\
^{sw}\Upsilon \lbrack \ ^{sw}\overline{q}_{0}],  \notag \\
&&\ ^{as}\Upsilon =\ _{0}^{P}\Upsilon +\ \ _{0}^{QC}\Upsilon +\ _{0}^{\psi
}\Upsilon +\ ^{sd}\Upsilon +\ _{0}^{sw}\Upsilon ,  \label{adsourcstat}
\end{eqnarray}%
the equation (\ref{e2a}) transforms into $\varpi ^{\diamond }\
h_{4}^{\diamond }=2h_{3}h_{4}\ \ ^{as}\Upsilon .$ This equation can be
integrated on $y^{3}=\varphi .$

Exact solutions for stationary configurations of the systems of nonlinear
PDEs (\ref{e1a})- (\ref{e4a}) can be constructed following the procedure
summarized in Table 2. We can generate such off-diagonal metrics and
generalized connections in general form for a generating function $h_{4}({%
r,\theta },\varphi )$ with Killing symmetry on $\partial _{t}$ determined by
sources $(\ _{h}\Upsilon ,\ ^{as}\Upsilon )$ and effective cosmological
constant $\ ^{as}\Lambda :=\ _{0}^{P}\Lambda +\ _{0}^{QC}\Lambda +\
_{0}^{\psi }\Lambda +\ ^{sd}\Lambda +\ _{0}^{sw}\Lambda $ related to $\
^{as}\Upsilon \ $ via nonlinear symmetry transforms (\ref{nnsymads}). The
corresponding class of quadratic elements defining stationary solutions can
be written in the form
\begin{eqnarray}
ds^{2} &=&e^{\ \psi (x^{k})}[(dx^{1})^{2}+(dx^{2})^{2}]-\frac{%
(h_{4}{}^{\diamond })^{2}}{|\int d\varphi \ \ \ ^{as}\Upsilon
h_{4}{}^{\diamond }|\ h_{4}}[dy^{3}+\frac{\partial _{i}(\int d\varphi \ \ \
^{as}\Upsilon \ h_{4}{}^{\diamond }])}{\ \ ^{as}\Upsilon \ h_{4}{}^{\diamond
}}dx^{i}]  \label{stasdm} \\
&&+h_{4}[dt+(\ _{1}n_{k}+4\ _{2}n_{k}\int d\varphi \frac{(h_{4}{}^{\diamond
})^{2}}{|\int dy^{3}\ ^{as}\Upsilon h_{4}^{\diamond }|\ (h_{4})^{5/2}}%
)dx^{k}].  \notag
\end{eqnarray}%
Such solutions are, in general, with nontrivial nonholonomically induced
torsion (\ref{dtors}). They can be re-defined equivalently in terms of
generating functions $\Psi ({r,\theta },\varphi )$ or $\Phi ({r,\theta }%
,\varphi ),$ see (\ref{gensolstat}).

LC-configurations in GR determined by quaisperiodic sources can be extracted
for additional zero torsion constraints resulting in a more special class of
"integrable" generating functions $(\check{h}_{4},$ and $\check{\Psi}({%
r,\theta },\varphi )$ and/or $\check{\Phi}({r,\theta },\varphi ))$ for
respective sources $\ ^{as}\check{\Upsilon}$ and $\ ^{as}\Lambda $ (\ref%
{lcsolstat}), \
\begin{equation}
ds^{2}=e^{\ \psi (x^{k})}[(dx^{1})^{2}+(dx^{2})^{2}]-\frac{(\check{h}%
_{4}{}^{\diamond })^{2}}{|\int d\varphi {\ ^{as}}\Upsilon \check{h}%
_{4}^{\diamond }|\ \check{h}_{4}}[d\varphi +(\partial _{i}\check{A})dx^{i}]+%
\check{h}_{4}\left[ dt+(\partial _{k}n)dx^{k}\right] .  \label{stasdmlc}
\end{equation}

Above classes of solutions define stationary off-diagonal gravitational
solutions generated by quasiperiodic/aperiodic additive sources $\
^{as}\Upsilon .$ The term $\ _{0}^{\psi }\Upsilon $ can be used for standard
and/or dark matter fields but other ones ($\ _{0}^{P}\Upsilon ,\ \
_{0}^{QC}\Upsilon ,\ ^{sd}\Upsilon ,\ _{0}^{sw}\Upsilon $) may model dark
matter stationary distributions with respective quasiperiodic/ aperiodic /
solitonic configurations. Considering smooth classes of generating /
integration functions and sources, we can construct various classes of
nonsingular exact solutions. Applying similar methods, we can generate, for
instance, generalizations of stationary models with nonlinear diffusion,
fractional, self-organizing and other type processes Refs. \cite%
{vcsf12,vepjc13}.

\paragraph{Stationary solutions with nonlinear functional sources: \newline
}

We can work with general nonlinear quasiperiodic / aperiodic / soliton
functionals for effective source of type $\ ^{qp}\Upsilon ({r,\theta }%
,\varphi )=\ ^{qp}\Upsilon \lbrack \ ^{P}\overline{q}_{0},\ ^{QC}\overline{q}%
_{0},\mathbf{\psi }_{0},\ ^{sd}q,\ ^{sw}\overline{q}_{0}]$ (\ref{nonfqps})
with nonlinear symmetries (\ref{nnsyms}). Stationary solutions of the
nonlinear system PDEs (\ref{e1a})- (\ref{e4a}) can be written as in Table 2,
\begin{eqnarray}
ds^{2} &=&e^{\ \psi (x^{k})}[(dx^{1})^{2}+(dx^{2})^{2}]-\frac{%
(h_{4}{}^{\diamond })^{2}}{|\int d\varphi \ ^{as}\Upsilon h_{4}{}^{\diamond
}|\ h_{4}}[dy^{3}+\frac{\partial _{i}(\int d\varphi \ \ \ ^{qp}\Upsilon \
h_{4}{}^{\diamond }])}{\ \ ^{qp}\Upsilon \ h_{4}{}^{\diamond }}dx^{i}
\label{stasdmn} \\
&&+h_{4}[dt+(\ _{1}n_{k}+4\ _{2}n_{k}\int d\varphi \frac{(h_{4}{}^{\diamond
})^{2}}{|\int dy^{3}\ ^{qp}\Upsilon h_{4}^{\diamond })|\ (h_{4})^{5/2}}%
)dx^{k}].  \notag
\end{eqnarray}%
This formula is similar to (\ref{stasdm}) but with another type of nonlinear
generation functions for (effective) sources for dark and/or usual matter
sources, when $\ ^{as}\Upsilon \rightarrow \ ^{qp}\Upsilon $ and $\
^{as}\Lambda \rightarrow $ $\ ^{qp}\Lambda .$ Similar re-definitions of
additive sources and cosmological constants in (\ref{stasdmlc}) into
nonlinear functionals generate nonlinear LC-configurations \
\begin{equation}
ds^{2}=e^{\ \psi (x^{k})}[(dx^{1})^{2}+(dx^{2})^{2}]-\frac{(\check{h}%
_{4}{}^{\diamond })^{2}}{|\int d\varphi {\ ^{qp}}\Upsilon \check{h}%
_{4}^{\diamond }|\ \check{h}_{4}}[d\varphi +(\partial _{i}\check{A})dx^{i}]+%
\check{h}_{4}\left[ dt+(\partial _{k}n)dx^{k}\right] .  \label{stasdmlcn}
\end{equation}%
\ We note that formulas (\ref{stasdmn}) and (\ref{stasdmlcn}) provide
respective generalizations of some classes of solutions (\ref{stasdm}) and (%
\ref{stasdmlc}) considering general "functionals of functionals" with
quasiperiodic (effective) real and dark matter structures which can be
organized by corresponding parameters in certain forms with cosmic webs,
filaments, quasiperiodic/ aperiodic and or/ solitonic distributions etc.

\subsubsection{Nonsingular solutions for stationary quasiperiodic generating
functions}

\paragraph{Stationary solutions with additive generating functions: \newline
}

For this class of solutions, the quasiperiodic / aperiodic / solitonic
structure is stated via generating functions (\ref{adgfs}) for nonlinear
gravitational field interactions without similar prescriptions for
(effective) sources as in the previous subsection. We write in brief
\begin{eqnarray}
&&\ _{0}^{P}\Phi ^{2}:=\ ^{P}\Phi ^{2}[\ ^{P}\overline{q}_{0}],\
_{0}^{QC}\Phi ^{2}:=\ ^{QC}\Phi ^{2}[\ ^{QC}\overline{q}_{0}],\ _{0}^{\psi
}\Phi ^{2}:=\ ^{\psi }\Phi ^{2}[\mathbf{\psi }_{0}],\ ^{sd}\Phi ^{2}=\
^{sd}\Phi ^{2}[\ ^{sd}q],\ _{0}^{sw}\Phi ^{2}:=\ ^{sw}\Phi ^{2}[\ ^{sw}%
\overline{q}_{0}],  \notag \\
&&\ ^{a}\Phi ^{2}=\ _{0}^{P}\Phi ^{2}+\ _{0}^{QC}\Phi ^{2}+\ _{0}^{\psi
}\Phi ^{2}\ +\ ^{sd}\Phi ^{2}+\ _{0}^{sw}\Phi ^{2}  \label{adstatgf}
\end{eqnarray}%
for additive generating functions subjected to nonlinear symmetries (\ref%
{adgfns1}).

The equation (\ref{e2a}) transforms into a functional equation $\varpi
^{\diamond }[\ ^{a}\Phi ,\Lambda ]\ h_{4}^{\diamond }[\ ^{a}\Phi ,\Lambda
]=2h_{3}[\ ^{a}\Phi ,\Lambda ]h_{4}[\ ^{a}\Phi ,\Lambda ]\Upsilon $, which
can be written in terms of functionals of type $h_{3,4}[\ ^{a}\Psi ,\Lambda
] $, and respective nonlinear functionals for coefficients in (\ref{e3a})
and (\ref{e4a}). The solutions of such equations can be parameterized in the
form (see the third parametrization in (\ref{gensolstat}))
\begin{eqnarray}
ds^{2} &=&e^{\ \psi (x^{k})}[(dx^{1})^{2}+(dx^{2})^{2}]-\frac{(\ ^{a}\Phi
^{2})^{\diamond }[(\ ^{a}\Phi ^{2})^{\diamond }]}{|\Lambda \int dy^{3}\
\Upsilon \lbrack (\ ^{a}\Phi )^{2}]^{\diamond }|\ (h_{4}^{[0]}-\ ^{a}\Phi
^{2}/4\Lambda )}[dy^{3}+\frac{\partial _{i}(\int dy^{3}\ \Upsilon \ (\
^{a}\Phi ^{2})^{\diamond })}{\Upsilon \ (\ ^{a}\Phi ^{2})^{\diamond }}dx^{i}]
\notag \\
&&+(h_{4}^{[0]}(x^{k})-\frac{\ ^{a}\Phi ^{2}}{4\Lambda })[dt+(_{1}n_{k}+\
_{2}n_{k}\int dy^{3}\frac{(\ ^{a}\Phi ^{2})^{\diamond }[(\ ^{a}\Phi
^{2})^{\diamond }]}{|\Lambda \int dy^{3}\ \Upsilon \lbrack (\ ^{a}\Phi
)^{2}]^{\diamond }|}|h_{4}^{[0]}(x^{k})-\frac{\ ^{a}\Phi ^{2}}{4\Lambda }%
|^{-5/2})dx^{k}],  \label{stasdmafa}
\end{eqnarray}

For LC-configurations, we obtain (see the third parametrization in (\ref%
{lcsolstat}))%
\begin{eqnarray}
ds^{2} &=&e^{\ \psi (x^{k})}[(dx^{1})^{2}+(dx^{2})^{2}]-\frac{(\ ^{a}\check{%
\Phi}^{2})^{\diamond }[(\ ^{a}\check{\Phi}^{2})^{\diamond }]}{|\Lambda \int
dy^{3}\ \Upsilon \lbrack (\ ^{a}\check{\Phi})^{2}]^{\diamond }|\
(h_{4}^{[0]}-\ ^{a}\check{\Phi}^{2}/4\Lambda )}[dy^{3}+(\partial _{i}\ ^{a}%
\check{A})dx^{i}]  \label{stasdmaflc} \\
&&+(h_{4}^{[0]}-\frac{\ ^{a}\check{\Phi}^{2}}{4\Lambda })\left[ dt+(\partial
_{k}n)dx^{k}\right] .  \notag
\end{eqnarray}

We emphasize that the (effective) source $\Upsilon $ in formulas (\ref%
{stasdmafa}) and (\ref{stasdmaflc}) is not obligatory quasiperiodic. Such a
source is considered for general matter fields (including both types of
standard and dark matter) and stationary distributions described by an
effective cosmological constant $\Lambda .$ In another turn, the
gravitational field distributions are with quasiperiodic / aperiodic /
solitonic structure. For such classes of solutions, the gravitational fields
encode certain dark energy nonlinear distributions with rich nonholonomic
geometric structure and generalized nonlinear symmetries. The coefficients
of this class of d--metrics can be chosen to be of necessary smooth class
(for instance, nonsingular ones).

\paragraph{Stationary solutions with nonlinear quasiperiodic functionals for
generating functions: \newline
}

Above classes of generic off-diagonal solutions can be generalized for
nonlinear quasiperiodic generating functionals $\ ^{qp}\Phi ^{2}[\ ^{P}%
\overline{q}_{0},\ ^{QC}\overline{q}_{0},\mathbf{\psi }_{0},\ ^{sd}q,\ ^{sw}%
\overline{q}_{0}]$ (\ref{nadgfs}) characterized by nonlinear symmetries of
type (\ref{nadgfns}). The equation (\ref{e2a}) transforms into a functional
equation $\varpi ^{\diamond }[\ ^{qp}\Phi ,\Lambda ]\ h_{4}^{\diamond }[\
^{qp}\Phi ,\Lambda ]=2h_{3}[\ ^{qp}\Phi ,\Lambda ]h_{4}[\ ^{qp}\Phi ,\Lambda
]\Upsilon ,$ which can be solved together with other equations form the
system (\ref{e1a})-(\ref{e4a}) following geometric methods summarized in
Table 2.

The solutions for such stationary configurations with general nonlinear
functionals for generating functions can be written the form (\ref%
{gensolstat}) (for simplicity, we consider only the third type
parametrization)
\begin{eqnarray}
ds^{2} &=&e^{\ \psi (x^{k})}[(dx^{1})^{2}+(dx^{2})^{2}]  \label{stasdmnf} \\
&&-\frac{(\ ^{qp}\Phi ^{2})^{\diamond }[(\ ^{qp}\Phi ^{2})^{\diamond }]}{%
|\Lambda \int dy^{3}\ \Upsilon \lbrack (\ ^{qp}\Phi )^{2}]^{\diamond }|\
(h_{4}^{[0]}-\ ^{qp}\Phi ^{2}/4\Lambda )}[dy^{3}+\frac{\partial _{i}(\int
dy^{3}\ \Upsilon \ (\ ^{qp}\Phi ^{2}]^{\diamond })}{\Upsilon \ (\ ^{qp}\Phi
^{2})^{\diamond }}dx^{i}]  \notag \\
&&+(h_{4}^{[0]}(x^{k})-\frac{\ ^{qp}\Phi ^{2}}{4\Lambda })[dt+(_{1}n_{k}+\
_{2}n_{k}\int dy^{3}\frac{(\ ^{qp}\Phi ^{2})^{\diamond }[(\ ^{qp}\Phi
^{2})^{\diamond }]}{|\Lambda \int dy^{3}\ \Upsilon \lbrack (\ ^{qp}\Phi
)^{2}]^{\diamond }|}|h_{4}^{[0]}(x^{k})-\frac{\ ^{qp}\Phi ^{2}}{4\Lambda }%
|^{-5/2})dx^{k}].  \notag
\end{eqnarray}%
We can impose additional zero torsion constraints and extract
LC-configurations as in (\ref{lcsolstat}),%
\begin{eqnarray}
ds^{2} &=&e^{\ \psi (x^{k})}[(dx^{1})^{2}+(dx^{2})^{2}]-\frac{(\ ^{qp}\check{%
\Phi}^{2})^{\diamond }[(\ ^{qp}\check{\Phi}^{2})^{\diamond }]}{|\Lambda \int
dy^{3}\ \Upsilon \lbrack (\ ^{qp}\check{\Phi})^{2}]^{\diamond }|\
(h_{4}^{[0]}-\ ^{qp}\check{\Phi}^{2}/4\Lambda )}[dy^{3}+(\partial _{i}\ ^{a}%
\check{A})dx^{i}]  \label{stasdmnflc} \\
&&+(h_{4}^{[0]}-\frac{\ ^{qp}\check{\Phi}^{2}}{4\Lambda })\left[
dt+(\partial _{k}n)dx^{k}\right] .  \notag
\end{eqnarray}

Considering additional assumptions and approximations for additive
functionals, the formulas (\ref{stasdmnf}) and (\ref{stasdmnflc}) transform
respectively into (\ref{stasdmafa}) and (\ref{stasdmaflc}). For such classes
of solutions, the gravitational fields encode certain dark energy nonlinear
distributions with a more rich nonholonomic geometric structure and
generalized nonlinear symmetries when quasiperiodicity is induced from
"quasiperiodicity" of matter fields. The coefficients of this class of
d--metrics can be chosen to be of necessary smooth class (for instance,
nonsingular ones) but can involve certain stochastic sources and fractional
derivative processes.

\subsubsection{Stationary solutions from nonlinear functionals for
quasiperiodic coefficients \& sources}

In a more general context, we can generate nonsingural stationary
off-diagonal generalized quasiperiodic solutions of the (modified) Einstein
equations determined both by nonlinear functionals for generating functions,
$\ ^{qp}\Phi $ (\ref{nadgfs}) and nonlinear functionals for (effective)
sources, $\ ^{qp}\Upsilon $ (\ref{nonfqps}). The quasiperiodic data (for
instance, scales, interaction constants and associated free energies)\ for
the generating functions are different from the quasiperiodic data for
sources. Nevertheless, such data can not be arbitrary independent ones but
subjected to nonlinear symmetries generalizing (\ref{nadgfns}) and (\ref%
{nnsyms}), $\ ^{qp}\Lambda \ \ ^{qp}\Psi ^{2}=\ ^{qp}\Phi ^{2}|\
^{qp}\Upsilon |-\int d\varphi \ \ ^{qp}\Phi ^{2}|\ ^{qp}\Upsilon |^{\diamond
}$. For additive functionals both in the gravitational and (effective
sources), such a nonlinear symmetry transforms into $\ ^{a}\Lambda \
^{a}\Psi ^{2}=\ ^{a}\Phi ^{2}|\ ^{a}\Upsilon |-\int d\varphi \ ^{a}\Phi
^{2}|\ ^{a}\Upsilon |^{\diamond }$, which is a generalization of (\ref%
{adgfns1}) and (\ref{nnsymads}).

Following again the procedure summarized in Table 2 but for the data $(\
^{qp}\Psi ,\ ^{qp}\Upsilon ),$ and/or, equivalently, $(\ \ ^{qp}\Phi ,\
^{qp}\Lambda ),$ the general multi-functional nonlinear generalization of
stationary solutions (\ref{stasdmn}) and (\ref{stasdmnf}) are constructed in
the form
\begin{eqnarray}
ds^{2} &=&e^{\ \psi (x^{k})}[(dx^{1})^{2}+(dx^{2})^{2}]  \label{stasdmafg} \\
&&-\frac{(\ ^{qp}\Phi ^{2})^{\diamond }[(\ ^{qp}\Phi ^{2})^{\diamond }]}{|\
^{qp}\Lambda \int dy^{3}\ \ ^{qp}\Upsilon \lbrack (\ ^{qp}\Phi
)^{2}]^{\diamond }|\ (h_{4}^{[0]}-\ ^{qp}\Phi ^{2}/4\ ^{qp}\Lambda )}[dy^{3}+%
\frac{\partial _{i}(\int dy^{3}\ \ ^{qp}\Upsilon \ (\ ^{qp}\Phi
^{2}]^{\diamond })}{\ ^{qp}\Upsilon \ (\ ^{qp}\Phi ^{2})^{\diamond }}dx^{i}]
\notag \\
&&+(h_{4}^{[0]}(x^{k})-\frac{\ ^{qp}\Phi ^{2}}{4\ ^{qp}\Lambda }%
)[dt+(_{1}n_{k}+\ _{2}n_{k}\int dy^{3}\frac{(\ ^{qp}\Phi ^{2})^{\diamond
}[(\ ^{qp}\Phi ^{2})^{\diamond }]}{|\ ^{qp}\Lambda \int dy^{3}\
^{qp}\Upsilon \lbrack (\ ^{qp}\Phi )^{2}]^{\diamond }|}|h_{4}^{[0]}(x^{k})-%
\frac{\ ^{qp}\Phi ^{2}}{4\ ^{qp}\Lambda }|^{-5/2})dx^{k}].  \notag
\end{eqnarray}%
For LC-configurations, we obtain multi-functional nonlinear generalizations
of (\ref{stasdmlcn}) and (\ref{stasdmnflc}), for stationary solutions in GR,
\begin{eqnarray}
ds^{2} &=&e^{\ \psi (x^{k})}[(dx^{1})^{2}+(dx^{2})^{2}]  \label{stasdmaflcg}
\\
&&-\frac{(\ ^{qp}\check{\Phi}^{2})^{\diamond }[(\ ^{qp}\check{\Phi}%
^{2})^{\diamond }]}{|\ ^{qp}\Lambda \int dy^{3}\ \ ^{qp}\Upsilon \lbrack (\
^{qp}\check{\Phi})^{2}]^{\diamond }|\ (h_{4}^{[0]}-\ ^{qp}\check{\Phi}%
^{2}/4\ ^{qp}\Lambda )}[dy^{3}+(\partial _{i}\ ^{a}\check{A}%
)dx^{i}]+(h_{4}^{[0]}-\frac{\ ^{qp}\check{\Phi}^{2}}{4\ ^{qp}\Lambda })\left[
dt+(\partial _{k}n)dx^{k}\right] .  \notag
\end{eqnarray}

The class of solutions (\ref{stasdmafg}) describes off-diagonal stationary
configurations determined by multi-functional nonlinear quasiperiodic
structures both for the dark energy (nonlinear gravitational distributions)
and for the dark (and standard) matter fields. In explicit form, such data
can be stated to be compatible with observations in modern astrophysics and
cosmology.

\subsection{BHs in (off-) diagonal quasiperiodic media}

Various classes of generic off-diagonal stationary quasiperiodic solutions
can be described in terms of $\eta $--polarization functions as in appendix %
\ref{ass31} and following the geometric method summarized in Tables 1 and 2.
In this section, we consider a primary BH d-metric $\mathbf{\mathring{g}}$ (%
\ref{primedm}) defined by data $[\mathring{g}_{i}(r,\theta ,\varphi ),
\mathring{g}_{a}=\mathring{h}_{a}(r,\theta ,\varphi );\mathring{N}_{k}^{3}=%
\mathring{w}_{k}(r,\theta ,\varphi ),\mathring{N}_{k}^{4}=\mathring{n}%
_{k}(r,\theta ,\varphi )]$ which can be diagonalized (for simplicity, we
consider the Schwarzschild metric) by frame/ coordinate transforms. The
stationary quasiperiodic solutions will be determined by target metrics $%
\widehat{\mathbf{g}}$ generated by nonholonomic deformations $\mathbf{%
\mathring{g}}\rightarrow \widehat{\mathbf{g}}\mathbf{=}[g_{i}(x^{k})=\eta
_{i}\mathring{g}_{i},g_{b}(x^{k},y^{3})=\eta _{b}\mathring{g}%
_{b},N_{i}^{a}(x^{k},y^{3})=\ \eta _{i}^{a}\mathring{N}_{i}^{a}].$ The
quadratic elements corresponding to by $\widehat{\mathbf{g}}$ are
parameterized in some forms similar to (\ref{dme}),
\begin{equation}
ds^{2}=\eta _{i}(r,\theta ,\varphi )\mathring{g}_{i}(r,\theta ,\varphi
)[dx^{i}(r,\theta ,\varphi )]^{2}+\eta _{a}(r,\theta ,\varphi )\mathring{g}%
_{a}(r,\theta ,\varphi )[d\varphi +\eta _{k}^{a}(r,\theta ,\varphi )%
\mathring{N}_{k}^{a}(r,\theta ,\varphi )dx^{k}(r,\theta ,\varphi )]^{2},
\label{statsingpf}
\end{equation}%
with summation on repeating contracted low-up indices.

\subsubsection{Singular solutions generated by stationary quasiperiodic
sources}

We consider qusiperiodic sources of type $\ ^{qp}\Upsilon \lbrack \ ^{P}%
\overline{q}_{0},\ ^{QC}\overline{q}_{0},\mathbf{\psi }_{0},\ ^{sd}q,\ ^{sw}%
\overline{q}_{0}]$ as in (\ref{stasdmn}) and compute the coefficients of (%
\ref{statsingpf}) following formulas.%
\begin{eqnarray}
\eta _{i} &=&e^{\ \psi (x^{k})}/\mathring{g}_{i};\eta _{3}=-\frac{4[(|\eta
_{4}\mathring{h}_{4}|^{1/2})^{\diamond }]^{2}}{\mathring{h}_{3}|\int dy^{3}\
^{qp}\Upsilon (\eta _{4}\mathring{h}_{4})^{\diamond }|\ };\eta _{4}=\eta
_{4}(r,\theta ,\varphi )\mbox{ as a generating
function};  \label{statsingpfqp} \\
\eta _{i}^{3} &=&\frac{\partial _{i}\ \int dy^{3}\ ^{qp}\Upsilon (\eta _{4}\
\mathring{h}_{4})^{\diamond }}{\mathring{w}_{i}\ ^{qp}\Upsilon \ (\eta _{4}\
\mathring{h}_{4})^{\diamond }};\eta _{k}^{4}\ =\frac{\ _{1}n_{k}}{\mathring{n%
}_{k}}+16\ \ \frac{\ _{2}n_{k}}{\mathring{n}_{k}}\int dy^{3}\frac{\left(
[(\eta _{4}\mathring{h}_{4})^{-1/4}]^{\diamond }\right) ^{2}}{|\int dy^{3}\
^{qp}\Upsilon (\eta _{4}\ \mathring{h}_{4})^{\diamond }|\ }.  \notag
\end{eqnarray}

In these formulas, $\eta _{4}(r,\theta ,\varphi )$ is taken as a (non)
singular generating function. Other types of generating functions are
determined with nonlinear symmetries (\ref{nsym1a}) and functionals of $\eta
_{4}(r,\theta ,\varphi )$ and data for the prime d-metric, {\small
\begin{equation*}
\Phi ^{2}=-4\ ^{qp}\Lambda h_{4}=-4\ ^{qp}\Lambda \lbrack \eta _{4}({%
r,\theta },\varphi )\mathring{h}_{4}({r,\theta },\varphi )],\ (\Psi
^{2})^{\diamond }=-\int d\varphi \ \ ^{qp}\Upsilon h_{4}^{\ \diamond }=-\int
d\varphi \ \ ^{qp}\Upsilon ({r,\theta },\varphi )[\eta _{4}({r,\theta }%
,\varphi )\mathring{h}_{4}({r,\theta },\varphi )]^{\diamond }.
\end{equation*}%
}

We can constrain the coefficients (\ref{statsingpfqp}) to a subclass of data
generating target stationary off-diagonal metrics of type (\ref{lcsolstat})
with zero torsion,
\begin{equation*}
\eta _{i}=e^{\ \psi (x^{k})}/\mathring{g}_{i};\eta _{3}=-\frac{4[(|\check{%
\eta}_{4}\mathring{h}_{4}|^{1/2})^{\diamond }]^{2}}{\mathring{h}_{3}|\int
d\varphi \ ^{qp}\check{\Upsilon}(\check{\eta}_{4}\mathring{h}_{4})^{\diamond
}|\ };\mbox{  generating function }\eta _{4}=\check{\eta}_{4}(r,\theta
,\varphi );\ \eta _{i}^{3}=\partial _{i}\check{A}/\mathring{w}_{k},\eta
_{k}^{4}=\frac{\ \partial _{k}n}{\mathring{n}_{k}}.
\end{equation*}

In above formulas, the nonlinear functionals for the quasiperiodic v-source
and (effective) cosmological constant can be changed into additive
functionals $\ ^{qp}\Upsilon \rightarrow \ ^{as}\Upsilon $ and $\
^{qp}\Lambda \rightarrow $ $\ ^{as}\Lambda .$ The singular behaviour of such
solutions is generated by some prime BH data. For certain classes of
generating functions and sources, the same type of singularity is preserved.
Such examples have been studied in some general forms for small parametric
deformations. Similar stationary configurations can be computed for general
quasiperiodic structures. The constructions depend on the type of explicit
model we construct (for instance, with certain web / filament / solitonic
stationary distributions). We can treat such generic off-diagonal stationary
solutions as certain conventional nonholonomically deformed BH
configurations imbedded into certain (non) singular media determined by
stationary dark and usual matter quasiperiodic distributions.

\subsubsection{BH solutions deformed by stationary quasiperiodic generating
functions}

Let us consider another class of solutions (\ref{statsingpf}) when the
coefficients of the d-metrics are determined by nonlinear generating
functionals $\ ^{qp}\Phi ^{2}[\ ^{P}\overline{q}_{0},\ ^{QC}\overline{q}_{0},%
\mathbf{\psi }_{0},\ ^{sd}q,\ ^{sw}\overline{q}_{0}]$ (\ref{nadgfs}).
Similarly, we can consider additive functionals $\ ^{a}\Phi ^{2}$ (\ref%
{adstatgf}) including terms with possible itnegration functions$\ h_{4}^{[0]}
$ for $h_{4}$ and defined by some prescribed data $\Upsilon ({r,\theta }%
,\varphi )$ and $\Lambda .$ The formulas for nonlinear symmetries (\ref%
{nadgfns}) allow us to find (recurrently) corresponding nonlinear
functionals, $\ ^{qp}\eta _{4}({r,\theta },\varphi )$, or additive
functionals, $\ ^{a}\eta _{4}({r,\theta },\varphi ),$ and related
polarization functions, {\small
\begin{equation*}
\ \ ^{qp}\eta _{4}=-\ ^{qp}\Phi ^{2}({r,\theta },\varphi )/4\Lambda
\mathring{h}_{4}({r,\theta },\varphi ),\ (\ \ ^{qp}\Psi ^{2})^{\diamond
}=-\int d\varphi \ \ ^{qp}\Upsilon h_{4}^{\ \diamond }=-\int d\varphi \ \
\Upsilon ({r,\theta },\varphi )[\ \ ^{qp}\eta _{4}({r,\theta },\varphi )%
\mathring{h}_{4}({r,\theta },\varphi )]^{\diamond }.
\end{equation*}%
}

The coefficients of (\ref{statsingpf}) are computed {\small
\begin{eqnarray}
\eta _{i} &=&e^{\ \psi (x^{k})}/\mathring{g}_{i};\eta _{3}=-\frac{4[(|\
^{qp}\eta _{4}\mathring{h}_{4}|^{1/2})^{\diamond }]^{2}}{\mathring{h}%
_{3}|\int dy^{3}\Upsilon (\ ^{qp}\eta _{4}\mathring{h}_{4})^{\diamond }|\ }%
;\eta _{4}=\ ^{qp}\eta _{4}(r,\theta ,\varphi )%
\mbox{ as a generating
function};  \label{statsingpfqp1} \\
\eta _{i}^{3} &=&\frac{\partial _{i}\ \int d\varphi \ \Upsilon (\ ^{qp}\eta
_{4}\ \mathring{h}_{4})^{\diamond }}{\mathring{w}_{i}\ \Upsilon \ (\
^{qp}\eta _{4}\ \mathring{h}_{4})^{\diamond }};\eta _{k}^{4}=\frac{\
_{1}n_{k}}{\mathring{n}_{k}}+16\ \ \frac{\ _{2}n_{k}}{\mathring{n}_{k}}\int
d\varphi \frac{\left( \lbrack (\ ^{qp}\eta _{4}\mathring{h}%
_{4})^{-1/4}]^{\diamond }\right) ^{2}}{|\int dy^{3}\Upsilon (\ ^{qp}\eta
_{4}\ \mathring{h}_{4})^{\diamond }|\ }.  \notag
\end{eqnarray}%
}

Target stationary off-diagonal metrics (\ref{lcsolstat}) with zero torsion
can be generated by polarization functions subjected to additional
integrability conditions,
\begin{equation*}
\eta _{i}=e^{\ \psi (x^{k})}/\mathring{g}_{i};\ \eta _{3}=-\frac{4[(|\ ^{qp}%
\check{\eta}_{4}\mathring{h}_{4}|^{1/2})^{\diamond }]^{2}}{\mathring{h}%
_{3}|\int dy^{3}\check{\Upsilon}(\ ^{qp}\check{\eta}_{4}\mathring{h}%
_{4})^{\diamond }|\ };\mbox{ generating function }\eta _{4}=\check{\eta}%
_{4}(r,\theta ,\varphi );\eta _{i}^{3}=\frac{\partial _{i}\ ^{qp}\check{A}}{%
\mathring{w}_{k}},\eta _{k}^{4}=\frac{\ \partial _{k}n}{\mathring{n}_{k}}.
\end{equation*}

The solutions determined in this subsection describe certain
nonholonomically deformed BH configurations self-consistently imbedded into
a quasiperiodic gravitational (dark energy) media.

\subsubsection{Stationary BH deformations by quasiperiodic sources and
generating functions}

The most general class of nonholonomic stationary quasiperiodic deformations
of BHs is determined by corresponding nonlinear quaisperiodic functionals
both for the generating functions and (effective) sources. Nonlinear
superpositions of solutions of type (\ref{statsingpf}) and (\ref%
{statsingpfqp1}) are determined by coefficients of (\ref{statsingpf})
computed {\small
\begin{eqnarray}
\eta _{i} &=&e^{\ \psi (x^{k})}/\mathring{g}_{i};\eta _{3}=-\frac{4[(|\
^{qp}\eta _{4}\mathring{h}_{4}|^{1/2})^{\diamond }]^{2}}{\mathring{h}%
_{3}|\int d\varphi \ ^{qp}\Upsilon (\ ^{qp}\eta _{4}\mathring{h}%
_{4})^{\diamond }|\ };\eta _{4}=\ ^{qp}\eta _{4}(r,\theta ,\varphi )%
\mbox{ as a generating
function};  \label{statsingpfqp12} \\
\eta _{i}^{3} &=&\frac{\partial _{i}\ \int d\varphi \ \ ^{qp}\Upsilon (\
^{qp}\eta _{4}\ \mathring{h}_{4})^{\diamond }}{\mathring{w}_{i}\ \
^{qp}\Upsilon \ (\ ^{qp}\eta _{4}\ \mathring{h}_{4})^{\diamond }};\ \eta
_{k}^{4}=\frac{\ _{1}n_{k}}{\mathring{n}_{k}}+16\ \ \frac{\ _{2}n_{k}}{%
\mathring{n}_{k}}\int d\varphi \frac{\left( \lbrack (\ ^{qp}\eta _{4}%
\mathring{h}_{4})^{-1/4}]^{\diamond }\right) ^{2}}{|\int dy^{3}\
^{qp}\Upsilon (\ ^{qp}\eta _{4}\ \mathring{h}_{4})^{\diamond }|\ }.  \notag
\end{eqnarray}%
} In such formulas, there are considered nonlinear generating functionals $\
^{qp}\Phi ^{2}[\ ^{P}\overline{q}_{0},\ ^{QC}\overline{q}_{0},\mathbf{\psi }%
_{0},\ ^{sd}q,\ ^{sw}\overline{q}_{0}]$ (\ref{nadgfs}). Similarly, we can
consider additive functionals $\ ^{a}\Phi ^{2}$ (\ref{adstatgf}) for some
prescribed nonlinear functionals $\ ^{qp}\Upsilon ({r,\theta },\varphi )$
and $\ ^{qp}\Lambda $ (in particular, additive nonlinear functionals, $\
^{as}\Upsilon $ (\ref{adsourcstat}) and $\ ^{as}\Lambda $ can be taken). All
such data are related via nonlinear symmetries generalizing (\ref{nadgfns})
which allows to find (recurrently) corresponding nonlinear functionals, $\
^{qp}\eta _{3}({r,\theta },\varphi )$, , or additive functionals, $\
^{a}\eta _{3}({r,\theta },\varphi ),$ for the polarization function, {\small
\begin{equation*}
\ \ ^{qp}\eta _{4}=-\ ^{qp}\Phi ^{2}({r,\theta },\varphi )/4\ ^{qp}\Lambda
\mathring{h}_{4}({r,\theta },\varphi ),\ (\ \ ^{qp}\Psi ^{2})^{\diamond
}=-\int d\varphi \ \ ^{qp}\Upsilon h_{4}^{\ \diamond }=-\int d\varphi \ \ \
^{qp}\Upsilon ({r,\theta },\varphi )[\ \ ^{qp}\eta _{4}({r,\theta },\varphi )%
\mathring{h}_{4}({r,\theta },\varphi )]^{\diamond }.
\end{equation*}%
}

Imposing additional conditions for zero torsion, target stationary
off-diagonal metrics (\ref{lcsolstat}) are generated
\begin{equation*}
\eta _{i}=\frac{e^{\ \psi (x^{k})}}{\mathring{g}_{i}};\ \eta _{3}=-\frac{%
4[(|\ ^{qp}\check{\eta}_{4}\mathring{h}_{4}|^{1/2})^{\diamond }]^{2}}{%
\mathring{h}_{3}|\int d\varphi \ ^{qp}\check{\Upsilon}(\ _{4}^{qp}\check{\eta%
}\mathring{h}_{4})^{\diamond }|\ };\mbox{  generating function }\eta _{4}=%
\check{\eta}_{4}(r,\theta ,\varphi );\eta _{i}^{3}=\frac{\partial _{i}\ ^{qp}%
\check{A}}{\mathring{w}_{k}},\eta _{k}^{4}=\frac{\ \partial _{k}n}{\mathring{%
n}_{k}}.
\end{equation*}

Finally, we conclude that the class of stationary solutions of type (\ref%
{statsingpfqp12}) describes nonholonomic deformations of a BH
self-consistently imbedded into quasiperiodic gravitational (dark energy)
backgrounds and quasiperiodic dark/ standard matter. Such distributions can
be with a very rich geometric structure (in general, with singular, or
nonsingular configurations) and described by respective nonlinear symmetries.

\subsection{Nonstationary deformations of BH metrics into quasiperiodic
cosmological solutions}

Prime BH metrics can be deformed nonholonomically into certain classes of
exact quasiperiodic solutions in MGTs and GR depending in explicit form on a
time like coordinate $t$. For such configurations, nonlinear quasiperiodic
interactions can transform black hole spacetimes into locally anisotropic
cosmological ones. In order to construct such exact solutions, we consider a
primary BH d-metric $\mathbf{\mathring{g}}$ (\ref{primedm}) which via
coordinate transforms is parameterized by data $[\mathring{g}_{i}(x^{k}),%
\mathring{g}_{a}=\mathring{h}_{a}(x^{k});\mathring{N}_{k}^{3}=\mathring{w}
_{k}(x^{i}),\mathring{N}_{k}^{4}=\mathring{n}_{k}(x^{i})].$ The
non-stationary quasiperiodic solutions will be determined by target metrics $%
\widehat{\mathbf{g}}$ generated by nonholonomic deformations $\ $%
\begin{equation*}
\mathbf{\mathring{g}}\rightarrow \widehat{\mathbf{g}}\mathbf{=}%
[g_{i}(x^{k})=\eta _{i}\mathring{g}_{i},g_{b}(x^{k},y^{4}=t)=\overline{\eta }%
_{b}\mathring{g}_{b},N_{i}^{a}(x^{k},y^{4}=t)=\ \overline{\eta }_{i}^{a}%
\mathring{N}_{i}^{a}].
\end{equation*}%
The quadratic elements generated by such $\widehat{\mathbf{g}}$ are
parameterized,
\begin{equation}
ds^{2}=\eta _{i}(x^{k})\mathring{g}_{i}(x^{k})[dx^{i}]^{2}+\overline{\eta }%
_{a}(x^{k},t)\mathring{g}_{a}(x^{i})[dt+\overline{\eta }_{k}^{a}(x^{i},t)%
\mathring{N}_{k}^{a}(x^{i})dx^{k}]^{2},  \label{dmbhcosm}
\end{equation}%
with summation on low-up indices.

The quasiperiodic sources are given by nonlinear functionals for effective
sources,
\begin{equation*}
\ ^{qp}\overline{\Upsilon }(x^{i},t)=\ ^{qp}\overline{\Upsilon }[\ \
^{P}\Upsilon \lbrack \ ^{P}\overline{q}(x^{i},t)],\ ^{QC}\Upsilon \lbrack \
^{QC}\overline{q}(x^{i},t)],\ ^{\psi }\Upsilon \lbrack \mathbf{\psi }%
(x^{i},t)],\ ^{sd}\Upsilon \lbrack \ ^{sd}q(x^{i})],\ \ ^{sw}\Upsilon
\lbrack \ ^{sw}\overline{q}(x^{i},t)]],
\end{equation*}%
and nonlinear functionals on cosmological constants $\ ^{qp}\overline{%
\Lambda }=\ \overline{\Lambda }[\ ^{P}\Lambda ,\ \ ^{QC}\Lambda ,\ ^{\psi
}\Lambda ,\ ^{sd}\Lambda ,\ ^{sw}\Lambda ].$ In general, we can consider
some nonlinear functionals for generating functions $\ ^{qp}\overline{\Phi }%
^{2}[^{P}\overline{q}(x^{i},t),\ ^{QC}\overline{q}(x^{i},t),\ \mathbf{\psi }%
(x^{i},t),\ ^{sd}q(x^{i}),\ ^{sw}\overline{q}(x^{i},t)],$ when the nonlinear
symmetries are stated by formulas {\small
\begin{equation*}
\ ^{qp}\overline{\eta }_{3}=\ \ ^{qp}\overline{\Phi }^{2}({x}^{i},t)/4\ ^{qp}%
\overline{\Lambda }\overline{\mathring{h}}_{3}(x^{i},t),\ (\ ^{qp}\overline{%
\Psi }^{2})^{\ast }=-\int dt\ \ ^{qp}\overline{\Upsilon }\ [\ ^{qp}\overline{%
\eta }_{3}(x^{i},t)\overline{\mathring{h}}_{3}(x^{i},t)]^{\ast }.
\end{equation*}%
}

The equation (\ref{e2a}) transforms into $\overline{\varpi }^{\ast }\
\overline{h}_{3}^{\ast }=2h_{3}h_{4}\ \ ^{qp}\overline{\Upsilon }.$ This
equation can be integrated on $y^{4}=t.$ Exact solutions for non-stationary
configurations of the systems of nonlinear PDEs (\ref{e1a})- (\ref{e4a}) can
be constructed following the procedure summarized as an interference in
Tables 2 and 3 and appendices \ref{ascs} and \ref{ass32}. The respective
coefficients for (\ref{dmbhcosm}) are computed {\small
\begin{eqnarray}
\eta _{i} &=&e^{\ \psi (x^{k})}/\mathring{g}_{i};\overline{\eta }_{3}=\ ^{qp}%
\overline{\eta }_{3}(x^{i},t)\mbox{ as a generating
function};\overline{\eta }_{4}=-\frac{4[(|\ ^{qp}\overline{\eta }_{3}%
\mathring{h}_{3}|^{1/2})^{\ast }]^{2}}{\mathring{h}_{4}|\int dt\ ^{qp}%
\overline{\Upsilon }(\ ^{qp}\overline{\eta }_{3}\mathring{h}_{3})^{\ast }|\ }%
;  \label{dmbhcosmdc} \\
\overline{\eta }_{k}^{3} &=&\frac{\ _{1}\overline{n}_{k}}{\overline{%
\mathring{n}}_{k}}+4\frac{\ _{2}\overline{n}_{k}}{\overline{\mathring{n}}_{k}%
}\int dt\frac{\left( [(\ ^{qp}\overline{\eta }_{3}\mathring{h}%
_{3})^{-1/4}]^{\ast }\right) ^{2}}{|\int dt\ ^{qp}\overline{\Upsilon }(\
^{qp}\overline{\eta }_{3}\mathring{h}_{3})^{\ast }|\ };\ \overline{\eta }%
_{i}^{4}=\frac{\partial _{i}\ \int dt\ ^{qp}\overline{\Upsilon }(\ ^{qp}%
\overline{\eta }_{3}\mathring{h}_{3})^{\ast }}{\overline{\mathring{w}}_{i}\
^{qp}\overline{\Upsilon }(\ ^{qp}\overline{\eta }_{3}\mathring{h}_{3})^{\ast
}}.  \notag
\end{eqnarray}%
}

For additional zero torsion conditions, we generate target non-stationary
off-diagonal metrics (\ref{lcsolcosm}) defined by coefficients
\begin{equation*}
\eta _{i}=\frac{e^{\ \psi (x^{k})}}{\mathring{g}_{i}};\mbox{ generating
function }\overline{\eta }_{3}=\overline{\check{\eta}}_{3}(x^{i},t);%
\overline{\eta }_{4}=-\frac{4[(|\ ^{qp}\overline{\eta }_{3}\mathring{h}%
_{3}|^{1/2})^{\ast }]^{2}}{\mathring{h}_{4}|\int dt\ ^{qp}\overline{\check{%
\Upsilon}}(\ ^{qp}\overline{\eta }_{3}\mathring{h}_{3})^{\ast }|\ };%
\overline{\eta }_{k}^{3}=\frac{\ \partial _{k}n}{\mathring{n}_{k}},\overline{%
\eta }_{i}^{4}=\frac{\partial _{i}\ ^{qp}\overline{\check{A}}}{\mathring{w}%
_{k}}.
\end{equation*}

Non-stationary solutions generated from BH prime metrics (with zero or
non-zero nonholonomically induced torsions) can be constructed in a similar
manner for additive functionals for (effective) sources and generating
functions which are similar to those for locally anisotropic spacetime. The
integration functions and primary data are taken for BH stationary
configurations.

Off-diagonal solutions of type (\ref{dmbhcosmdc}) describe nonholonomic
deformations of a BH solution by imbedding it self-consistently in a
dynamical quasiperiodic gravitational (dark energy) background and
quasiperiodic dark/ standard matter interacting fields. Such locally
anisotropic cosmological like spacetimes contain BH like structures are
described by relativistic evolution of geometric structure, singular, or
nonsingular, and described by respective nonlinear symmetries. For certain
special parameterizations, we can model "dissipation" of singular BH
structure into certain types (locally anisotropic) cosmological spacetimes.

\section{Off-diagonal Quasiperiodic Cosmological Spacetimes}

\label{scc}Locally anisotropic and inhomogeneous cosmological solutions and
accelerating universe scenarios (in MGTs, GR, and geometric flow theories)
were studied in a series of works \cite%
{vmon06,vepjc14a,gvvcqg15,vplb16,ruchin17,rajpoot17}, see also references
therein. Various classes of generic off-diagonal cosmological metrics were
constructed by applying the AFDM as a geometric alternative to numeric
methods \cite{macpherson17}. Recently, the approach was developed by
constructing quasiperiodic cosmological solutions with small parametric
deformations \cite{abvfirs16p,aabovw16p}.

The goal of this section is to study physical implications of (non)
stationary generic off-diagonal solutions describing deformations of some
prime cosmological spacetimes by quasiperiodic / aperiodic / solitonic
and/or pattern forming structures. The necessary geometric formalism is
summarized in Table 3 and appendices \ref{ascs} and \ref{ass32}. We
emphasize that in this work the cosmological solutions are constructed for
general nonlinear or additive nonlinear functionals for generating functions
and (effective) sources without additional assumptions on modelling small
parameter configurations (see footnote \ref{fnsmallp}).

\subsection{Nonlinear PDEs for quasiperiodic/ aperiodic cosmological
configurations}

There are two possibilities to transform the (modified) Einstein equations (%
\ref{qgreq}) into systems of nonlinear PDEs (\ref{eq1b})--(\ref{eq4b}) with
quasiperiodic solutions depending in explicit form on a time like variable.
In the first case, one considers quasiperiodic sources determined by some
additive or general nonlinear functionals. In the second case, respective
nonlinear functionals determining quasiperiodic solutions are prescribed for
generating functions. It is also possible to construct certain classes of
locally anisotropic and inhomogeneous cosmological solutions by considering
nonlinear / additive functionals both for generating functions and
(effective) sources.

\subsubsection{Cosmological solutions for quasiperiodic sources}

\paragraph{Cosmological solutions generated by additive functionals for
sources: \newline
}

Let us consider an additive functional for a quasiperiodic source of type $%
\overline{\Upsilon }(x^{i},t)$ (\ref{qpfunctc}),%
\begin{equation}
\ ^{as}\overline{\Upsilon }=\ ^{P}\overline{\Upsilon }+\ _{0}^{QC}\overline{%
\Upsilon }+\ ^{\psi }\overline{\Upsilon }+\ _{0}^{sd}\overline{\Upsilon }[\
^{sd}q_{0}]+\ ^{sw}\overline{\Upsilon }[\ ^{sw}\overline{q}],
\label{adsourccosm}
\end{equation}%
for $\ _{0}^{QC}\overline{\Upsilon }=\ ^{QC}\overline{\Upsilon }[\ ^{QC}%
\overline{q}_{0}]$ and $\ _{0}^{sd}\overline{\Upsilon }=\ ^{sd}\overline{%
\Upsilon }[\ ^{sd}q_{0}],$ and an associated additive cosmological constant $%
\ ^{as}\overline{\Lambda }=\ ^{P}\overline{\Lambda }+\ ^{QC}\overline{%
\Lambda }+\ ^{\psi }\overline{\Lambda }+\ ^{sd}\overline{\Lambda }+\ ^{sw}%
\overline{\Lambda }.$ Such values are related to different types of
generating functions via nonlinear symmetries of type (\ref{nnsymadc}). The
equation (\ref{e2}) transforms into ${\overline{\varpi }}^{\ast }\ \overline{%
h}_{3}^{\ast }=2\overline{h}_{3}\overline{h}_{4}\ ^{as}\overline{\Upsilon }$
which can be integrated on time like variable $y^{4}=t.$ The systems of
nonlinear PDEs (\ref{e1})- (\ref{e4}) can be integrated following the
procedure summarized in Table 3. Such generic off-diagonal solutions are
parameterized in the form
\begin{eqnarray}
ds^{2} &=&e^{\ \psi (x^{k})}[(dx^{1})^{2}+(dx^{2})^{2}]+\overline{h}%
_{3}[dy^{3}+(\ _{1}n_{k}+4\ _{2}n_{k}\int dt\frac{(\overline{h}_{3}{}^{\ast
})^{2}}{|\int dt\ \ ^{as}\overline{\Upsilon }\overline{h}_{3}{}^{\ast }|(%
\overline{h}_{3})^{5/2}})dx^{k}]  \notag \\
&&-\frac{(\overline{h}_{3}{}^{\ast })^{2}}{|\int dt\ ^{as}\overline{\Upsilon
}\overline{h}_{3}{}^{\ast }|\ \overline{h}_{3}}[dt+\frac{\partial _{i}(\int
dt\ \ ^{as}\overline{\Upsilon }\ \overline{h}_{3}{}^{\ast }])}{\ \ ^{as}%
\overline{\Upsilon }\ \overline{h}_{3}{}^{\ast }}dx^{i}].  \label{cosmasdm}
\end{eqnarray}%
For local pseudo-Riemannian configurations, we have to fix respective sign
of the coefficient $\overline{h}_{4}({x}^{k},t)$ which can be considered as
a generating function with Killing symmetry on $\partial _{3}$ determined by
sources $(\ _{h}\overline{\Upsilon },\ ^{as}\overline{\Upsilon }).$ Such
solutions are of type (\ref{gensolcosm}) and can be re-written equivalently
with coefficients stated as functionals of $\ ^{as}\overline{\Phi }$ and $\
^{as}\overline{\Psi }$ define by nonlinear symmetry formulas determined by
additive (effective) sources and cosmological constants.

We can extract from (\ref{cosmasdm}) locally anisotropic generic
off-diagonal cosmological LC-configurations in GR determined by
quaisperiodic sources by imposing additional zero torsion constraints. This
restricts the respective classes of "integrable" generating functions $(%
\overline{\check{h}}_{4},\overline{\check{\Psi}}(x^{i},t)$ and/or $\overline{%
\check{\Phi}}(x^{i},t))$ for sources $\ ^{as}\overline{\check{\Upsilon}}$
and $\ ^{as}\overline{\Lambda }$ (\ref{lcsolcosm}),
\begin{equation}
ds^{2}=e^{\ \psi (x^{k})}[(dx^{1})^{2}+(dx^{2})^{2}]+\overline{\check{h}}_{3}%
\left[ dy^{3}+(\partial _{k}\overline{n})dx^{k}\right] -\frac{(\overline{%
\check{h}}_{3}{}^{\ast })^{2}}{|\int dt\ ^{as}\overline{\check{\Upsilon}}%
\overline{\check{h}}_{3}{}^{\ast }|\ \overline{\check{h}}_{3}}[dt+(\partial
_{i}\overline{\check{A}})dx^{i}.  \label{cosmadmlc}
\end{equation}

Above linear quadratic elements define off-diagonal cosmological solutions
generated by quasiperiodic additive sources $\ ^{as}\overline{\Upsilon }.$
The term $\ ^{\psi }\overline{\Upsilon }$ encodes contributions of standard
and/or dark matter fields. The sources $(\ ^{P}\overline{\Upsilon }, \
_{0}^{QC}\overline{\Upsilon },\ _{0}^{sd}\overline{\Upsilon },\ ^{sw}%
\overline{\Upsilon }$) can be chosen to model dark matter stationary
distributions with respective quasiperiodic/ aperiodic / solitonic nonlinear
wave interactions. Considering smooth classes of generating / integration
functions and sources, we can generate nonsingular cosmological exact
solutions. Applying similar methods, we can study, for instance,
generalizations of cosmological models to effects determined by nonlinear
diffusion, fractional, self-organizing and other type processes as we proved
in Refs. \cite{vcsf12,vepjc13}.

\paragraph{Cosmological solutions generated by nonlinear functionals for
quasiperiodic sources: \newline
}

Exact solutions can be generated by nonlinear quasiperiodic functionals for
effective sources, $\ ^{qp}\overline{\Upsilon }(x^{i},t)=\ ^{qp}\overline{%
\Upsilon }[\ ^{P}\overline{\Upsilon },\ _{0}^{QC}\overline{\Upsilon },\
^{\psi }\overline{\Upsilon },\ _{0}^{sd}\overline{\Upsilon },\ ^{sw}%
\overline{\Upsilon }]$ (\ref{nonfqpc}) with nonlinear symmetries (\ref%
{nnsymc}). Using the AFDM summarized in Table 3, we construct cosmological
solutions of the nonlinear system PDEs (\ref{e1})- (\ref{e4}),
\begin{eqnarray}
ds^{2} &=&e^{\ \psi (x^{k})}[(dx^{1})^{2}+(dx^{2})^{2}]+\overline{h}%
_{3}[dy^{3}+(\ _{1}n_{k}+4\ _{2}n_{k}\int dt\frac{(\overline{h}_{3}{}^{\ast
})^{2}}{|\int dt\ \ ^{qp}\overline{\Upsilon }\overline{h}_{3}{}^{\ast }|(%
\overline{h}_{3})^{5/2}})dx^{k}]  \notag \\
&&-\frac{(\overline{h}_{3}{}^{\ast })^{2}}{|\int dt\ ^{qp}\overline{\Upsilon
}\ \overline{h}_{3}{}^{\ast }|\ \overline{h}_{3}}[dt+\frac{\partial
_{i}(\int dt\ \ ^{qp}\overline{\Upsilon }\ \overline{h}_{3}{}^{\ast }])}{\ \
^{qp}\overline{\Upsilon }\ \overline{h}_{3}{}^{\ast }}dx^{i}].
\label{cosmnfdm}
\end{eqnarray}%
This formula is similar to (\ref{stasdm}) but with another type of nonlinear
generation functions for (effective) sources for dark and/or usual matter
sources, when $\ ^{as}\Upsilon \rightarrow \ ^{qp}\Upsilon $ and $\
^{as}\Lambda \rightarrow $ $\ ^{qp}\Lambda .$ Similar re-definitions of
additive sources and cosmological constants in (\ref{stasdmlc}) into
nonlinear functionals generate nonlinear LC-configurations \
\begin{equation}
ds^{2}=e^{\ \psi (x^{k})}[(dx^{1})^{2}+(dx^{2})^{2}]+\overline{\check{h}}_{3}%
\left[ dy^{3}+(\partial _{k}\overline{n})dx^{k}\right] -\frac{(\overline{%
\check{h}}_{3}{}^{\ast })^{2}}{|\int dt\ ({\ ^{qp}}\overline{\check{\Upsilon}%
})\overline{h}_{3}{}^{\ast }|\ \overline{\check{h}}_{3}}[dt+(\partial _{i}%
\overline{\check{A}})dx^{i}].  \label{cosmnfdmlcn}
\end{equation}%
For additive functionals for cosmological sources, the formulas (\ref%
{cosmnfdm}) and (\ref{cosmnfdmlcn}) transforms respectively into quadratic
linear elements (\ref{cosmasdm}) and (\ref{cosmadmlc}). Considering small
parametric deformations for 4-d cosmological solutions, we can reproduce the
results from \cite{abvfirs16p,aabovw16p}.

\subsubsection{Cosmological solutions for nonstationary quasiperiodic
generating functions}

In this subsection, the sources are with arbitrary data $\overline{\mathbf{%
\Upsilon }}_{\ \nu }^{\mu }=[~\ _{h}\overline{\Upsilon }({x}^{k}),\overline{%
\Upsilon }({x}^{k},t)]$ but the generating functions are considered for some
additive or general nonlinear functionals with quasiperiodic structure.

\paragraph{Locally anisotropic and inhomogeneous cosmological metrics with
additive generating functions: \newline
}

For this class of solutions, the quasiperiodic / aperiodic / solitonic
structure is stated via generating functions (\ref{adgfc}) for nonlinear
quasiperiodic gravitational field interactions buth without explicit
prescriptions on any quasiperiodic structure for (effective) sources. In
brief, such additive functionals are written
\begin{equation*}
\ ^{a}\overline{\Phi }^{2}(x^{i},t)=\ ^{P}\overline{\Phi }^{2}[\ ^{P}%
\overline{q}]+\ ^{QC}\overline{\Phi }^{2}[\ ^{QC}\overline{q}]+\ ^{\psi }%
\overline{\Phi }^{2}[\mathbf{\psi }]+\ _{0}^{sd}\overline{\Phi }^{2}[\
^{sd}q]+\ ^{sw}\overline{\Phi }^{2}[\ ^{sw}\overline{q}]
\end{equation*}%
with nonlinear symmetries (\ref{adgfnc1}), where $\ _{0}^{sd}\overline{\Phi }%
^{2}[\ ^{sd}q]$ is taken for any functions $\ ^{sd}q(x^{k})$ but other
components are considered as functionals on respective functions $\ ^{P}%
\overline{q},\ ^{QC}\overline{q},\mathbf{\psi ,}$ and $\ ^{sw}\overline{q}$
on $(x^{i},t).$ In result, the equation (\ref{e2}) transforms into a
functional equation $\overline{\varpi }^{\ast }[\ ^{a}\overline{\Phi },%
\overline{\Lambda }]\ \overline{h}_{3}^{\ast }[\ ^{a}\overline{\Phi },%
\overline{\Lambda }]=2\overline{h}_{3}[\ ^{a}\overline{\Phi },\overline{%
\Lambda }]\overline{h}_{4}[\ ^{a}\overline{\Phi },\overline{\Lambda }]%
\overline{\Upsilon }$, which can be written equivalently in different forms
with functionals of type $\overline{h}_{3,4}[\ ^{a}\overline{\Psi },%
\overline{\Lambda }]$ and respective nonlinear functionals for coefficients
in (\ref{e3}) and (\ref{e4}). The solutions of such equations can be
parameterized in a form (\ref{gensolcosm}), {\small
\begin{eqnarray}
ds^{2} &=&e^{\ \psi (x^{k})}[(dx^{1})^{2}+(dx^{2})^{2}]  \notag \\
&&+(h_{3}^{[0]}-\ ^{a}\overline{\Phi }^{2}/4\overline{\Lambda })[dy^{3}+(\
_{1}n_{k}+\ _{2}n_{k}\int dt\frac{[(\ ^{a}\overline{\Phi }^{2}){}^{\ast
}]^{2}}{4|\overline{\Lambda }\int dt\ \overline{\Upsilon }(\ ^{a}\overline{%
\Phi }^{2})^{\ast }|(h_{3}^{[0]}-\ ^{a}\overline{\Phi }^{2}/4\overline{%
\Lambda })^{5/2}})dx^{k}]  \notag \\
&&-\frac{[(\ ^{a}\overline{\Phi }^{2}){}^{\ast }]^{2}}{4|\overline{\Lambda }%
\int dt\ \overline{\Upsilon }(\ ^{a}\overline{\Phi }^{2})^{\ast }|\
(h_{3}^{[0]}-\ ^{a}\overline{\Phi }^{2}/4\overline{\Lambda })}[dt+\frac{%
\partial _{i}(\int dt\ \ \overline{\Upsilon }\ (\ ^{a}\overline{\Phi }%
^{2}){}^{\ast }])}{\ \overline{\Upsilon }\ (\ ^{a}\overline{\Phi }%
^{2}){}^{\ast }}dx^{i}].  \label{cosmdmaf}
\end{eqnarray}%
}

For LC-configurations, we obtain (see the third parametrization in (\ref%
{lcsolcosm}))%
\begin{equation}
ds^{2}=e^{\ \psi (x^{k})}[(dx^{1})^{2}+(dx^{2})^{2}]+(h_{3}^{[0]}-\frac{%
\overline{\check{\Phi}}^{2}}{4\overline{\Lambda }})[dy^{3}+(\partial _{k}%
\overline{n})dx^{k}]-\frac{[(\ ^{a}\overline{\check{\Phi}}^{2}){}^{\ast
}]^{2}}{4|\overline{\Lambda }\int dt\ \overline{\check{\Upsilon}}(\ ^{a}%
\overline{\check{\Phi}}^{2}){}^{\ast }|\ (h_{3}^{[0]}-\ (\ ^{a}\overline{%
\check{\Phi}}^{2})/4\overline{\Lambda })}[dt+(\partial _{i}\overline{\check{A%
}})dx^{i}].  \label{cosmdmaflc}
\end{equation}

We can consider small parametric decompositions and frame/coordinate
transforms in order to related such solutions to some well known (off)
diagonal cosmological metrics.

\paragraph{Cosmological solutions with nonlinear quasiperiodic functionals
for generating functions: \newline
}

The formulas (\ref{cosmdmaf}) and (\ref{cosmdmaflc}) can be generalized in
order to generate solutions from nonlinear functionals for generating
functions when nonlinear quasiperiodic generating functionals $\ ^{qp}%
\overline{\Phi }^{2}[\ ^{P}\overline{\Phi },\ ^{QC}\overline{\Phi },\ ^{\psi
}\overline{\Phi },\ _{0}^{sd}\overline{\Phi }^{2},\ ^{sw}\overline{\Phi }%
^{2}]$ (\ref{nadgfc}) characterized by nonlinear symmetries of type (\ref%
{nadgfnc}). The equation (\ref{e2}) transforms into a more general
functional equation, $\overline{\varpi }^{\ast }[\ ^{qp}\overline{\Phi },%
\overline{\Lambda }]\ \overline{h}_{3}^{\ast }[\ ^{qp}\overline{\Phi },%
\overline{\Lambda }]=2\overline{h}_{3}[\ ^{qp}\overline{\Phi },\overline{%
\Lambda }]\overline{h}_{4}[\ ^{qp}\overline{\Phi },\overline{\Lambda }]%
\overline{\Upsilon }$. Such a nonlinear system of PDEs can be solved
together with other equations form the system (\ref{e1})-(\ref{e4})
following the steps summarized in Table 3. We obtain such solutions: {\small
\begin{eqnarray}
ds^{2} &=&e^{\ \psi (x^{k})}[(dx^{1})^{2}+(dx^{2})^{2}]  \notag \\
&&+(h_{3}^{[0]}-\ ^{qp}\overline{\Phi }^{2}/4\overline{\Lambda })[dy^{3}+(\
_{1}n_{k}+\ _{2}n_{k}\int dt\frac{[(\ ^{qp}\overline{\Phi }^{2}){}^{\ast
}]^{2}}{4|\overline{\Lambda }\int dt\ \overline{\Upsilon }(\ ^{qp}\overline{%
\Phi }^{2})^{\ast }|(h_{3}^{[0]}-\ ^{qp}\overline{\Phi }^{2}/4\overline{%
\Lambda })^{5/2}})dx^{k}]  \notag \\
&&-\frac{[(\ ^{qp}\overline{\Phi }^{2}){}^{\ast }]^{2}}{4|\overline{\Lambda }%
\int dt\ \overline{\Upsilon }(\ ^{qp}\overline{\Phi }^{2})^{\ast }|\
(h_{3}^{[0]}-\ ^{qp}\overline{\Phi }^{2}/4\overline{\Lambda })}[dt+\frac{%
\partial _{i}(\int dt\ \ \overline{\Upsilon }\ (\ ^{qp}\overline{\Phi }%
^{2}){}^{\ast }])}{\ \overline{\Upsilon }\ (\ ^{qp}\overline{\Phi }%
^{2}){}^{\ast }}dx^{i}].  \label{cosmsdmnf}
\end{eqnarray}%
} For zero torsion constraints in order to extract LC-configurations,%
\begin{eqnarray}
ds^{2} &=&e^{\ \psi (x^{k})}[(dx^{1})^{2}+(dx^{2})^{2}]+(h_{3}^{[0]}-\ ^{qp}%
\overline{\check{\Phi}}^{2}/4\overline{\Lambda })[dy^{3}+(\partial _{k}%
\overline{n})dx^{k}]-  \label{cosmdmnflc} \\
&&\frac{[(\ ^{qp}\overline{\check{\Phi}}^{2}){}^{\ast }]^{2}}{4|\overline{%
\Lambda }\int dt\ \overline{\Upsilon }(\ ^{qp}\overline{\check{\Phi}}%
^{2})^{\ast }|\ (h_{3}^{[0]}-\ ^{qp}\overline{\check{\Phi}}^{2}/4\overline{%
\Lambda })}[dt+(\partial _{i}\overline{\check{A}})dx^{i}].  \notag
\end{eqnarray}%
Here we emphasize that there is certain duality between formulas when $%
y^{4}=t\leftrightarrow y^{3\text{ }}$and, respectively, "overlined" values
are changed into "not overlined" ones, and inversely. For such nonholonomic
dual transforms, the formulas (\ref{stasdmnf}) and (\ref{stasdmnflc})
transform into corresponding (\ref{cosmsdmnf}) and (\ref{cosmdmnflc})
[inverse maps can be also considered]. The coefficients of this class of
d--metrics can be chosen to be of necessary smooth class and involve certain
stochastic sources and fractional derivative processes. Such nonholonomic
deformation and generalized transform may change the topological spacetime
structure and encode dark energy and dark matter effects.

\subsubsection{Cosmology from nonstationary functionals for quasiperiodic
coefficients \& sources}

Conventionally, all classes of considered above cosmological solutions can
be formulated in terms of generalized quasiperiodic nonlinear functionals
both for generating functions, $\ ^{qp}\Phi $ (\ref{nadgfc}) and nonlinear
functionals for (effective) sources, $\ ^{qp}\Upsilon $ (\ref{nonfqpc}).
Such data are subjected to conditions of nonlinear symmetries generalizing (%
\ref{nadgfnc}) and (\ref{nnsymc}), when $\ ^{qp}\overline{\Lambda }\ \ ^{qp}%
\overline{\Psi }^{2}=\ ^{qp}\overline{\Phi }^{2}|\ ^{qp}\overline{\Upsilon }%
|-\int dt\ \ ^{qp}\overline{\Phi }^{2}|\ ^{qp}\overline{\Upsilon }|^{\ast }.$
Similar nonlinear symmetries can be considered for additive functionals both
for the gravitational fields and (effective) sources, when (\ref{adgfnc1})
and (\ref{nnsymadc}) transform into $\ ^{a}\overline{\Lambda \ }\ ^{a}%
\overline{\Psi }^{2}=\ ^{a}\overline{\Phi }^{2}|\ ^{a}\overline{\Upsilon }%
|-\int dt\ ^{a}\overline{\Phi }^{2}|\ ^{a}\overline{\Upsilon }|^{\ast }.$
The procedure summarized in Table 3 and generalized for the data $(\ ^{qp}%
\overline{\Psi },\ ^{qp}\overline{\Upsilon }),$ and/or, equivalently, $(\ \
^{qp}\overline{\Phi },\ ^{qp}\overline{\Lambda }),$ allows us to construct
multi-functional nonlinear quasiperiodic cosmological configurations,
{\small
\begin{eqnarray}
ds^{2} &=&e^{\ \psi (x^{k})}[(dx^{1})^{2}+(dx^{2})^{2}]+(h_{3}^{[0]}-\ ^{qp}%
\overline{\Phi }^{2}/4\ ^{qp}\overline{\Lambda })  \notag \\
&&[dy^{3}+(\ _{1}n_{k}+\ _{2}n_{k}\int dt\frac{[(\ ^{qp}\overline{\Phi }%
^{2}){}^{\ast }]^{2}}{4\ ^{qp}|\overline{\Lambda }\int dt\ \ ^{qp}\overline{%
\Upsilon }(\ ^{qp}\overline{\Phi }^{2})^{\ast }|(h_{3}^{[0]}-\ ^{qp}%
\overline{\Phi }^{2}/4\ ^{qp}\overline{\Lambda })^{5/2}})dx^{k}]  \notag \\
&&-\frac{[(\ ^{qp}\overline{\Phi }^{2}){}^{\ast }]^{2}}{4|\ ^{qp}\overline{%
\Lambda }\int dt\ \ ^{qp}\overline{\Upsilon }(\ ^{qp}\overline{\Phi }%
^{2})^{\ast }|\ (h_{3}^{[0]}-\ ^{qp}\overline{\Phi }^{2}/4\ ^{qp}\overline{%
\Lambda })}[dt+\frac{\partial _{i}(\int dt\ \ \ ^{qp}\overline{\Upsilon }\
(\ ^{qp}\overline{\Phi }^{2}){}^{\ast }])}{\ \ ^{qp}\overline{\Upsilon }\ (\
^{qp}\overline{\Phi }^{2}){}^{\ast }}dx^{i}].  \label{cosmsdmnfg}
\end{eqnarray}%
} For LC-configurations, we obtain multi-functional nonlinear
generalizations of (\ref{cosmsdmnf}) and (\ref{cosmdmnflc}) which result in
more general classes of locally anisotropic and inhomogeneous solutions in
GR,%
\begin{eqnarray}
ds^{2} &=&e^{\ \psi (x^{k})}[(dx^{1})^{2}+(dx^{2})^{2}]+(h_{3}^{[0]}-\frac{\
^{qp}\overline{\check{\Phi}}^{2}}{4\ ^{qp}\overline{\Lambda }}%
)[dy^{3}+(\partial _{k}\overline{n})dx^{k}]  \label{cosmdmnflcg} \\
&&-\frac{[(\ ^{qp}\overline{\check{\Phi}}^{2}){}^{\ast }]^{2}}{4|\ ^{qp}%
\overline{\Lambda }\int dt\ \ ^{qp}\overline{\Upsilon }(\ ^{qp}\overline{%
\check{\Phi}}^{2})^{\ast }|\ (h_{3}^{[0]}-\ ^{qp}\overline{\check{\Phi}}%
^{2}/4\ ^{qp}\overline{\Lambda })}[dt+(\partial _{i}\overline{\check{A}}%
)dx^{i}].  \notag
\end{eqnarray}

The classes (\ref{cosmsdmnfg}), of cosmological solutions in MGTs, and (\ref%
{cosmdmnflcg}), for cosmological solutions in GR, describe off-diagonal
non-stationary configurations determined by multi-functional nonlinear
quasiperiodic structures. Such rich geometric nonholonomically dynamical
structures can be described both for the dark energy (nonlinear
gravitational distributions) and for the dark (and standard) matter fields.
In general, such cosmological solutions may not have smooth limits to
well-known cosmological metric (for instance, FLRW, or any Bianchi type; we
shall study such configurations in next subsections). In explicit form, the
geometric data for such generic off-diagonal cosmological solutions can be
stated to explain various observations in modern cosmology.

\subsection{Cosmological metrics evolving in (off-) diagonal quasiperiodic
media}

In this subsection, generic off-diagonal quasiperiodic cosmological
solutions are constructed in terms of $\eta $--polarization functions as in
appendix \ref{ass32} and following the geometric method summarized in Tables
1 and 3. We consider a primary cosmological d-metric $\mathbf{\mathring{g}}$
(\ref{primedm}) defined by data $[\mathring{g}_{i}(x^{k},t),\mathring{g}_{a}=%
\mathring{h}_{a}(x^{k},t);\mathring{N}_{k}^{3}= \mathring{n}_{k}(x^{i},t),%
\mathring{N}_{k}^{4}=\mathring{w}_{k}(x^{i},t)]$ which can be diagonalized
for a FLRW cosmological metric (in general, we can consider off-diagonal
Bianchi anisotropic metrics) by frame/ coordinate transforms. The
cosmological quasiperiodic solutions will be determined by target metrics $%
\widehat{\mathbf{g}}$ generated by nonholonomic deformations $\ \mathbf{%
\mathring{g}}\rightarrow \overline{\mathbf{g}}\mathbf{=}[\overline{g}%
_{i}(x^{k})=\overline{\eta }_{i}\mathring{g}_{i},\overline{g}_{b}(x^{k},t)=%
\overline{\eta }_{b}\mathring{g}_{b},\overline{N}_{i}^{a}(x^{k},t)=\
\overline{\eta }_{i}^{a}\mathring{N}_{i}^{a}].$ The quadratic elements
corresponding to cosmological metrics $\widehat{\mathbf{g}}$ are similar to (%
\ref{dm}) but with a corresponding parametrization in terms of polarization
functions,
\begin{eqnarray*}
ds^{2} &=&\overline{\eta }_{1}(x^{i},t)\mathring{g}_{1}(x^{i},t)[dx^{1}]^{2}+%
\overline{\eta }_{2}(x^{i},t)\mathring{g}_{2}(x^{i},t)[dx^{1}]^{2}+ \\
&&\overline{\eta }_{3}(x^{i},t)\mathring{h}_{3}(x^{i},t)[dy^{3}+\overline{%
\eta }_{i}^{3}(x^{i},t)\mathring{N}_{i}^{3}(x^{k},t)dx^{i}]^{2}+\overline{%
\eta }_{4}(x^{i},t)\mathring{h}_{4}(x^{i},t)[dt+\overline{\eta }%
_{i}^{4}(x^{k},t)\mathring{N}_{i}^{4}(x^{k},t)dx^{i}]^{2}.
\end{eqnarray*}
The target d-metrics $\widehat{\mathbf{g}}=\overline{\mathbf{g}}(x^{k},t)$
are characterized by N-adapted coefficients
\begin{equation*}
\overline{g}_{i}(x^{k})= \overline{\eta }_{i}\mathring{g}_{i},h_{a}(x^{i},t)=%
\overline{\eta}_{a}\mathring{h}_{a},\overline{N}_{i}^{3}= \overline{\eta}%
_{i}^{3}(x^{i},t)\mathring{N}_{i}^{3}(x^{k},t)=\overline{n}_{i}(x^{k},t),
\overline{N}_{i}^{4}=\overline{\eta }_{i}^{4}(x^{i},t)\mathring{N}%
_{i}^{4}(x^{k},t)=\overline{w}_{i}(x^{k},t).
\end{equation*}

\subsubsection{Cosmological evolutions generated by nonstationary
quasiperiodic sources}

We consider qusiperiodic cosmological sources of type $\ ^{qp}\overline{%
\Upsilon }(x^{i},t)=\ ^{qp}\overline{\Upsilon }[\ ^{P}\overline{\Upsilon },\
_{0}^{QC}\overline{\Upsilon },\ ^{\psi }\overline{\Upsilon },\ _{0}^{sd}%
\overline{\Upsilon },\ ^{sw}\overline{\Upsilon }]$ (\ref{nonfqpc})$\ $as in (%
\ref{cosmnfdm}) and (\ref{cosmnfdmlcn}) and compute the $\eta $%
--polarization functions following formulas
\begin{eqnarray}
\overline{\eta }_{i} &=&e^{\ \psi }/\mathring{g}_{i};\overline{\eta }_{3}=%
\overline{\eta }_{3}(x^{i},t)\mbox{  as a generating function};\overline{%
\eta }_{4}=-\frac{4[(|\overline{\eta }_{3}\mathring{h}_{3}|^{1/2})^{\ast
}]^{2}}{\mathring{h}_{4}|\int dt\ ^{qp}\overline{\Upsilon }(\overline{\eta }%
_{3}\mathring{h}_{3})^{\ast }|\ };  \label{cosmpfqp} \\
\overline{\eta }_{i}^{3} &=&\frac{_{1}n_{k}}{\mathring{n}_{k}}+4\frac{\
_{2}n_{k}}{\mathring{n}_{k}}\int dt\frac{\left( [(\overline{\eta }_{3}%
\mathring{h}_{3})^{-1/4}]^{\ast }\right) ^{2}}{|\int dt\ ^{qp}\overline{%
\Upsilon }(\overline{\eta }_{3}\mathring{h}_{3})^{\ast }|\ };\overline{\eta }%
_{k}^{4}=\frac{\partial _{i}\ \int dt\ ^{qp}\overline{\Upsilon }(\overline{%
\eta }_{3}\mathring{h}_{3})^{\ast }}{\overline{\mathring{w}}_{i}\ ^{qp}%
\overline{\Upsilon }(\overline{\eta }_{3}\mathring{h}_{3})^{\ast }}.  \notag
\end{eqnarray}

There are other types of f generating functions determined determined by $%
\overline{\eta }_{3}(x^{i},t),$ data for the prime d-metric, subjected to
nonlinear symmetries (\ref{nsym1b}),
\begin{equation*}
\overline{\Phi }^{2}=4|\ ^{qp}\Lambda \lbrack h_{3}^{[0]}(x^{k})-\overline{%
\eta }_{3}({x}^{i},t)\mathring{h}_{3}({x}^{k},t)]|,\ (\overline{\Psi }%
^{2})^{\ast }=-\int dt\ \ ^{qp}\overline{\Upsilon }\ [\overline{\eta }%
_{3}(x^{i},t)\overline{\mathring{h}}_{3}(x^{i},t)]^{\ast }.
\end{equation*}

Restricting the classes of generating functionals and sources to subclasses
of solutions of (\ref{lccondb}) (see Table 3 and appendix \ref{ass32}) for
target off-diagonal cosmological metrics (\ref{lcsolcosm}) with zero
torsion, we obtain
\begin{eqnarray*}
\overline{\eta }_{i} &=&e^{\ \psi }/\mathring{g}_{i};\overline{\eta }_{3}=%
\overline{\check{\eta}}_{3}({x}^{i},t)\mbox{  as
a generating function}; \\
\overline{\eta }_{4} &=&-\frac{4[(|\ \overline{\check{\eta}}_{3}\mathring{h}%
_{3}|^{1/2})^{\ast }]^{2}}{\mathring{h}_{4}|\int dt\ ^{qp}\overline{\check{%
\Upsilon}}(\overline{\check{\eta}}_{3}\mathring{h}_{3})^{\ast }|\ };%
\overline{\eta }_{k}^{3}=(\partial _{k}\overline{n})/\mathring{n}_{k};%
\overline{\eta }_{k}^{4}=\partial _{k}\overline{\check{A}}/\mathring{w}_{k}.
\end{eqnarray*}%
In (\ref{cosmpfqp}) and next formula for LC-configurations, the nonlinear
functionals for the quasiperiodic v-source and (effective) cosmological
constant can be changed into additive functionals $\ ^{qp}\overline{\Upsilon
}\rightarrow \ ^{as}\overline{\Upsilon }$ and $\ ^{qp}\overline{\Lambda }%
\rightarrow \ ^{as}\overline{\Lambda }.$ We generate solutions of type (\ref%
{cosmasdm}) and, respectively, (\ref{cosmadmlc}) but with a prime
cosmological structure modified by certain classes of quasiperiodic
generating functions and sources. For cosmological scenarios, we can
generate solutions without singularities. Such examples have been studied in
details for small parametric deformations in \cite%
{vmon06,vepjc14a,gvvcqg15,vplb16,ruchin17,rajpoot17} and, for quasiperiodic
configurations, in \cite{abvfirs16p,aabovw16p}. In this work, cosmological
configurations are studied for general quasiperiodic structures. The
constructions depend on the type of model we construct (for instance,
different classes of generating functionals and sources have to be
prescribed for certain web / filament / solitonic stationary distributions).
We can treat such generic off-diagonal locally anisotropic solutions as
nonholonomically deformed prime cosmological configurations imbedded into
certain (non) singular media determined by evolving dark and usual matter
with quasiperiodic interactions.

\subsubsection{Cosmology from nonstationary quasiperiodic generating
functions}

We can construct other classes of locally anisotropic and inhomogeneous
cosmological solutions as nonholonomic deformations of some prime
cosmological metrics when the coefficients of the d-metrics are determined
nonlinear generating functionals $\ ^{qp}\overline{\Phi }^{2}[\ ^{P}%
\overline{q},\ ^{QC}\overline{q},\mathbf{\psi },\ _{0}^{sd}q,\ ^{sw}%
\overline{q}]$ (\ref{nadgfc}). In a similar manner, we can generate similar
cosmological metrics by additive functionals $\ ^{a}\overline{\Phi }^{2}$ (%
\ref{adgfs}) for a prescribed effective source $\overline{\Upsilon }(x^{i},t)
$ and cosmological constant $\overline{\Lambda }.$ The formulas for
nonlinear symmetries (\ref{nadgfnc}) allow us to find (recurrently)
corresponding nonlinear functionals, $\ ^{qp}\overline{\eta }_{3}({x}^{i},t),
$ or additive functionals, $\ \ \ ^{a}\overline{\eta }_{3}({x}^{i},t),$ and
related polarization functions,%
\begin{equation*}
\ ^{qp}\overline{\Phi }^{2}=4|\ \Lambda \lbrack h_{3}^{[0]}(x^{k})-\ ^{qp}%
\overline{\eta }_{3}({x}^{i},t)\mathring{h}_{3}({x}^{k},t)]|,\ (\ ^{qp}%
\overline{\Psi }^{2})^{\ast }=-\int dt\ \ \overline{\Upsilon }\ [\ ^{qp}%
\overline{\eta }_{3}(x^{i},t)\overline{\mathring{h}}_{3}(x^{i},t)]^{\ast }.
\end{equation*}

The coefficients of quadratic elements of type (\ref{gensolcosm}) are
recurrently computed,{\small
\begin{eqnarray}
\overline{\eta }_{i} &=&e^{\ \psi }/\mathring{g}_{i};\overline{\eta }_{3}=\
^{qp}\overline{\eta }_{3}(x^{i},t)\mbox{  as a generating
function};\overline{\eta }_{4}=-\frac{4[(|\ ^{qp}\overline{\eta }_{3}%
\mathring{h}_{3}|^{1/2})^{\ast }]^{2}}{\mathring{h}_{4}|\int dt\overline{%
\Upsilon }(\ ^{qp}\overline{\eta }_{3}\mathring{h}_{3})^{\ast }|\ };
\label{cosmpfqp1} \\
\overline{\eta }_{i}^{3} &=&\frac{_{1}n_{k}}{\mathring{n}_{k}}+4\frac{\
_{2}n_{k}}{\mathring{n}_{k}}\int dt\frac{\left( [(\ ^{qp}\overline{\eta }_{3}%
\mathring{h}_{3})^{-1/4}]^{\ast }\right) ^{2}}{|\int dt\overline{\Upsilon }%
(\ ^{qp}\overline{\eta }_{3}\mathring{h}_{3})^{\ast }|\ };\overline{\eta }%
_{k}^{4}=\frac{\partial _{i}\ \int dt\ \overline{\Upsilon }(\ ^{qp}\overline{%
\eta }_{3}\mathring{h}_{3})^{\ast }}{\overline{\mathring{w}}_{i}\ \overline{%
\Upsilon }(\ ^{qp}\overline{\eta }_{3}\mathring{h}_{3})^{\ast }}.  \notag
\end{eqnarray}%
}

Target off-diagonal cosmological metrics (\ref{lcsolcosm}) with zero torsion
can be generated by polarization functions%
\begin{equation*}
\overline{\eta }_{i}=e^{\ \psi }/\mathring{g}_{i};\mbox{  generating
function }\overline{\eta }_{3}=\ ^{qp}\overline{\check{\eta}}_{3}({x}^{i},t);%
\overline{\eta }_{4}=-\frac{4[(|\ ^{qp}\ \overline{\check{\eta}}_{3}%
\mathring{h}_{3}|^{1/2})^{\ast }]^{2}}{\mathring{h}_{4}|\int dt\overline{%
\check{\Upsilon}}(\ ^{qp}\overline{\check{\eta}}_{3}\mathring{h}_{3})^{\ast
}|\ };\overline{\eta }_{k}^{3}=(\partial _{k}\overline{n})/\mathring{n}_{k};%
\overline{\eta }_{k}^{4}=\partial _{k}\overline{\check{A}}/\mathring{w}_{k},
\end{equation*}%
see appendix \ref{ass32} for explanations on conventions and nonholonomic
constraints for functions with "inverse hats" and transforming general
quasiperiodic functionals into additive ones.

The solutions generated in this subsection describe certain nonholonomically
deformed prime cosmological configurations (for instance, a FLRW, or
Bianchi, type metric) self-consistently imbedded into a quasiperiodic
gravitational (dark energy) media.

\subsubsection{Cosmological solutions for nonstationary quasiperiodic
sources \& generating functions}

We can construct very general classes of nonholonomic deformations of prime
cosmological metrics generated by nonlinear quaisperiodic functionals both
for the generating functions and (effective) sources. Nonlinear
superpositions of cosmological solutions (\ref{cosmpfqp}) and (\ref%
{cosmpfqp1}) are determined by coefficients of (\ref{cosmnfdm}) computed,%
{\small
\begin{eqnarray}
\overline{\eta }_{i} &=&e^{\ \psi }/\mathring{g}_{i};\overline{\eta }_{3}=\
^{qp}\overline{\eta }_{3}(x^{i},t)\mbox{  as a
generating function};\overline{\eta }_{4}=-\frac{4[(|\ ^{qp}\overline{\eta }%
_{3}\mathring{h}_{3}|^{1/2})^{\ast }]^{2}}{\mathring{h}_{4}|\int dt\ ^{qp}%
\overline{\Upsilon }(\ ^{qp}\overline{\eta }_{3}\mathring{h}_{3})^{\ast }|\ }%
;  \label{cosmpfqp12} \\
\overline{\eta }_{i}^{3} &=&\frac{_{1}n_{k}}{\mathring{n}_{k}}+4\frac{\
_{2}n_{k}}{\mathring{n}_{k}}\int dt\frac{\left( [(\ ^{qp}\overline{\eta }_{3}%
\mathring{h}_{3})^{-1/4}]^{\ast }\right) ^{2}}{|\int dt\ ^{qp}\overline{%
\Upsilon }(\ ^{qp}\overline{\eta }_{3}\mathring{h}_{3})^{\ast }|\ };\
\overline{\eta }_{k}^{4}=\frac{\partial _{i}\ \int dt\ \ ^{qp}\overline{%
\Upsilon }(\ ^{qp}\overline{\eta }_{3}\mathring{h}_{3})^{\ast }}{\overline{%
\mathring{w}}_{i}\ ^{qp}\overline{\Upsilon }(\ ^{qp}\overline{\eta }_{3}%
\mathring{h}_{3})^{\ast }}.  \notag
\end{eqnarray}%
} In such formulas, there are considered nonlinear generating functionals $\
^{qp}\overline{\Phi }^{2}[\ ^{P}\overline{\Phi },\ ^{QC}\overline{\Phi },\
^{\psi }\overline{\Phi },\ _{0}^{sd}\overline{\Phi }^{2},\ ^{sw}\overline{%
\Phi }^{2}]$ (\ref{nadgfc}) characterized by nonlinear symmetries of type (%
\ref{nadgfnc}) for some prescribed nonlinear functionals $\ ^{qp}\overline{%
\Upsilon }(x^{i},t)$ and $\ ^{qp}\overline{\Lambda }.$ Instead of $(\ ^{qp}%
\overline{\Phi },\ ^{qp}\overline{\Upsilon },\ ^{qp}\overline{\Lambda }),$
we can consider additive data $(\ ^{a}\overline{\Phi },\ ^{a}\overline{%
\Upsilon },\ ^{a}\overline{\Lambda }).$ $\ $Using formulas for nonlinear
symmetries, we can define general nonlinear, $\ ^{qp}\overline{\eta }%
_{3}(x^{i},t),$ or additive functionals, $\ ^{a}\overline{\eta }%
_{3}(x^{i},t),$ for the polarization function,%
\begin{equation*}
\ ^{qp}\overline{\Phi }^{2}=4|\ \ ^{qp}\Lambda \lbrack h_{3}^{[0]}(x^{k})-\
^{qp}\overline{\eta }_{3}({x}^{i},t)\mathring{h}_{3}({x}^{k},t)]|\ ,(\ ^{qp}%
\overline{\Psi }^{2})^{\ast }=-\int dt\ \ \ ^{qp}\overline{\Upsilon }\ [\
^{qp}\overline{\eta }_{3}(x^{i},t)\overline{\mathring{h}}_{3}(x^{i},t)]^{%
\ast }.
\end{equation*}

LC-configurations with zero torsion for target off-diagonal cosmological
metrics (\ref{lcsolcosm}) are generated%
\begin{equation*}
\overline{\eta }_{i}=e^{\ \psi }/\mathring{g}_{i};\mbox{ generating function
}\overline{\eta }_{3}=\ ^{qp}\overline{\check{\eta}}_{3}({x}^{i},t);%
\overline{\eta }_{4}=-\frac{4[(|\ ^{qp}\overline{\check{\eta}}_{3}\mathring{h%
}_{3}|^{1/2})^{\ast }]^{2}}{\mathring{h}_{4}|\int dt\ ^{qp}\overline{\check{%
\Upsilon}}(\ ^{qp}\overline{\check{\eta}}_{3}\mathring{h}_{3})^{\ast }|\ };%
\overline{\eta }_{k}^{3}=(\partial _{k}\overline{n})/\mathring{n}_{k};%
\overline{\eta }_{k}^{4}=\partial _{k}\overline{\check{A}}/\mathring{w}_{k}.
\end{equation*}

We note that the class of cosmological solutions (\ref{cosmpfqp12}) is
"dual" on $y^{3}$ and $y^{4}$ coordinates to stationary solutions (\ref%
{statsingpfqp12}). For certain classes of parameterizations, such classes
describe nonholonomic deformations of cosmological spacetimes, or BHs,
self-consistently imbedded into quasiperiodic cosmological (dark energy)
backgrounds and quasiperiodic dark/ standard matter.

\section{Discussion and Concluding Remarks}

\label{sconcl}

Cosmology theories rely on the cosmological principle stating that our
universe is sufficiently homogeneous and isotropic on large scales.
Geometric models and computation algorithms in numerical relativity encode
observational data and theoretical assumptions that expansion is that for
the FLWR model governed by (modified) Friedmann equations and employing a
Newtonian approximation of gravity. The transition to cosmic homogeneity
begins on scales $\sim 80h^{-1}$ Mpc when the Universe is inhomogeneous and
anisotropic on smaller scales, see Planck2015 data \cite%
{planck15a13,planck15a14,planck15a20,planck15a31}. Modern telescopes reached
a precision which shows that nonlinear general relativistic effects from
inhomogeneities could be important.

There are more extreme hypotheses that inhomogeneities may provide an
alternative explanation for the accelerating expansion of the Universe (for
instance, it is considered the back reaction or replaced the role assigned
to dark energy in the standard $\Lambda $CDM model). Such alternative
approaches are based on a number of cosmological observations during the
last 20 years emphasizing important phenomena of the accelerating Universe
and dark energy and dark matter. The dark energy physics yields a late time
acceleration of the spacetime. In its turn, the dark matter physics is
"hidden" as an invisible matter (there are various models of dust, cold and
hot matter, gravitational polarizations etc.) which favour the process of
gravitational clustering. It is not clear how such intriguing physical
effects (substances and (non) linear interactions) could emerge in the
general relativity theory, GR, or should complement such a theory. The
search for alternative modified gravity theories, MGTs, has become an active
area for theoretical and experimental investigations. A very important task
is to elaborate on new methods of constructing exact and parametric
solutions of motion and evolution field equations in MGT and GR describing
nonlinear gravitational and matter field interactions (as we motivated in
the Introduction section). A large number of MGTs is available and studied
in various details in modern literature \cite%
{starob,capoz,nojod1,clifton,gorbunov,linde2,bambaodin,sami,mimet2,odints2,guend1,guend2,vepjc14a,vplb16,grvvap16,rajpoot17a,gvvepjc14,gvvcqg15}%
.

The predominantly accepted $\Lambda $CDM paradigm predicts for the
cosmological structure that dark matter is organized as a cosmic web
structure of walls, filaments and halos \cite%
{crystalinks,starob,rucklidge13,rucklidge15,ruchin17,rajpoot17,diemer17,gurzadyan13,cross09,kooistra17}%
. Numerous images of the visually striking cosmic web have been created and
studied theoretically in a framework of research of the large-scale
structure of the universe. Recently, an advanced observational and numerical
simulation technique was elaborated for pseudo-3d visualisations aimed to
minimal loss of information and accurate representation of cosmic web
fundamental shapes and components. In the cosmic web, one observes large
scale filaments with lengths that can reach tens of Mps, see \cite%
{kooistra17} and references therein. Such filaments are observed indirectly
through the galaxies positions (using large galaxy surveys), or through
absorption features in the spectra of hight red shift sources (with direct
detecting of intergalactic medium filaments through their emission on the HI
21cm line). One estimates that gas in filaments of length $l\geq 15h^{-1}$
Mpc with relatively small inclinations to the line of sight ($\leq 10^{\circ
}$) can be observed during 40-100 hours with modern (radio) telescopes. It
is revealed from observations of the local Universe that galaxies reside in
a complex network of filamentary structures (with cosmic web). Such
structures can be explained in the $\Lambda $CDM framework as resulting from
nonlinear gravitational evolution. The dark matter halos (within which
galaxies reside) are connected to each other through a patchwork of
filaments and sheets and quasiperiodic/ aperiodic structures that constitute
the structure of the intergalactic medium, IGM. In modern astronomy it is
explored the possibility of using the HI 21 cm line to directly observe ICM
filaments.

Various alternative approaches to structure formation in modern cosmology
were elaborated using numerical relativity \cite%
{pretorius05,lehner15,campanelli06,baker06}. This involves modelling by
numerical methods which began with evolutions of planar and spherically
symmetric spacetime following the Arnowitt-Deser-Misner, ADM, formalism \cite%
{misner73}. Latter, there were considered generalizations with Kasner and
matter-fields, and for the propagation and collision and gravitational wave
perturbations and linearised perturbations to homogeneous spacetimes. In
order to simplify the numerical calculations there are included linear and
nonlinear symmetries. Simulations with (non) linear symmetries have been
performed in order to study the evolution of small perturbations to an FLRW
spacetime, to explore observational implications showed differential
expansion in an inhomogeneous universe. Such works indicate that the effects
of nonlinear inhomogeneities may be significant.

Modern cosmological observational data and related phenomenological models
emphasize a crucial importance of nonlinear physics and related mathematical
methods. Distinguishing general relativistic effects determined by nonlinear
structures requires more general classes of solutions of Einstein's
equations in GR and application of advanced geometric and numeric methods
elaborated recently for theoretical studies in MGT, non standard particle
physics, astrophysics and cosmology. Post-Newtonian, small parametric
(perturbative and non-perturbative) numeric approximations consist a
worthwhile approach extended with methods including density perturbations in
a highly nonlinear form. However the validity of nonlinear effects must be
checked against more precise exact and parametric solutions constructed in
analytic form.

In this work, we performed a brief review and feasibility study of geometric
methods of constructing exact off-diagonal solutions to the (modified)
Einstein equations for stationary, locally anisotropic BH and cosmological
(in)homogeneous cosmology by geometric modeling and comparing with numeric
simulating the growths of quasiperiodic / aperiodic structures and comparing
to known analytic solutions. We also presented a study on the evolution of
nonlinear stationary and/ or locally anisotropic configurations and analysed
the resulting new classes of exact solutions in MGTs and GR. Such a research
was performed by developing the anholonomic frame method, AFDM, for
constructing new classes of stationary and nonstationary (cosmological)
solutions of (modified) Einstein equations. Such solutions are described by
generic off-diagonal metrics and generalized connections and depend, in
principle, on all possible 4-d and extra dimensional space coordinates; see
details, examples and various applications in \cite%
{vexsol98,vjhep01,vpcqg01,vjmp05,vmon06,vijgmmp07,vijgmmp11,vgrg12,vcsf12,vepjp12,vepjc13,vepjc14a,gvvepjc14,gvvcqg15}%
.

A crucial difference from former approaches to constructing exact solutions
\cite{misner73,hawking73,wald82,kramer03,griffith09} is that the AFDM allows
us to work with generating and integration functions for coefficients of
generic off-diagonal metrics, generalized connections and (effective)
sources transforming motion and geometric evolution equations into nonlinear
systems of partial differential equations, PDEs, with decoupling properties.
Following this geometric method, we perform such nonholonomic deformations
of some prime stationary and nonstationary solutions (for instance, BH
and/or (an) isotropic cosmological solutions ) when the (generalized)
Einstein equations can be decoupled in general forms and integrated for
various classes of metrics $g_{\alpha \beta}(x^{i},y^{3},t).$ In particular,
we can reproduce former results for diagonalizable ansatz with dependence on
radial and warping coordinates as solutions of ordinary differential
equations, ODEs. Nevertheless, the AFDM, is more than a constructive
interference of geometric and analytic methods for constructing exact
solutions for certain classes of important nonlinear systems of PDEs, in
mathematical relativity and cosmology. It reflects new and formerly unknown
properties and nonlinear symmetries of the (modified) Einstein equations
when generic off--diagonal interactions and mixed continuous and discrete
structures (quasiperiodic / aperiodic / pattern forming / solitonic ones)
are considered for vacuum and non-vacuum gravitational interactions. The
AFDM is appealing in some sense it is "economical and very efficient";
allowing us to proceed in the same manner but with fractional / random /
noncommutative sources and their respective interaction parameters.

Finally, we emphasize that there are a number of directions in (modified)
gravity, cosmology and astrophysics which can be pursued using as starting
points the methods and solutions elaborated in this works.\footnote{%
During the review process of this work, there were elaborated the review
article \cite{v18a}, on axiomatic formulation and historical remarks of
gravity and matter field theories with modified dispersion relations, and
\cite{v18b}, and on space-time quasicrystal structures and inflationary and
late time evolution dynamics in accelerating cosmology. In those works, the
AFDM and related nonholonomy methods are completed with an analysis of main
physical principles for formulating locally anisotropic theories and new
classes of cosmological solutions are provided. The version 3 in arXiv.org
of this paper contains a system of notations which is different from
previous arXiv versions 1 and 2, and the published journal version. The
version 3 allows an extension to an unified system of notations both for
nonholonomic manifolds and tangent Lorentz bundles with
Finsler-Lagrange-Hamilton structures.} This includes noncommutative and
nonassociative generalizations defined by possible modified dispersion
relations and/ or extra dimension contributions to dark energy/matter
physics and/or quantum models with quasiperiodic and pattern forming
structures.

\vskip3pt \textbf{Acknowledgments:} SV research is based on program IDEI,
PN-II-ID-PCE-2011-3-0256;\ a program for visitors at CERN; and DAAD projects
in Germany for M. Planck Institute for Physics (W. Heisenberg) in Munich,
and Institute of Theoretical Physics of Leibnitz University of Hannover.

\appendix

\setcounter{equation}{0} \renewcommand{\theequation}
{A.\arabic{equation}} \setcounter{subsection}{0}
\renewcommand{\thesubsection}
{A.\arabic{subsection}}

\section{MGTs in Nonholonomic 2+2 N-adapted Variables}

\label{as1}We summarize most important definitions and formulas necessary
for generating solutions of gravitational field equations following the
anholonomic frame deformation method, AFDM, see details and proofs in \cite%
{vexsol98,vjhep01,vpcqg01,vjmp05,vmon06,vijgmmp07,vijgmmp11,vgrg12,vcsf12,vepjp12,vepjc13,vepjc14a,gvvepjc14,gvvcqg15}%
.

\subsection{N-adapted coefficients for curvatures and torsions}

Using standard formulas, we can define and compute both in abstract and
coordinate forms the torsion, $\mathcal{T},$ the nonmetricity, $\mathcal{Q},$
and the curvature, $\mathcal{R}$, tensors for any d--connection $\mathbf{D}%
=(hD,vD),$
\begin{equation*}
\mathcal{T}(\mathbf{X,Y}):=\mathbf{D}_{\mathbf{X}}\mathbf{Y}-\mathbf{D}_{%
\mathbf{Y}}\mathbf{X}-[\mathbf{X,Y}],\mathcal{Q}(\mathbf{X}):=\mathbf{D}_{%
\mathbf{X}}\mathbf{g,}\mathcal{R}(\mathbf{X,Y}):=\mathbf{D}_{\mathbf{X}}%
\mathbf{D}_{\mathbf{Y}}-\mathbf{D}_{\mathbf{Y}}\mathbf{D}_{\mathbf{X}}-%
\mathbf{D}_{\mathbf{[X,Y]}}.
\end{equation*}%
In literature \cite%
{vexsol98,vjhep01,vpcqg01,vjmp05,vmon06,vijgmmp07,vijgmmp11}, there are used
terms like distinguished tensor, d-tensor, and distinguished (geometric)
object, d-object, (also d-metric, d-connection) etc. for geometric and
physical values defined in N-adapted form, i.e. when all values are defined
in some coefficient forms preserving under parallelism the N--connection
splitting (\ref{ncon})). Using N-adapted coefficients of a d-connection, $%
\mathbf{D}=\{\mathbf{\Gamma }_{\ \alpha \beta }^{\gamma
}=(L_{jk}^{i},L_{bk}^{a},C_{jc}^{i},C_{bc}^{a})\},$ we can compute with
respect to N--adapted frames (\ref{nfr}) and (\ref{ndfr}) corresponding
N-adapted coefficients
\begin{equation*}
\mathcal{T}=\{\mathbf{T}_{\ \alpha \beta }^{\gamma }=\left( T_{\
jk}^{i},T_{\ ja}^{i},T_{\ ji}^{a},T_{\ bi}^{a},T_{\ bc}^{a}\right) \},%
\mathcal{Q}=\mathbf{\{Q}_{\ \alpha \beta }^{\gamma }\},\mathcal{R}\mathbf{=}%
\mathbf{\{R}_{\ \beta \gamma \delta }^{\alpha }\mathbf{=}\left( R_{\ hjk}^{i}%
\mathbf{,}R_{\ bjk}^{a}\mathbf{,}R_{\ hja}^{i}\mathbf{,}R_{\ bja}^{c},R_{\
hba}^{i},R_{\ bea}^{c}\right) \}.
\end{equation*}
We omit such cumbersome formulas which can be found in above mentioned
references.

The coefficients of the canonical d-connection $\widehat{\mathbf{D}}=\{%
\widehat{\mathbf{\Gamma }}_{\ \alpha \beta }^{\gamma }=(\widehat{L}_{jk}^{i},%
\widehat{L}_{bk}^{a},\widehat{C}_{jc}^{i},\widehat{C}_{bc}^{a})\}$ in (\ref%
{twocon}) are
\begin{eqnarray}
\widehat{L}_{jk}^{i} &=&\frac{1}{2}g^{ir}\left( \mathbf{e}_{k}g_{jr}+\mathbf{%
e}_{j}g_{kr}-\mathbf{e}_{r}g_{jk}\right) ,\widehat{C}_{bc}^{a}=\frac{1}{2}%
g^{ad}\left( e_{c}g_{bd}+e_{b}g_{cd}-e_{d}g_{bc}\right) ,  \label{candcon} \\
\widehat{C}_{jc}^{i} &=&\frac{1}{2}g^{ik}e_{c}g_{jk},\ \widehat{L}%
_{bk}^{a}=e_{b}(N_{k}^{a})+\frac{1}{2}g^{ac}\left( \mathbf{e}%
_{k}g_{bc}-g_{dc}\ e_{b}N_{k}^{d}-g_{db}\ e_{c}N_{k}^{d}\right) .  \notag
\end{eqnarray}%
The coefficients of the distortion d--tensor, $\widehat{\mathbf{Z}}_{\
\alpha \beta }^{\gamma }=\widehat{\mathbf{\Gamma }}_{\ \alpha \beta
}^{\gamma }-\Gamma _{\ \alpha \beta }^{\gamma }$ (\ref{candistr}) can be
written in N--adapted form using (\ref{candcon})\ and LC-connection $\nabla
=\{\Gamma _{\ \alpha \beta }^{\gamma }\},$ all computed with respect to (\ref%
{nfr}) and (\ref{ndfr}). Using such values, we find the nontrivial
d--torsion coefficients\ $\widehat{\mathbf{T}}_{\ \alpha \beta }^{\gamma }$
of $\widehat{\mathbf{D}},$
\begin{equation}
\widehat{T}_{\ jk}^{i}=\widehat{L}_{jk}^{i}-\widehat{L}_{kj}^{i},\,\,\,%
\widehat{T}_{\ ja}^{i}=\widehat{C}_{jb}^{i},\,\,\,\widehat{T}_{\
ji}^{a}=-\Omega _{\ ji}^{a},\,\,\,\widehat{T}_{aj}^{c}=\widehat{L}%
_{aj}^{c}-e_{a}(N_{j}^{c}),\,\,\,\widehat{T}_{\ bc}^{a}=\ \widehat{C}%
_{bc}^{a}-\ \widehat{C}_{cb}^{a}.  \label{dtors}
\end{equation}%
We note that the d-torsion coefficients (\ref{dtors}) vanish if in
N--adapted form there are satisfied the conditions
\begin{equation}
\widehat{L}_{aj}^{c}=e_{a}(N_{j}^{c}),\,\,\,\widehat{C}_{jb}^{i}=0,\Omega
_{\ ji}^{a}=0.  \label{lccond}
\end{equation}%
Following similar formulas, we can compute (see details in above references)
the nontrivial coefficients of the Riemann d-tensor, $\widehat{\mathbf{R}}%
_{\ \beta \gamma \delta }^{\alpha },$ the Ricci d-tensor, $\widehat{\mathbf{R%
}}_{\alpha \beta }$ (\ref{candricci}), and the Einstein d-tensor $\widehat{%
\mathbf{E}}_{\alpha \beta }:=\widehat{\mathbf{R}}_{\alpha \beta }-\frac{1}{2}%
\mathbf{g}_{\alpha \beta }\ \widehat{R}.$

\subsection{Decoupling property of (modified) Einstein equations}

For general assumptions and using frame/ coordinate transforms, any
d--metric $\widehat{\mathbf{g}}$ (\ref{dm}) can be parameterized {\small
\begin{eqnarray}
g_{i} &=&e^{\psi {(r,\theta )}},\,\,\,\,g_{a}=\omega ({r,\theta }%
,y^{b})h_{a}({r,\theta },\varphi ),\ N_{i}^{3}=w_{i}({r,\theta },\varphi
),\,\,\,\,N_{i}^{4}=n_{i}({r,\theta },\varphi ),\mbox{ for }\omega =1, %
\mbox{stationary conf.};  \label{data1st} \\
g_{i} &=&e^{\psi {(x^{k})}},\,\,\,\,g_{a}=\omega (x^{k},y^{b})\overline{h}%
_{a}(x^{k},t),\ N_{i}^{3}=\overline{n}_{i}(x^{k},t),\,\,\,\,N_{i}^{4}=%
\overline{w}_{i}(x^{k},t),\mbox{ for }\omega =1,\mbox{ cosmological conf.}
\label{data1c}
\end{eqnarray}%
} In order to write certain formulas in compact forms, we shall use also
brief notations of partial derivatives $\partial _{\alpha }q=\partial
q/\partial u^{\alpha }$ (for instance, for a function $q(x^{k},y^{a}))$
\begin{eqnarray*}
\partial _{1}q &=&q^{\bullet }=\partial q/\partial x^{1},\partial
_{2}q=q^{\prime }=\partial q/\partial x^{2},\partial _{3}q=\partial
q/\partial y^{3}=\partial q/\partial \varphi =q^{\diamond },\partial
_{4}q=\partial q/\partial t=\partial _{t}q=q^{\ast }, \\
\partial _{33}^{2} &=&\partial ^{2}q/\partial \varphi ^{2}=\partial
_{\varphi \varphi }^{2}q=q^{\diamond \diamond },\partial _{44}^{2}=\partial
^{2}q/\partial t^{2}=\partial _{tt}^{2}q=q^{\ast \ast }.
\end{eqnarray*}%
The sources (\ref{sourc}) for (effective) matter field configurations can be
parameterized via frame transforms in respective N--adapted forms
\begin{equation}
\mathbf{\Upsilon }_{\ \nu }^{\mu }=\mathbf{e}_{\ \mu ^{\prime }}^{\mu }%
\mathbf{e}_{\nu }^{\ \nu ^{\prime }}[~^{m}\mathbf{\Upsilon }_{\ \nu ^{\prime
}}^{\mu ^{\prime }}+~\widehat{\mathbf{\Upsilon }}_{\ \nu ^{\prime }}^{\mu
^{\prime }}]=\left\{
\begin{array}{cc}
\lbrack ~\ _{h}\Upsilon ({r,\theta })\delta _{j}^{i},\Upsilon ({r,\theta }%
,\varphi )\delta _{b}^{a}], & \mbox{ stationary configurations }; \\
\lbrack ~\ _{h}\overline{\Upsilon }(x^{i})\delta _{j}^{i},\overline{\Upsilon
}(x^{i},t)\delta _{b}^{a}], & \mbox{ cosmological configurations }.%
\end{array}%
\right.  \label{dsourcparam}
\end{equation}

In these formulas, there are considered necessary type vielbein transforms $%
\mathbf{e}_{\ \mu ^{\prime }}^{\mu }(u^{\gamma })$ and their duals $\mathbf{e%
}_{\nu }^{\ \nu ^{\prime }}(u^{\gamma }),$ when $\mathbf{e}_{\ }^{\mu }=%
\mathbf{e}_{\ \mu ^{\prime }}^{\mu }du^{\mu ^{\prime }},$ and $\mathbf{%
\Upsilon }_{\ \nu ^{\prime }}^{\mu ^{\prime }}=~^{m}\mathbf{\Upsilon }_{\
\nu ^{\prime }}^{\mu ^{\prime }}+~\widehat{\mathbf{\Upsilon }}_{\ \nu
^{\prime }}^{\mu ^{\prime }}.$ The values $[~\ _{h}\Upsilon ({r,\theta }%
),\Upsilon ({r,\theta },\varphi )]~\ $, or $[~\ _{h}\overline{\Upsilon }%
(x^{i}),\overline{\Upsilon }(x^{i},t)]$, are considered as generating
functions for (effective) matter sources imposing nonholonomic frame
constraints on stationary distributions or cosmological dynamics of
(effective) matter fields. For simplicity, we shall generate in explicit
form certain classes of generic off--diagonal solutions with $\omega =1$
(i.e. with at least one Killing symmetry on $\partial _{t}$ or $\partial
_{\varphi })$ when the frame/coordinate systems and transforms are
compatible with the conditions $\partial _{b}h_{a}\neq 0$ and $[~\
_{h}\Upsilon ({r,\theta }),\Upsilon ({r,\theta },\varphi )]\neq 0$,\ or $[~\
_{h}\overline{\Upsilon }(x^{i}),\overline{\Upsilon }(x^{i},t)]\neq 0.$ In
next subsections, we shall prove that above introduced parameterizations of
d--metrics and (effective) sources allows us to integrate explicitly the
gravitational field equations (\ref{qgreq}).\footnote{%
It is possible to construct various classes of physically important vacuum
and nonvacuum solutions if such conditions are violated with respect to
certain systems of references, or in some points, open regions of
nonholonomic spacetime models. This requests more cumbersome technical
considerations. We shall not study such solutions in this work, see examples
in Refs. \cite%
{vpcqg01,vjmp05,vmon06,vijgmmp07,vijgmmp11,vgrg12,vcsf12,vepjp12,vepjc13,vepjc14a,gvvepjc14,gvvcqg15}%
.}

\subsubsection{Off--diagonal stationary configurations}

\label{asst}

In this subsection, we outline key steps for proofs of general decoupling
and integrability of (modified) Einstein equations for general assumptions
on coefficients d--metrics and N--connections which do not depend on $y^{4}$
with respect to a class of N-adapted frames.

\paragraph{Nontrivial components of the Ricci d-tensor and (modified)
Einstein equations: \newline
}

Introducing d-metric data (\ref{data1st}) with $\omega =1$ into formulas (%
\ref{candcon}) and (\ref{candricci}) (with respective sources (\ref%
{dsourcparam}), and considering nontrivial N--adaped coefficients of the
Ricci d-tensor), we transform the modified Einstein equations (\ref{qgreq})
into such a system of nonlinear PDEs
\begin{eqnarray}
\widehat{R}_{1}^{1} &=&\widehat{R}_{2}^{2}=\frac{1}{2g_{1}g_{2}}[\frac{%
g_{1}^{\bullet }g_{2}^{\bullet }}{2g_{1}}+\frac{\left( g_{2}^{\bullet
}\right) ^{2}}{2g_{2}}-g_{2}^{\bullet \bullet }+\frac{g_{1}^{\prime
}g_{2}^{\prime }}{2g_{2}}+\frac{(g_{1}^{\prime })^{2}}{2g_{1}}-g_{1}^{\prime
\prime }]=-\ \ _{h}\Upsilon ,  \label{eq1a} \\
\widehat{R}_{3}^{3} &=&\widehat{R}_{4}^{4}=\frac{1}{2h_{3}h_{4}}[\frac{%
\left( h_{4}^{\diamond }\right) ^{2}}{2h_{4}}+\frac{h_{3}^{\diamond }\
h_{4}^{\diamond }}{2h_{3}}-h_{4}^{\diamond \diamond }]=-\Upsilon ,
\label{eq2a} \\
\widehat{R}_{3k} &=&-\frac{w_{k}}{2h_{4}}[\frac{\left( h_{4}^{\diamond
}\right) ^{2}}{2h_{4}}+\frac{h_{3}^{\diamond }\ h_{4}^{\diamond }}{2h_{3}}%
-h_{4}^{\diamond \diamond }]+\frac{h_{4}^{\diamond }}{4h_{4}}(\frac{\partial
_{k}h_{3}}{h_{3}}+\frac{\partial _{k}h_{4}}{h_{4}})-\frac{\partial
_{k}h_{4}^{\diamond }}{2h_{4}}=0,  \label{eq3a} \\
\widehat{R}_{4k} &=&\frac{h_{4}}{2h_{4}}n_{k}^{\diamond \diamond }+(\frac{3}{%
2}h_{4}^{\diamond }-\frac{h_{4}}{h_{3}}h_{3}^{\diamond })\frac{%
n_{k}^{\diamond }}{2h_{3}}=0.  \label{eq4a}
\end{eqnarray}%
In N-adapted frames, this system posses a \textsf{decoupling property:} \
The equations (\ref{eq1a}) allow us to find $g_{1}$ (or, inversely, $g_{2}$)
for any prescribed h-source $\ _{h}\Upsilon (r,\theta )$ and given
coefficient $g_{2}$ (or, inversely, $g_{1}$). Integrating on variable $y^{3}$
in (\ref{eq2a}), we can define $h_{3}({r,\theta },\varphi )$ as a solution
of first order PDE for any prescribed v-source $\Upsilon ({r,\theta }%
,\varphi )$ and given coefficient $h_{4}({r,\theta },\varphi )$ \ [we can
define $h_{4}({r,\theta },\varphi )$ \ if, inversely, $h_{3}({r,\theta }%
,\varphi )$ is given but in such cases we have to solve a second order PDE].
\ For well-defined values of $h_{3}$ and $h_{4},$ the equations (\ref{eq3a})
transform into an algebraic linear equation for $w_{k}({r,\theta },\varphi
). $ We have to integrate two times on $y^{3}$ in order to compute $n_{k}({%
r,\theta },\varphi )$ for any well-defined $h_{3}$ and $h_{4}.$ So, the
decoupling property of the system (\ref{eq1a})--(\ref{eq4a}) reflects the
possibility to integrate such PDEs step by step by defining the
h-coefficients, $g_{i},$ and v--coefficients, $h_{a},$ of a d-metric $[%
\mathbf{g}_{i}=g_{i}(x^{k}),\mathbf{g}_{a}=g_{a}(x^{k},y^{3})]$ (\ref{dm})
(with data (\ref{data1st})), and, finally, the N-connection coefficients, $%
N_{i}^{a}=[w_{i}(x^{k},y^{3}),n_{i}(x^{k},y^{3})].$

\paragraph{Extracting torsionless configurations: \newline
}

We note that the LC-conditions (\ref{lccond}) for stationary configurations
transform into a system of 1st order PDEs,
\begin{equation}
\partial _{\varphi }w_{i}=(\partial _{i}-w_{i}\partial _{\varphi })\ln \sqrt{%
|h_{3}|},(\partial _{i}-w_{i}\partial _{\varphi })\ln \sqrt{|h_{4}|}%
=0,\partial _{k}w_{i}=\partial _{i}w_{k},\partial _{\varphi
}n_{i}=0,\partial _{i}n_{k}=\partial _{k}n_{i},  \label{lcconda}
\end{equation}%
imposing additional constraints on off-diagonal coefficients of metrics of
type (\ref{mcoord}). Such conditions can be imposed on d-metric and
N--connection coefficients after a class of off-diagonal solutions for (\ref%
{qgreq}) has been constructed in an explicit form. For certain well-defined
N-adapted parameterizations, the equations (\ref{lcconda}) can be solved in
explicit forms.

\paragraph{Nonlinear gravitational PDEs with explicit decoupling: \newline
}

Let us show how the system of nonlinear PDEs (\ref{eq1a})--(\ref{eq4a}) can
be integrated in explicit form. We introduce the coefficients $\alpha
_{i}=(\partial _{\varphi }h_{4})\ (\partial _{i}\varpi ),\ \beta =(\partial
_{\varphi }h_{4})\ (\partial _{\varphi }\varpi ),\ \gamma =\partial
_{\varphi }\left( \ln |h_{4}|^{3/2}/|h_{3}|\right)$, where
\begin{equation}
\varpi {=\ln |\partial _{3}h_{4}/\sqrt{|h_{3}h_{4}|}|}.  \label{genf1a}
\end{equation}%
Using such nontrivial and nonsingular values for $\partial _{3}h_{a}\neq 0$
and $\partial _{t}\varpi \neq 0,$\footnote{%
we can construct nontrivial solutions if such conditions are not satisfied;
for simplicity, we omit a study of such more special cases} we obtain
\begin{eqnarray}
\psi ^{\bullet \bullet }+\psi ^{\prime \prime } &=&2~\ \ _{h}\Upsilon ,
\label{e1a} \\
\varpi ^{\diamond }\ h_{4}^{\diamond } &=&2h_{3}h_{4}\Upsilon ,  \label{e2a}
\\
\beta w_{i}-\alpha _{i} &=&0,  \label{e3a} \\
n_{k}^{\diamond \diamond }+\gamma n_{k}^{\diamond } &=&0.  \label{e4a}
\end{eqnarray}%
This system can be integrated in explicit form (see below) for any
generating function $\Psi ({r,\theta },\varphi ):=e^{\varpi }$ and
generating sources $\ _{h}\Upsilon (r,\theta )$ and $\Upsilon ({r,\theta }%
,\varphi ).$

\paragraph{Nonlinear symmetries for stationary generating functions and
effective cosmological constant: \newline
}

We emphasize that the system of two equations (\ref{genf1a}) and (\ref{eq2a}%
) relates four functions ($h_{3},h_{4},\Upsilon ,$ and $\Psi )$ and posses
an important nonlinear symmetry for re-defining generating functions, $(\Psi
,\Upsilon )\iff (\Phi ,\Lambda )$, when
\begin{equation}
\Lambda (\ \Psi ^{2})^{\diamond }=|\Upsilon |(\Phi ^{2})^{\diamond },%
\mbox{
or  }\Lambda \ \Psi ^{2}=\Phi ^{2}|\Upsilon |-\int dy^{3}\ \Phi
^{2}|\Upsilon |^{\diamond },  \label{nsym1a}
\end{equation}%
which allows to introduce a new generating function $\Phi ({r,\theta }%
,\varphi )$ and an (effective) cosmological constant $\Lambda \neq 0.$ The
value of $\Lambda $ can be chosen from certain physical considerations. It
can be positive or negative. Solutions with $\Lambda =0$ have to be studied
by special methods, see detils and examples in \cite%
{vpcqg01,vjmp05,vmon06,vijgmmp07,vijgmmp11,vgrg12,vcsf12,vepjp12,vepjc13,vepjc14a,gvvepjc14,gvvcqg15}%
. We can describe nonlinear systems of PDEs by two equivalent generating
data $(\Psi ,\Upsilon )$ or $(\Phi ,\Lambda )$ [for different classes of
solutions, one can be convenient to work with different types of such data].
Modules in such formulas are taken in certain forms which allows to work
with real functions.

\paragraph{Stationary solutions for off-diagonal metrics and
N--coefficients: \newline
}

By explicit verifications (see similar details and rigorous proofs in \cite%
{vexsol98,vjhep01,vpcqg01,vjmp05,vmon06,vijgmmp07,vijgmmp11,vgrg12,vcsf12,vepjp12,vepjc13,vepjc14a,gvvepjc14,gvvcqg15}%
), we can prove integrating "step by step" the system (\ref{e1a})--(\ref{e4a}%
) \ that there are generated generic off-diagonal solutions the solutions of
the \ nonlinear PDEs (\ref{eq1a})--(\ref{eq4a}), i.e. the gravitational
field equations (\ref{qgreq}), if the d--metric coefficients are computed
\begin{eqnarray}
\ g_{i} &=&e^{\ \psi (x^{k})}\mbox{ as a solution of 2-d Poisson eqs. }\psi
^{\bullet \bullet }+\psi ^{\prime \prime }=2~\ _{h}\Upsilon ;
\label{offdstat} \\
g_{3} &=&h_{3}({r,\theta },\varphi )=-\frac{(\Psi ^{\diamond })^{2}}{%
4\Upsilon ^{2}h_{4}}=-\frac{(\partial _{\varphi }\Psi )^{2}}{4\Upsilon
^{2}\left( h_{4}^{[0]}(x^{k})-\int dy^{3}(\Psi ^{2})^{\diamond }/4\Upsilon
\right) }  \notag \\
&=&-\frac{(\Phi ^{2})(\Phi ^{2})^{\diamond }}{h_{4}|\Lambda \int
dy^{3}\Upsilon \lbrack \Phi ^{2}]^{\diamond }|}=-\frac{[\partial _{\varphi
}(\Phi ^{2})]^{2}}{4[h_{4}^{[0]}(x^{k})-\Phi ^{2}/4\Lambda ]|\int d\varphi \
\Upsilon \partial _{\varphi }[\Phi ^{2}]|};  \notag \\
g_{4} &=&h_{4}({r,\theta },\varphi )=h_{4}^{[0]}(x^{k})-\int dy^{3}\frac{%
(\Psi ^{2})^{\diamond }}{4\Upsilon }=h_{4}^{[0]}(x^{k})-\Phi ^{2}/4\Lambda ;
\notag
\end{eqnarray}%
and the N--connection coefficients are%
\begin{eqnarray}
\ N_{i}^{3} &=&w_{i}({r,\theta },\varphi )=\frac{\partial _{i}\ \Psi }{%
\partial _{\varphi }\Psi }\ =\frac{\partial _{i}\ \Psi ^{2}}{\partial
_{\varphi }\Psi ^{2}}\ =\frac{\partial _{i}[\int dy^{3}\ \Upsilon (\Phi
^{2})^{\diamond }]}{\Upsilon (\Phi ^{2})^{\diamond }};\ \   \notag \\
N_{k}^{4} &=&n_{k}({r,\theta },\varphi )=\ _{1}n_{k}(x^{i})+\
_{2}n_{k}(x^{i})\int dy^{3}\frac{(\Psi ^{\diamond })^{2}}{\Upsilon
^{2}|h_{4}^{[0]}(x^{i})-\int dy^{3}(\Psi ^{2})^{\diamond }/4\Upsilon |^{5/2}}
\notag \\
&=&\ _{1}n_{k}(x^{i})+\ _{2}n_{k}(x^{i})\int dy^{3}\frac{(\Phi ^{\diamond
})^{2}}{4|\Lambda \int dy^{3}\Upsilon \lbrack \Phi ^{2}]^{\diamond
}||h_{4}|^{5/2}}.  \notag
\end{eqnarray}%
In these formulas, there are considered also integration functions $%
h_{3}^{[0]}(x^{k}),$ $\ _{1}n_{k}(x^{i}),$ and $\ _{2}n_{k}(x^{i})$ encoding
various possible sets of (non) commutative parameters and integration
constants. Such values, together with generating data $(\Psi ,\Upsilon ),$
or $(\Phi ,\Lambda ),$ related by nonlinear differential / integral
transforms (\ref{nsym1a}) can be stated in explicit form following certain
topology/ symmetry / asymptotic conditions for some classes of exact /
parametric solutions of gravitational field equations. The coefficients (\ref%
{offdstat}) define generic off-diagonal solutions if the corresponding
anholonomy coefficients $C_{\alpha \beta }^{\gamma }(u)$ (\ref{anhr}) are
not trivial. Such solutions are with nontrivial nonholonomically induced
d-torsion (\ref{dtors}) with N-adapted coefficients which can be computed in
explicit form.

\paragraph{Quadratic line elements for off-diagonal stationary
configurations: \newline
}

As a matter of principle, we can consider any coefficient $%
h_{4}=h_{4}^{[0]}(x^{k})-\Phi ^{2}/4\Lambda ,$ $h_{4}^{\diamond }\neq 0,$ as
a generating function. Using this formula with $h_{4}^{[0]}(x^{k})$
introduced in $\Phi ,$ we find $\Phi ^{2}=-4\Lambda h_{4}({r,\theta }%
,\varphi )$ and transform (\ref{nsym1a}) in $\ (\Psi ^{2})^{\diamond }=\int
d\varphi \ \Upsilon ({r,\theta },\varphi )h_{4}^{\diamond }({r,\theta }%
,\varphi ).$ Introducing such a functional $\Psi \lbrack
h_{4}^{[0]},h_{4},\Upsilon ]$ into respective formulas for $h_{a}$ and $%
\Upsilon $ in (\ref{offdstat}), we express possible generating functions and
the respective \ d--metric (\ref{dm}), with data (\ref{data1st}), in terms
of $h_{4},$ integration functions and effective sources. Respective
quadratic elements can be expressed in three equivalent forms{\small
\begin{eqnarray}
ds^{2} &=&e^{\ \psi (x^{k})}[(dx^{1})^{2}+(dx^{2})^{2}]  \label{gensolstat}
\\
&&+\left\{
\begin{array}{cc}
\begin{array}{c}
-\frac{(h_{4}{}^{\diamond })^{2}}{|\int d\varphi \ \Upsilon
h_{4}{}^{\diamond }|\ h_{4}}[dy^{3}+\frac{\partial _{i}(\int d\varphi
\Upsilon \ h_{4}^{\diamond }{})}{\ \ \Upsilon \ h_{4}^{\diamond }{}}dx^{i}]-
\\
h_{4}[dt+(\ _{1}n_{k}+4\ _{2}n_{k}\int d\varphi \frac{(h_{4}^{\diamond
}{})^{2}}{|\int dy^{3}\Upsilon h_{4}^{\diamond }|\ (h_{4})^{5/2}})dx^{k}],
\\
\mbox{ or }%
\end{array}
&
\begin{array}{c}
\mbox{gener.  funct.}h_{4}, \\
\mbox{ source }\Upsilon ,\mbox{ or }\Lambda ;%
\end{array}
\\
\begin{array}{c}
\frac{\partial _{\varphi }(\Psi ^{2})}{4\Upsilon ^{2}(h_{4}^{[0]}-\int dy^{3}%
\frac{(\Psi ^{2})^{\diamond }}{4\Upsilon })}[dy^{3}+\frac{\partial _{i}\
\Psi }{\ \partial _{\varphi }\Psi }dx^{i}]- \\
(h_{4}^{[0]}-\int dy^{3}\frac{(\Psi ^{2})^{\diamond }}{4\Upsilon }%
)[dt+(_{1}n_{k}+\ _{2}n_{k}\int dy^{3}\frac{(\Psi ^{\diamond })^{2}}{%
4\Upsilon ^{2}|h_{4}^{[0]}-\int dy^{3}\frac{(\Psi ^{2})^{\diamond }}{%
4\Upsilon }|^{5/2}})dx^{k}], \\
\mbox{ or }%
\end{array}
&
\begin{array}{c}
\mbox{gener.  funct.}\Psi , \\
\mbox{source }\Upsilon ;%
\end{array}
\\
\begin{array}{c}
-\frac{[(\Phi ^{2})^{\diamond }]^{2}}{4|\Lambda \int dy^{3}\Upsilon \lbrack
(\Phi )^{2}]^{\diamond }|\ (h_{4}^{[0]}-\Phi ^{2}/4\Lambda )}[dy^{3}+\frac{%
\partial _{i}[\int dy^{3}\ \ \Upsilon \ (\Phi ^{2})^{\diamond }]}{\ \Upsilon
\ (\Phi ^{2})^{\diamond }}dx^{i}]- \\
(h_{4}^{[0]}(x^{k})-\frac{\Phi ^{2}}{4\Lambda })[dt+(_{1}n_{k}+\
_{2}n_{k}\int dy^{3}\frac{[(\Phi ^{2})^{\diamond }]^{2}}{|\ 4\Lambda \int
dy^{3}\Upsilon \lbrack (\Phi )^{2}]^{\diamond }|}|h_{4}^{[0]}(x^{k})-\frac{%
\Phi ^{2}}{4\Lambda }|^{-5/2})dx^{k}],%
\end{array}
&
\begin{array}{c}
\mbox{gener.  funct.}\Phi  \\
\mbox{effective }\Lambda \mbox{ for }\Upsilon .%
\end{array}%
\end{array}%
\right.   \notag
\end{eqnarray}%
} If we consider nonholonomic deformations of a primary d-metric $\mathbf{%
\mathring{g}}$ into a target one $\ \widehat{\mathbf{g}}\mathbf{=}[g_{\alpha
}=\eta _{\alpha }\mathring{g}_{\alpha },\ \eta _{i}^{a}\mathring{N}_{i}^{a}]$
with Killing symmetry on $\partial _{t},$ we can re-write all formulas (\ref%
{offdstat}) \ and (\ref{gensolstat}) in terms of $\eta $--polarization
functions ($\eta _{\alpha }$ and $\eta _{i}^{a},$ determined by generation
and integration functions and respective sources) and encoding primary data $%
[\mathring{g}_{\alpha },\mathring{N}_{i}^{a}].$ For intance, $\mathbf{%
\mathring{g}}$ can be chosen for a BH in GR. Off-diagonal nonholonomic
deformations may preserve the singular structure of a primary metric (with
certain possible deformations of the horizons, for certain classes of
solutions), or (for more general classes of solutions) to eliminate the
singular structure, or to change the topology in the resulting target
solutions. In section \ref{sbh}, we analyze explicit examples for
quasiperiodic off-diagonal deformations.

\paragraph{Off-diagonal Levi-Civita stationary configurations: \newline
}

We can impose additional constraints on generating functions and sources in
order to extract solutions with zero torsion. The equations (\ref{lcconda})
can be solved for a special class of generating functions and sources when,
for instance, $\Psi =\check{\Psi}(x^{i},\varphi ),(\partial _{i}\check{\Psi}%
)^{\diamond }=\partial _{i}(\check{\Psi}^{\diamond })$ and $\Upsilon
(x^{i},\varphi )=\Upsilon \lbrack \check{\Psi}]=\check{\Upsilon},$ or $%
\Upsilon =const.$ If such conditions are imposed, the nonlinear symmetries (%
\ref{nsym1a}) result in formulas
\begin{equation*}
\Lambda \ \check{\Psi}^{2}=\check{\Phi}^{2}|\check{\Upsilon}|-\int dy^{3}\
\Phi ^{2}|\check{\Upsilon}|^{\diamond },\check{\Phi}^{2}=-4\Lambda \check{h}%
_{4}({r,\theta },\varphi ),\check{\Psi}^{2}=\int d\varphi \ \check{\Upsilon}(%
{r,\theta },\varphi )\check{h}_{4}^{\diamond }({r,\theta },\varphi ),
\end{equation*}%
where $h_{4}=\check{h}_{4}({r,\theta },\varphi )$ can be considered also as
generating function. For such LC--configurations, we find some functions $%
\check{A}(r,\theta ,\varphi )$ and $n(r,\theta )$ when the N--connection
coefficients are
\begin{equation*}
w_{i}=\partial _{i}\check{A}=\left\{
\begin{array}{c}
\frac{\partial _{i}(\int d\varphi \check{\Upsilon}\ \check{h}_{4}^{\diamond
}{}])}{\ \ \check{\Upsilon}\ \check{h}_{4}^{\diamond }{}}; \\
\frac{\partial _{i}\check{\Psi}}{\check{\Psi}^{\diamond }}; \\
\frac{\partial _{i}[\int dy^{3}\ \ \check{\Upsilon}(\check{\Phi}%
^{2})^{\diamond }]}{\ \check{\Upsilon}(\check{\Phi}^{2})^{\diamond }};%
\end{array}%
\right. \mbox{ and }n_{k}=\check{n}_{k}=\partial _{k}n(x^{i}).
\end{equation*}%
In result, we can construct new classes of off-diagonal stationary solutions
in GR defined as subclasses of solutions (\ref{gensolstat}) with zero
torsion, {\small
\begin{eqnarray}
ds^{2} &=&e^{\ \psi (x^{k})}[(dx^{1})^{2}+(dx^{2})^{2}]  \label{lcsolstat} \\
&&+\left\{
\begin{array}{cc}
\begin{array}{c}
-\frac{(\check{h}_{4}^{\diamond }{})^{2}}{|\int d\varphi \ \check{\Upsilon}%
\check{h}_{4}^{\diamond }|\ \check{h}_{4}}[dy^{3}+(\partial _{i}\check{A}%
)dx^{i}]+\check{h}_{4}\left[ dt+(\partial _{k}n)dx^{k}\right] , \\
\mbox{ or }%
\end{array}
&
\begin{array}{c}
\mbox{gener.  funct.}\check{h}_{4}, \\
\mbox{ source }\check{\Upsilon},\mbox{ or }\Lambda ;%
\end{array}
\\
\begin{array}{c}
-\frac{\partial _{\varphi }(\check{\Psi}^{2})}{4\check{\Upsilon}%
^{2}(h_{4}^{[0]}-\int dy^{3}\frac{(\check{\Psi}^{2})^{\diamond }}{4\check{%
\Upsilon}})}[dy^{3}+(\partial _{i}\check{A})dx^{i}]+(h_{4}^{[0]}-\int dy^{3}%
\frac{(\check{\Psi}^{2})^{\diamond }}{4\check{\Upsilon}})\left[ dt+(\partial
_{k}n)dx^{k}\right] , \\
\mbox{ or }%
\end{array}
&
\begin{array}{c}
\mbox{gener.  funct.}\check{\Psi}, \\
\mbox{source }\check{\Upsilon};%
\end{array}
\\
-\frac{[(\check{\Phi}^{2})^{\diamond }]^{2}}{4|\Lambda \int dy^{3}\Upsilon (%
\check{\Phi}^{2})^{\diamond }|\ (h_{4}^{[0]}-\Phi ^{2}/4\Lambda )}%
[dy^{3}+(\partial _{i}\check{A})dx^{i}]+(h_{4}^{[0]}-\frac{\check{\Phi}^{2}}{%
4\Lambda })\left[ dt+(\partial _{k}n)dx^{k}\right] , &
\begin{array}{c}
\mbox{gener.  funct.}\ \check{\Phi} \\
\mbox{effective }\Lambda \mbox{ for }\check{\Upsilon}.%
\end{array}%
\end{array}%
\right.   \notag
\end{eqnarray}%
}Such stationary metrics are generic off-diagonal and define new classes of
solutions different, for instance, from the Kerr metric (defined by rotation
coordinates, or other equivalent ones) if the anholonomy coefficients $%
C_{\alpha \beta }^{\gamma }=\{C_{ia}^{b}=\partial _{a}N_{i}^{b},C_{ji}^{a}=%
\mathbf{e}_{j}N_{i}^{a}-\mathbf{e}_{i}N_{j}^{a}\},$ see formulas (\ref{anhr}%
), are not zero for $N_{i}^{3}=\partial _{i}\check{A}$ and $%
N_{k}^{4}=\partial _{k}n.$ We can analyze certain nonholonomic
configurations determined, for instance, by data $(\check{\Upsilon},\check{%
\Psi},h_{4}^{[0]},\check{n}_{k}),$ when $w_{i}=\partial _{i}\check{A}%
\rightarrow 0$ and $\partial _{k}n\rightarrow 0$ (or $0$ values are
considered as certain additional constraints).

\subsubsection{Off--diagonal cosmological solutions}

\label{ascs}

In this paper, for simplicity, there are considered solutions $g_{\alpha
\beta }(x^{i},t)$ with Killing symmetry on $\partial _{3},$ i.e. with $%
\omega =1$ in (\ref{data1c}), which allows us to generate exact solutions in
explicit form. Solutions depending on all spacetime coordinates, $g_{\alpha
\beta }(x^{i},y^{3},t),$ can be constructed for nontrivial vertical
conformal factors $\omega (x^{i},y^{a}),$ see details and examples in \cite%
{vpcqg01,vjmp05,vmon06,vijgmmp07,vijgmmp11,vgrg12,vcsf12,vepjp12,vepjc13,vepjc14a,gvvepjc14,gvvcqg15}%
. It should be noted that if certain classes of off-diagonal solutions for
such nonholonomic cosmological configurations have been constructed in
explicit form, we can impose additional nonholonomic constraints, or limits
with necessary smooth classes of functions, when $g_{\alpha \beta
}(x^{i},t)\approx g_{\alpha \beta }(t)$ are related to Bianchi type, or
FLRW, like cosmological metrics.

\paragraph{Nontrivial components of the Ricci d-tensor for nonholonomic
cosmological configurations: \newline
}

Let us consider d-metric data (\ref{data1c}) with $\omega =1$ in order to
compute the N-adapted and nontrivial coefficients $\widehat{\mathbf{D}}=\{%
\widehat{\mathbf{\Gamma }}_{\ \alpha \beta }^{\gamma }\}$ (\ref{candcon})
and $\widehat{\mathbf{R}}_{\alpha \beta }$ (\ref{candricci}). For nontrivial
sources $[~\ _{h}\overline{\Upsilon }(x^{i})\delta _{j}^{i},\overline{%
\Upsilon }(x^{i},t)\delta _{b}^{a}]$ (\ref{dsourcparam}), we obtain from the
modified Einstein equations (\ref{qgreq}) a system of nonlinear PDEs%
\footnote{%
with partial derivatives $\partial _{t}q=$ $\partial _{4}q=q^{\ast }$ and $%
\partial _{i}q=(\partial _{1}q=q^{\bullet },\partial _{2}q=q^{\prime })$}
\begin{eqnarray}
\widehat{R}_{1}^{1} &=&\widehat{R}_{2}^{2}=\frac{1}{2g_{1}g_{2}}[\frac{%
g_{1}^{\bullet }g_{2}^{\bullet }}{2g_{1}}+\frac{\left( g_{2}^{\bullet
}\right) ^{2}}{2g_{2}}-g_{2}^{\bullet \bullet }+\frac{g_{1}^{\prime
}g_{2}^{\prime }}{2g_{2}}+\frac{(g_{1}^{\prime })^{2}}{2g_{1}}-g_{1}^{\prime
\prime }]=-\ ~\ _{h}\overline{\Upsilon },  \label{eq1b} \\
\widehat{R}_{3}^{3} &=&\widehat{R}_{4}^{4}=\frac{1}{2\overline{h}_{3}%
\overline{h}_{4}}[\frac{\left( \overline{h}_{3}^{\ast }\right) ^{2}}{2%
\overline{h}_{3}}+\frac{\overline{h}_{3}^{\ast }\overline{h}_{4}^{\ast }}{2%
\overline{h}_{4}}-\overline{h}_{3}^{\ast \ast }]=-\overline{\Upsilon }
\label{eq2b} \\
\widehat{R}_{3k} &=&\frac{\overline{h}_{3}}{2\overline{h}_{4}}\overline{n}%
_{k}^{\ast \ast }+(\frac{3}{2}\overline{h}_{3}^{\ast }-\frac{\overline{h}_{3}%
}{\overline{h}_{4}}\overline{h}_{4}^{\ast })\frac{\overline{n}_{k}^{\ast }}{2%
\overline{h}_{4}}=0,  \label{eq3b} \\
\widehat{R}_{4k} &=&-\frac{\overline{w}_{k}}{2\overline{h}_{3}}[\frac{\left(
\overline{h}_{3}^{\ast }\right) ^{2}}{2\overline{h}_{3}}+\frac{\overline{h}%
_{3}^{\ast }\overline{h}_{4}^{\ast }}{2\overline{h}_{4}}-\overline{h}%
_{3}^{\ast \ast }]+\frac{\overline{h}_{3}^{\ast }}{4\overline{h}_{3}}(\frac{%
\partial _{k}\overline{h}_{3}}{\overline{h}_{3}}+\frac{\partial _{k}%
\overline{h}_{4}}{\overline{h}_{4}})-\frac{\partial _{k}\overline{h}%
_{3}^{\ast }}{2\overline{h}_{3}}=0.  \label{eq4b}
\end{eqnarray}%
These equations can be transformed, respectively, into the system (\ref{eq1a}%
)-(\ref{eq4a}) if $~$%
\begin{eqnarray*}
\ _{h}\overline{\Upsilon }(x^{i}) &\rightarrow &\ _{h}\Upsilon (x^{i}),%
\newline
\overline{\Upsilon }(x^{i},t)\rightarrow \Upsilon (x^{i},y^{3}=\varphi ),%
\overline{h}_{3}(x^{i},t)\rightarrow h_{4}(x^{i},\varphi ),\overline{h}%
_{4}(x^{i},t)\rightarrow h_{3}(x^{i},\varphi ), \\
\overline{h}_{3}^{\ast }(x^{i},t) &\rightarrow &h_{4}^{\diamond
}(x^{i},\varphi ),\overline{h}_{4}^{\ast }(x^{i},t)\rightarrow
h_{3}^{\diamond }(x^{i},\varphi ),\newline
\overline{w}_{k}(x^{i},t)\rightarrow n_{k}(x^{i},\varphi ),\overline{n}%
_{k}(x^{i},t)\rightarrow w_{k}(x^{i},\varphi )\mbox{ etc.}
\end{eqnarray*}
The AFDM allows to redefine the procedure considered in the previous
subsection for stationary nonholonomic configurations in order to generate
solutions with explicit dependence on time like coordinate.

\paragraph{Extracting torsionless locally anisotropic cosmological
configurations: \newline
}

The LC--conditions (\ref{lccond}) for data (\ref{data1c}) transform into
\begin{equation}
\partial _{t}\overline{w}_{i}=(\partial _{i}-\overline{w}_{i}\partial
_{t})\ln \sqrt{|\overline{h}_{4}|},(\partial _{i}-\overline{w}_{i}\partial
_{t})\ln \sqrt{|\overline{h}_{3}|}=0,\partial _{k}\overline{w}_{i}=\partial
_{i}\overline{w}_{k},\partial _{t}\overline{n}_{i}=0,\partial _{i}\overline{n%
}_{k}=\partial _{k}\overline{n}_{i}.  \label{lccondb}
\end{equation}%
Such nonlinear first order PDEs can be solved in explicit form by imposing
additional nonholonomic constraints on cosmological d--metrics and
N-coefficients of (modified) Einstein equations.

\paragraph{Decoupling of nonlinear PDEs for off-diagonal cosmological
solutions: \newline
}

The system of nonlinear PDE (\ref{eq1b})--(\ref{eq4b}) can be decoupled and
integrated following the AFDM. Let us introduce the coefficients
\begin{eqnarray}
\overline{\alpha }_{i} &=&(\partial _{t}\overline{h}_{3})\ (\partial _{i}%
\overline{\varpi }),\ \overline{\beta }=(\partial _{t}\overline{h}_{3})\
(\partial _{t}\overline{\varpi }),\ \overline{\gamma }=\partial _{t}\left(
\ln |\overline{h}_{3}|^{3/2}/|\overline{h}_{4}|\right) ,  \label{abc} \\
&&\mbox{ where }\overline{\varpi }={\ln |\partial _{t}}\overline{{h}}{_{3}/%
\sqrt{|\overline{h}_{3}\overline{h}_{4}|}|}.  \label{genf1b}
\end{eqnarray}%
For $\partial _{t}h_{a}\neq 0$ and $\partial _{t}\varpi \neq 0,$\footnote{%
we have to consider other special methods for generating solutions if such
conditions are not satisfied} we rewrite the equations in the form
\begin{eqnarray}
\psi ^{\bullet \bullet }+\psi ^{\prime \prime } &=&2\ ~\ _{h}\overline{%
\Upsilon }  \label{e1} \\
{\overline{\varpi }}^{\ast }\ \overline{h}_{3}^{\ast } &=&2\overline{h}_{3}%
\overline{h}_{4}\overline{\Upsilon }  \label{e2} \\
\overline{n}_{i}^{\ast \ast }+\overline{\gamma }\overline{n}_{i}^{\ast }
&=&0,  \label{e3} \\
\overline{\beta }\overline{w}_{i}-\overline{\alpha }_{i} &=&0.  \label{e4}
\end{eqnarray}%
We can integrate this system for any generating function $\overline{\Psi }%
(x^{i},t):=e^{\overline{\varpi }}$ and sources $~\ _{h}\overline{\Upsilon }%
(x^{i})$ and $\overline{\Upsilon }(x^{k},t).$

\paragraph{Nonlinear symmetries for generating functions and sources with
effective cosmological constant: \newline
}

The system of two equations (\ref{genf1b}) and (\ref{e2}) relates four
functions ($\overline{h}_{3},\overline{h}_{4},\overline{\Upsilon },$ and $%
\overline{\Psi })$ which emphasizes an important nonlinear symmetry for
locally anisotropic cosmological solutions and respective generating
functions, $(\overline{\Psi },\overline{\Upsilon })\iff (\overline{\Phi }%
,\Lambda ),$ when
\begin{equation}
\overline{\Lambda }(\ \overline{\Psi }^{2})^{\ast }=|\overline{\Upsilon }|(%
\overline{\Phi }^{2})^{\ast },\mbox{
or  }\overline{\Lambda }\overline{\ \Psi }^{2}=\overline{\Phi }^{2}|%
\overline{\Upsilon }|-\int dt\ \overline{\Phi }^{2}|\overline{\Upsilon }%
|^{\ast }.  \label{nsym1b}
\end{equation}%
This allows to introduce a new generating function $\overline{\Phi }({x}%
^{i},t)$ and an (effective) cosmological constant $\overline{\Lambda }\neq 0$%
, which can be applyied for generating exact off-diagonal solutions in
explicit forms.

\paragraph{Off-diagonal metrics and N--coefficients for locally anisotropic
cosmological solutions: \newline
}

Integrating "step by step" the system (\ref{e1})--(\ref{e4}), we generate
exact solutions of the \ nonlinear PDEs (\ref{eq1b})--(\ref{eq4b}), i.e. for
the (modified) Einstein equations (\ref{qgreq}), determined by d--metric
coefficients,
\begin{eqnarray}
\ g_{i} &=&e^{\ \psi (x^{k})}\mbox{ as a solution of 2-d Poisson eqs. }\psi
^{\bullet \bullet }+\psi ^{\prime \prime }=2~\ _{h}\overline{\Upsilon };
\notag \\
g_{3} &=&\overline{h}_{3}({x}^{i},t)=h_{3}^{[0]}(x^{k})-\int dt\frac{(%
\overline{\Psi }^{2})^{\ast }}{4\overline{\Upsilon }}=h_{3}^{[0]}(x^{k})-%
\overline{\Phi }^{2}/4\overline{\Lambda };  \notag \\
g_{4} &=&\overline{h}_{4}({x}^{i},t)=-\frac{(\overline{\Psi }^{2})^{\ast }}{4%
\overline{\Upsilon }^{2}\overline{h}_{3}}=-\frac{(\overline{\Psi }%
^{2})^{\ast }}{4\overline{\Upsilon }^{2}(h_{3}^{[0]}(x^{k})-\int dt(%
\overline{\Psi }^{2})^{\ast }/4\overline{\Upsilon })}  \label{offdcosm} \\
&=&-\frac{[(\overline{\Phi }^{2})^{\diamond }]^{2}}{4\overline{h}_{3}|%
\overline{\Lambda }\int dt\overline{\Upsilon }[\overline{\Phi }%
^{2}]^{\diamond }|}=-\frac{[(\overline{\Phi }^{2})^{\ast }]^{2}}{%
4[h_{3}^{[0]}(x^{k})-\overline{\Phi }^{2}/4\overline{\Lambda }]|\int dt\
\overline{\Upsilon }[\overline{\Phi }^{2}]^{\ast }|};  \notag
\end{eqnarray}%
and N--connection coefficients,
\begin{eqnarray*}
N_{k}^{3} &=&\overline{n}_{k}({x}^{i},t)=\ _{1}n_{k}(x^{i})+\
_{2}n_{k}(x^{i})\int dt\frac{(\overline{\Psi }^{\ast })^{2}}{\overline{%
\Upsilon }^{2}|h_{3}^{[0]}(x^{i})-\int dt(\overline{\Psi }^{2})^{\ast }/4%
\overline{\Upsilon }|^{5/2}} \\
&=&\ _{1}n_{k}(x^{i})+\ _{2}n_{k}(x^{i})\int dt\frac{(\overline{\Phi }^{\ast
})^{2}}{4|\overline{\Lambda }\int dt\overline{\Upsilon }[\overline{\Phi }%
^{2}]^{\ast }||\overline{h}_{3}|^{5/2}}; \\
N_{i}^{4} &=&\overline{w}_{i}({x}^{i},t)=\frac{\partial _{i}\ \overline{\Psi
}}{\ \overline{\Psi }^{\ast }}=\frac{\partial _{i}\ \overline{\Psi }^{2}}{(%
\overline{\Psi }^{2})^{\ast }}\ =\frac{\partial _{i}[\int dt\ \overline{%
\Upsilon }(\ \overline{\Phi }^{2})^{\ast }]}{\overline{\Upsilon }(\
\overline{\Phi }^{2})^{\ast }},\ \
\end{eqnarray*}%
where $h_{3}^{[0]}(x^{k}),$ $\ _{1}n_{k}(x^{i}),$ and $\ _{2}n_{k}(x^{i})$
are integration functions encoding various possible sets of (non)
commutative parameters and integration constants. We can chose some
generating data $(\overline{\Psi },\overline{\Upsilon }),$ or $(\overline{%
\Phi },\overline{\Lambda }),$ related by nonlinear differential / integral
transforms (\ref{nsym1b}), and respective integration functions in explicit
form following certain topology/ symmetry / asymptotic conditions for some
classes of exact / parametric cosmological solutions. The coefficients (\ref%
{offdcosm}) define generic off-diagonal cosmological solutions if the
corresponding anholonomy coefficients $C_{\alpha \beta }^{\gamma }(x^{i},t)$
(\ref{anhr}) are not trivial. Such locally cosmological solutions are with
nontrivial nonholonomically induced d-torsion (\ref{dtors}) with N-adapted
coefficients which can be computed in explicit form. In order to generate as
particular cases some well-known cosmological FLRW, or Bianchi, type
metrics, we have to consider data of type $(\overline{\Psi }(t),\overline{%
\Upsilon }(t)),$ or $(\overline{\Phi }(t),\overline{\Lambda }),$ with
integration functions which allow frame/ coordinate transforms to respective
(off-) diagonal configurations $g_{\alpha \beta }(t).$

\paragraph{Quadratic line elements for off-diagonal locally anisotropic
cosmological configurations: \newline
}

Any coefficient $\overline{h}_{3}=h_{3}^{[0]}(x^{k})-\overline{\Phi }%
^{2}/4\Lambda ,$ $\overline{h}_{3}^{\ast }\neq 0,$ can be considered also as
a generating function. Using (\ref{offdcosm}), we find $\overline{\Phi }%
^{2}=-4\overline{\Lambda }\overline{h}_{3}({r,\theta },t)$ transforming (\ref%
{nsym1b}) in $(\overline{\Psi }^{2})^{\ast }=\int dt\ \overline{\Upsilon }({%
r,\theta },t)\overline{h}_{3}^{\ast }({r,\theta },t).$ AIntroducing such
values into respective formulas for $\overline{h}_{a}$ and $\overline{%
\Upsilon }$ in (\ref{offdstat}), we construct locally anisotropic
cosmological solutions of type \ d--metric (\ref{dm}), with data (\ref%
{data1c}), {\small
\begin{eqnarray}
ds^{2} &=&e^{\ \psi (x^{k})}[(dx^{1})^{2}+(dx^{2})^{2}]  \label{gensolcosm}
\\
&&+\left\{
\begin{array}{cc}
\begin{array}{c}
+\overline{h}_{3}[dy^{3}+(\ _{1}n_{k}+4\ _{2}n_{k}\int dt\frac{(\overline{h}%
_{3}^{\ast }{})^{2}}{|\int dy^{4}\overline{\Upsilon }\overline{h}_{3}^{\ast
}|\ (\overline{h}_{3})^{5/2}})dx^{k}] \\
-\frac{(\overline{h}_{3}^{\ast }{})^{2}}{|\int dt\ \overline{\Upsilon }%
\overline{h}_{3}^{\ast }|\ \overline{h}_{3}}[dt+\frac{\partial _{i}(\int dt%
\overline{\Upsilon }\ \overline{h}_{3}^{\ast }{})}{\ \ \overline{\Upsilon }%
\overline{h}_{3}^{\ast }\ {}}dx^{i}], \\
\mbox{ or }%
\end{array}
&
\begin{array}{c}
\mbox{gener.  funct.}\overline{h}_{3}, \\
\mbox{ source }\overline{\Upsilon },\mbox{ or }\overline{\Lambda };%
\end{array}
\\
\begin{array}{c}
+(h_{3}^{[0]}-\int dt\frac{(\overline{\Psi }^{2})^{\ast }}{4\overline{%
\Upsilon }})[dy^{3}+(_{1}n_{k}+\ _{2}n_{k}\int dt\frac{(\overline{\Psi }%
^{\ast })^{2}}{4\overline{\Upsilon }^{2}|h_{3}^{[0]}-\int dy^{4}\frac{(%
\overline{\Psi }^{2})^{\ast }}{4\overline{\Upsilon }}|^{5/2}})dx^{k}] \\
-\frac{(\overline{\Psi }^{2})^{\ast }}{4\overline{\Upsilon }%
^{2}(h_{3}^{[0]}-\int dt\frac{(\overline{\Psi }^{2})^{\ast }}{4\overline{%
\Upsilon }})}[dt+\frac{\partial _{i}\ \overline{\Psi }}{\ \overline{\Psi }%
^{\ast }}dx^{i}], \\
\mbox{ or }%
\end{array}
&
\begin{array}{c}
\mbox{gener.  funct.}\overline{\Psi }, \\
\mbox{source }\Upsilon ;%
\end{array}
\\
\begin{array}{c}
+(h_{3}^{[0]}-\frac{\overline{\Phi }^{2}}{4\overline{\Lambda }}%
)[dy^{3}+(_{1}n_{k}+\ _{2}n_{k}\int dt\frac{[(\overline{\Phi }^{2})^{\ast
}]^{2}}{|\ 4\overline{\Lambda }\int dy^{4}\overline{\Upsilon }(\overline{%
\Phi }^{2})^{\ast }|}|h_{4}^{[0]}(x^{k})-\frac{\overline{\Phi }^{2}}{4%
\overline{\Lambda }}|^{-5/2})dx^{k}] \\
-\frac{[(\overline{\Phi }^{2})^{\ast }]^{2}}{|\ 4\overline{\Lambda }\int
dy^{4}\overline{\Upsilon }(\overline{\Phi }^{2})^{\ast }|(h_{3}^{[0]}-%
\overline{\Phi }^{2}/4\overline{\Lambda })}[dt+\frac{\partial _{i}[\int dt\
\ \overline{\Upsilon }(\overline{\Phi }^{2})^{\ast }]}{\ \overline{\Upsilon }%
(\overline{\Phi }^{2})^{\ast }}dx^{i}],%
\end{array}
&
\begin{array}{c}
\mbox{gener.  funct.}\overline{\Phi } \\
\mbox{effective }\overline{\Lambda }\mbox{ for }\overline{\Upsilon }.%
\end{array}%
\end{array}%
\right.   \notag
\end{eqnarray}%
} Such solutions posses a Killing symmetry on $\partial _{3}$ and can be
re-written in terms of $\eta $--polarization function functions for target
locally anistropic cosmological metrics $\ \widehat{\mathbf{g}}\mathbf{=}%
[g_{\alpha }=\eta _{\alpha }\mathring{g}_{\alpha },\ \eta _{i}^{a}\mathring{N%
}_{i}^{a}]$ encoding primary cosmological data $[\mathring{g}_{\alpha },%
\mathring{N}_{i}^{a}].$ In section \ref{scc}, we analyze explicit examples
for quasiperiodic off-diagonal nonholonomic deformations of cosmological
solutions.

\paragraph{Off-diagonal Levi-Civita locally anisotropic cosmological
configurations: \newline
}

We can extract cosmological spacetimes in GR, with zero torsion, if the
equations (\ref{lccondb}) are solved for a special class of generating
functions and sources when, for instance, $\overline{\Psi }=\overline{\check{%
\Psi}}(x^{i},t),$ when $(\partial _{i}\overline{\check{\Psi}})^{\ast
}=\partial _{i}(\overline{\check{\Psi}}^{\ast })$ and $\overline{\Upsilon }%
(x^{i},t)=\overline{\Upsilon }[\overline{\check{\Psi}}]=\overline{\check{%
\Upsilon}},$ or $\overline{\Upsilon }=const.$ For such classes of generting
functions and sources, the nonlinear symmetries (\ref{nsym1b}) result in
formulas
\begin{equation*}
\overline{\Lambda }\ \overline{\check{\Psi}}^{2}=\overline{\check{\Phi}}^{2}|%
\overline{\check{\Upsilon}}|-\int dt\ \overline{\check{\Phi}}^{2}|\overline{%
\check{\Upsilon}}|^{\ast },\overline{\check{\Phi}}^{2}=-4\overline{\Lambda }%
\overline{\check{h}_{3}}({r,\theta },t),\overline{\check{\Psi}}^{2}=\int dt\
\overline{\check{\Upsilon}}({r,\theta },t)\overline{\check{h}}_{3}^{\ast }({%
r,\theta },t),
\end{equation*}%
where $\overline{h}_{4}=\overline{\check{h}}_{4}(x^{i},t)$ can be considered
also as generating function for cosmological solutions. For such
LC--configurations, there are some functions $\overline{\check{A}}(x^{i},t)$
and $n(x^{i})$ when the N--connection coefficients are
\begin{equation*}
\overline{n}_{k}=\overline{\check{n}}_{k}=\partial _{k}\overline{n}(x^{i})%
\mbox{ and }\overline{w}_{i}=\partial _{i}\overline{\check{A}}=\left\{
\begin{array}{c}
\frac{\partial _{i}(\int dt\ \overline{\check{\Upsilon}}\ \overline{\check{h}%
}_{3}^{\ast }{}{}])}{\ \ \overline{\check{\Upsilon}}\ \overline{\check{h}}%
_{3}^{\ast }{}}; \\
\frac{\partial _{i}\overline{\check{\Psi}}}{\overline{\check{\Psi}}^{\ast }};
\\
\frac{\partial _{i}[\int dt\ \ \overline{\check{\Upsilon}}(\overline{\check{%
\Phi}}^{2})^{\ast }]}{\ \ \overline{\check{\Upsilon}}(\overline{\check{\Phi}}%
^{2})^{\ast }};%
\end{array}%
\right.
\end{equation*}%
In result, we can construct new classes of off-diagonal locally anisotropic
cosmological solutions in GR defined as subclasses of solutions (\ref%
{gensolcosm}) with zero torsion, {\small
\begin{eqnarray}
ds^{2} &=&e^{\ \psi (x^{k})}[(dx^{1})^{2}+(dx^{2})^{2}]  \label{lcsolcosm} \\
&&+\left\{
\begin{array}{cc}
\begin{array}{c}
+\overline{\check{h}}_{3}\left[ dy^{3}+(\partial _{k}\overline{n})dx^{k}%
\right] -\frac{(\ \overline{\check{h}}_{3}^{\ast }{}{}{})^{2}}{|\int dt\ \
\overline{\check{\Upsilon}}\ \overline{\check{h}}_{3}^{\ast }{}|\ \
\overline{\check{h}}_{3}}[dt+(\partial _{i}\overline{\check{A}})dx^{i}], \\
\mbox{ or }%
\end{array}
&
\begin{array}{c}
\mbox{gener.  funct.}\overline{\check{h}}_{3}, \\
\mbox{ source }\overline{\check{\Upsilon}},\mbox{ or }\overline{\Lambda };%
\end{array}
\\
\begin{array}{c}
+(h_{3}^{[0]}-\int dt\frac{(\overline{\check{\Psi}}^{2})^{\ast }}{4\overline{%
\check{\Upsilon}}})[dy^{3}+(\partial _{k}\overline{n})dx^{k}]-\frac{(%
\overline{\check{\Psi}}^{2})^{\ast }}{4\overline{\check{\Upsilon}}%
^{2}(h_{3}^{[0]}-\int dt\frac{(\overline{\check{\Psi}}^{2})^{\ast }}{4%
\overline{\check{\Upsilon}}})}[dt+(\partial _{i}\overline{\check{A}})dx^{i}],
\\
\mbox{ or }%
\end{array}
&
\begin{array}{c}
\mbox{gener.  funct.}\overline{\check{\Psi}}, \\
\mbox{source }\overline{\check{\Upsilon}};%
\end{array}
\\
+(h_{3}^{[0]}-\frac{\overline{\check{\Phi}}^{2}}{4\overline{\Lambda }}%
)[dy^{3}+(\partial _{k}\overline{n})dx^{k}]-\frac{[(\overline{\check{\Phi}}%
^{2})^{\ast }]^{2}}{|\ 4\overline{\Lambda }\int dy^{4}\overline{\check{%
\Upsilon}}(\overline{\check{\Phi}}^{2})^{\ast }|(h_{3}^{[0]}-\overline{%
\check{\Phi}}^{2}/4\overline{\Lambda })}[dt+(\partial _{i}\overline{\check{A}%
})dx^{i}], &
\begin{array}{c}
\mbox{gener.  funct.}\ \overline{\check{\Phi}} \\
\mbox{effective }\Lambda \mbox{ for }\overline{\check{\Upsilon}}.%
\end{array}%
\end{array}%
\right.  \notag
\end{eqnarray}%
}Such cosmological metrics are generic off-diagonal and define new classes
of solutions if the anholonomy coefficients $C_{\alpha \beta }^{\gamma }$,
see formulas (\ref{anhr}), are not zero for $N_{k}^{3}=\partial _{k}%
\overline{n}$ and $N_{i}^{4}=\partial _{i}\overline{\check{A}}.$ We can
analyze certain nonholonomic cosmological configurations determined, for
instance, by data $(\overline{\check{\Upsilon}},\overline{\check{\Psi}}%
,h_{3}^{[0]},\overline{\check{n}}_{k}),$ when $\partial _{k}\overline{n}%
\rightarrow 0$ and $\overline{w}_{i}=\partial _{i}\overline{\check{A}}%
\rightarrow 0$ (we notat that $0$ values can be fixed by certain additional
constraints). Choosing data $(\overline{\check{\Upsilon}}(t),\overline{%
\check{\Psi}}(t),h_{3}^{[0]}=const,\overline{\check{n}}_{k}=const),$ we can
generate (off-) diagonal metrics of Bianchi, or FLRW, types and
generalizations to other type configurations $g_{\alpha \beta }(t)$ in GR.

\subsection{General polarization functions for stationary and cosmological
solutions}

\label{ass3}

Quadratic linear elements for exact off-diagonal solutions constructed in
previous section can be parameterized in the form (\ref{dme}) [in terms of
polarization functions $\eta _{\alpha }=(\eta _{i},\eta _{a})$ and $\eta
_{i}^{a}$] defining nonholonomic deformations of a prime d-metric, $\mathbf{%
\mathring{g},}$ into a target d-metric, $\widehat{\mathbf{g}}=[g_{\alpha
}=\eta _{\alpha }\mathring{g}_{\alpha },\ \eta _{i}^{a}\mathring{N}%
_{i}^{a}]\rightarrow \mathbf{\mathring{g}}$. Such parameterizations are
useful for analyzing possible physical implications of general off-diagonal
deformations of some physically important solutions when, for instance, $%
\mathbf{\mathring{g}}$ is taken for a standard black hole, BH, or
cosmological, solution in GR, or a MGT.

\subsubsection{Stationary polarization functions}

\label{ass31}

We write the stationary d-metrics in the forms%
\begin{eqnarray*}
&&ds^{2} =\eta _{1}(r,\theta )\mathring{g}_{1}(r,\theta )[dx^{1}(r,\theta
)]^{2}+\eta _{2}(r,\theta )\mathring{g}_{2}(r,\theta )[dx^{2}(r,\theta
)]^{2}+ \\
&&\eta _{3}(r,\theta ,\varphi )\mathring{g}_{3}(r,\theta )[d\varphi +\eta
_{i}^{3}(r,\theta ,\varphi )\mathring{N}_{i}^{3}(r,\theta )dx^{i}(r,\theta
)]^{2}+\eta _{4}(r,\theta ,\varphi )\mathring{g}_{4}(r,\theta )[dt+\eta
_{i}^{4}(r,\theta ,\varphi )\mathring{N}_{i}^{4}(r,\theta )dx^{i}(r,\theta
)]^{2},
\end{eqnarray*}%
where data $[\mathring{g}_{i}(r,\theta ),\mathring{g}_{a}=\mathring{h}%
_{a}(r,\theta );\mathring{N}_{k}^{3}=\mathring{w}_{k}(r,\theta ),\mathring{N}%
_{k}^{4}=\mathring{n}_{k}(r,\theta )]$ define a BH metric diagonalizable by
frame/ coordinate transforms.

The polarization functions for a general target stationary d--metric (\ref%
{gensolstat}) can be parameterized{\small
\begin{eqnarray*}
\eta _{i} &=&e^{\ \psi (x^{k})}/\mathring{g}_{i};\eta _{3}=-\frac{4[(|\eta
_{4}\mathring{h}_{4}|^{1/2})^{\diamond }]^{2}}{\mathring{h}_{3}|\int
d\varphi \ ^{qp}\Upsilon (\ ^{qp}\eta _{4}\mathring{h}_{4})^{\diamond }|\ }%
;\eta _{4}=\eta _{4}(r,\theta ,\varphi )\mbox{ as a generating
function}; \\
\eta _{i}^{3} &=&\frac{\partial _{i}\ \int d\varphi \ \Upsilon (\eta _{4}\
\mathring{h}_{4})^{\diamond }}{\mathring{w}_{i}\ \Upsilon \ (\eta _{4}\
\mathring{h}_{4})^{\diamond }};\ \eta _{k}^{4}=\frac{\ _{1}n_{k}}{\mathring{n%
}_{k}}+16\ \ \frac{\ _{2}n_{k}}{\mathring{n}_{k}}\int d\varphi \frac{\left(
\lbrack (\eta _{4}\mathring{h}_{4})^{-1/4}]^{\diamond }\right) ^{2}}{|\int
dy^{3}\Upsilon (\eta _{4}\ \mathring{h}_{4})^{\diamond }|\ }.
\end{eqnarray*}%
}Other type generating functions with nonlinear symmetries (\ref{nsym1a})
are functionals of $\eta _{3}(r,\theta ,\varphi )$ and data for the prime
d-metric,{\small
\begin{equation*}
\ \eta _{4}=-\ \Phi ^{2}({r,\theta },\varphi )/4\Lambda \mathring{h}_{4}({%
r,\theta },\varphi ),\ (\Psi ^{2})^{\diamond }=-\int d\varphi \ \Upsilon
h_{4}^{\ \diamond }=-\int d\varphi \ \Upsilon ({r,\theta },\varphi )[\eta
_{4}({r,\theta },\varphi )\mathring{h}_{4}({r,\theta },\varphi )]^{\diamond
}.
\end{equation*}%
}

Target stationary off-diagonal metrics (\ref{lcsolstat}) with zero torsion
can be generated by polarization functions
\begin{equation*}
\check{\eta}_{i}=e^{\ \psi (x^{k})}/\mathring{g}_{i};\ \check{\eta}_{3}=-%
\frac{4[(|\check{\eta}_{4}\mathring{h}_{4}|^{1/2})^{\diamond }]^{2}}{%
\mathring{h}_{3}|\int d\varphi \check{\Upsilon}(\check{\eta}_{4}\mathring{h}%
_{4})^{\diamond }|\ };\eta _{4}=\check{\eta}_{4}(r,\theta ,\varphi )%
\mbox{ as a generating
function};\check{\eta}_{i}^{3}=\frac{\partial _{i}\check{A}}{\mathring{w}_{k}%
},\check{\eta}_{k}^{4}=\frac{\ \partial _{k}n}{\mathring{n}_{k}}.
\end{equation*}%
Above formulas can be simplified by choosing the integration functions $%
h_{3}^{[0]}=\mathring{h}_{3}$ and $_{1}n_{k}=\mathring{n}_{k}=\partial
_{k}n. $

\subsubsection{Locally anisotropic cosmological polarization functions}

\label{ass32}

The locally anisotropic cosmological metrics are parameterized in the forms
\begin{eqnarray*}
ds^{2} &=&\overline{\eta }_{1}(x^{i},t)\mathring{g}_{1}(x^{i},t)[dx^{1}]^{2}+%
\overline{\eta }_{2}(x^{i},t)\mathring{g}_{2}(x^{i},t)[dx^{1}]^{2}+ \\
&&\overline{\eta }_{3}(x^{i},t)\mathring{h}_{3}(x^{i},t)[dy^{3}+\overline{%
\eta }_{i}^{3}(x^{i},t)\mathring{N}_{i}^{3}(x^{k},t)dx^{i}]^{2}+\overline{%
\eta }_{4}(x^{i},t)\mathring{h}_{4}(x^{i},t)[dt+\overline{\eta }%
_{i}^{4}(x^{k},t)\mathring{N}_{i}^{4}(x^{k},t)dx^{i}]^{2},
\end{eqnarray*}%
where data $[\mathring{g}_{i}(x^{i},t),\mathring{g}_{a}=\mathring{h}%
_{a}(x^{i},t);\mathring{N}_{k}^{3}=\mathring{n}_{k}(x^{i}),\mathring{N}%
_{k}^{4}=\mathring{w}_{k}(x^{i},t)]$ define, in general, a locally
cosmological solution (in particular, it can be a frame/coordinate transform
of the FLRW, or a Bianchi type metric, of a metric $\mathring{g}_{\alpha
\beta }(t)$ in GR). The target d-metrics $\overline{\mathbf{g}}(x^{k},t)$
are characterized by N-adapted coefficients $\overline{g}_{i}(x^{k})=%
\overline{\eta }_{i}\mathring{g}_{i},h_{a}(x^{i},t)=\overline{\eta }_{a}%
\mathring{h}_{a},\overline{N}_{i}^{3}=\overline{\eta }_{i}^{3}(x^{i},t)%
\mathring{N}_{i}^{3}(x^{k},t)=\overline{n}_{i}(x^{k}),\overline{N}_{i}^{4}=%
\overline{\eta }_{i}^{4}(x^{i},t)\mathring{N}_{i}^{4}(x^{k},t)=\overline{w}%
_{i}(x^{k},t).$

The polarization functions for a  target locally anisotropic cosmological
d--metric (\ref{gensolcosm}) can be parameterized{\small
\begin{eqnarray*}
\overline{\eta }_{i} &=&e^{\ \psi }/\mathring{g}_{i};\overline{\eta }_{3}=%
\overline{\eta }_{3}(x^{i},t)\mbox{  as a
generating function};\overline{\eta }_{4}=-\frac{4[(|\overline{\eta }_{3}%
\mathring{h}_{3}|^{1/2})^{\ast }]^{2}}{\mathring{h}_{4}|\int dt\ \overline{%
\Upsilon }(\overline{\eta }_{3}\mathring{h}_{3})^{\ast }|\ }; \\
\overline{\eta }_{i}^{3} &=&\frac{_{1}n_{k}}{\mathring{n}_{k}}+4\frac{\
_{2}n_{k}}{\mathring{n}_{k}}\int dt\frac{\left( [(\overline{\eta }_{3}%
\mathring{h}_{3})^{-1/4}]^{\ast }\right) ^{2}}{|\int dt\ \overline{\Upsilon }%
(\overline{\eta }_{3}\mathring{h}_{3})^{\ast }|\ };\ \overline{\eta }%
_{k}^{4}=\frac{\partial _{i}\ \int dt\ \ \overline{\Upsilon }(\overline{\eta
}_{3}\mathring{h}_{3})^{\ast }}{\overline{\mathring{w}}_{i}\ \overline{%
\Upsilon }(\overline{\eta }_{3}\mathring{h}_{3})^{\ast }}.
\end{eqnarray*}%
}Corresponding generating functions subjected to transforms (\ref{nsym1b})
are \ expressed in terms of polarization coefficient $\overline{\eta }_{4}:\
\overline{\Phi }^{2}=4|\ \overline{\Lambda }[h_{3}^{[0]}(x^{k})-\overline{%
\eta }_{3}({x}^{i},t)\mathring{h}_{3}({x}^{k},t)]|\ ,\ (\overline{\Psi }%
^{2})^{\ast }=-\int dt\ \overline{\Upsilon }\ [\overline{\eta }_{3}(x^{i},t)%
\overline{\mathring{h}}_{3}(x^{i},t)]^{\ast }.$

Target off-diagonal cosmological metrics (\ref{lcsolcosm}) with zero torsion
can be generated by polarization functions%
\begin{equation*}
\overline{\eta }_{i}=e^{\ \psi }/\mathring{g}_{i};\mbox{  generating
function }\overline{\eta }_{3}=\overline{\check{\eta}}_{3}({x}^{i},t);%
\overline{\eta }_{4}=-\frac{4[(|\overline{\check{\eta}}_{3}\mathring{h}%
_{3}|^{1/2})^{\ast }]^{2}}{\mathring{h}_{4}|\int dt\ \overline{\check{%
\Upsilon}}(\ \overline{\check{\eta}}_{3}\mathring{h}_{3})^{\ast }|\ };%
\overline{\eta }_{k}^{3}=(\partial _{k}\overline{n})/\mathring{n}_{k};%
\overline{\eta }_{k}^{4}=\partial _{k}\overline{\check{A}}/\mathring{w}_{k}.
\end{equation*}%
Above formulas can be simplified by choosing the integration functions in
the form $h_{3}^{[0]}=\mathring{h}_{3}$ and $_{1}n_{k}=\mathring{n}%
_{k}=\partial _{k}\overline{n}.$ We use symbols with inverse "hat" following
the conventions on integrability of generating functions which are similar
to proofs of formula (\ref{lcsolstat}).

\end{document}